\documentclass[11pt]{article}

\usepackage{graphicx,epsfig}
\usepackage{multicol}
\usepackage{amsmath,amssymb,cite,color,hyperref}

\newcommand{\sqrts}{\sqrt{s}}
\newcommand{\pp}{$p$-$p$}
\newcommand{\ppbar}{$p$-$\bar p$}

\newcommand{\qqbar}{$q \bar q$}

\newcommand{\pt}{p_{_{\rm T}}}
\newcommand{\ET}{E_{_{\rm T}}}
\newcommand{\ETg}{E_{_{\rm{T}}}^{\gamma}}
\newcommand{\ETgSq}{E_{_{\rm T}}^{\gamma\;2}}
\newcommand{\xT}{x_{_{\rm T}}}

\newcommand{\spps}{Sp$\bar{\rm p}$S}

\newcommand{\jetphox}{{\sc jetphox}}

\newcommand{\lhapdf}{{\sc lhapdf}}

\def\cO#1{{{\cal{O}}}\left(#1\right)}

\def\smallfrac#1#2{\hbox{${{#1}\over {#2}}$}}
\newcommand{\be}{\begin{equation}}
\newcommand{\ee}{\end{equation}}
\newcommand{\bea}{\begin{eqnarray}}
\newcommand{\eea}{\end{eqnarray}}
\newcommand{\bi}{\begin{itemize}}
\newcommand{\ei}{\end{itemize}}
\newcommand{\ben}{\begin{enumerate}}
\newcommand{\een}{\end{enumerate}}
\newcommand{\la}{\left\langle}
\newcommand{\ra}{\right\rangle}

\newcommand{\lp}{\left(}
\newcommand{\rp}{\right)}
\newcommand{\as}{\alpha_s}

\def\frac#1#2{{{#1}\over {#2}}}
\def\gsim{\mathrel{\rlap{\lower4pt\hbox{\hskip1pt$\sim$}}
    \raise1pt\hbox{$>$}}}         %greater than or approx. symbol
\def\lsim{\mathrel{\rlap{\lower4pt\hbox{\hskip1pt$\sim$}}
    \raise1pt\hbox{$<$}}}         %less than or approx. symbol

\newcommand{\draft}[1]{}

\definecolor{grey}{rgb}{0.5,0.5,0.5}

\begin{document}
\hfill {\sf CERN-PH-TH/2011-302}
%\begin{flushright}
%CERN-PH-TH/2011-302\\
%\end{flushright}
\vspace{1.cm}

\begin{center}
{\Large\bf Quantitative constraints on the gluon distribution function\\ in the proton from collider isolated-photon data}
\end{center}
\vspace{0.5cm}

\begin{center}
David~d'Enterria$^{1,2}$ and Juan~Rojo$^3$
\vspace{0.5cm}

{\it $^1$ CERN, PH Department, CH-1211 Geneva 23, Switzerland\\}
{\it $^2$ ICREA \& ICC-UB, Universitat de Barcelona, 08028 Barcelona, Catalonia\\}
{\it $^3$ CERN, PH Department, TH Unit, CH-1211 Geneva 23, Switzerland\\}
\end{center}

\vspace{0.8cm}

\begin{center}
{\bf \large Abstract}
\end{center}
%Isolated photon production in proton-(anti)proton collisions is directly sensitive to the gluon content of the proton. 
%The possibility of including isolated photon data from proton-(anti)proton collisions
%at RHIC, \spps, Tevatron and LHC energies, into global PDF analyses is studied in detail. 
The impact of isolated-photon data from proton-(anti)proton collisions at RHIC, \spps, 
Tevatron and LHC energies, on the parton distribution functions of the proton is studied 
using a recently developed Bayesian reweighting method.
The impact on the gluon density of the 35 existing isolated-$\gamma$ measurements
is quantified using next-to-leading order (NLO) perturbative QCD calculations complemented 
with the NNPDF2.1 parton densities. The NLO predictions are found to describe well
most of the datasets from 200~GeV up to 7~TeV centre-of-mass energies.
%, with a few exceptions that can be traced back to experimentally underestimated uncertainties. 
The isolated-photon spectra recently measured at the LHC are precise enough to constrain the gluon distribution
and lead to a moderate reduction (up to 20\%) of its uncertainties around fractional momenta $x\approx$~0.02. 
As a particular case, we show that the improved gluon density reduces the PDF uncertainty for the Higgs boson
production cross section in the gluon-fusion channel by more than 20\% at the LHC.
We conclude that present and future isolated-photon measurements %from the LHC 
constitute an interesting addition to coming global PDF analyses.

\clearpage

\tableofcontents

%%%%%%%%%%%%%%%%%%%%%%%%%%%%%%%%%%%%%%%%%%%%%%%%%%%%%%%%%%%%%%%%%%%%%%%%%%%%%%%%%%%%%%%%

\section{Introduction}

The accurate determination of the parton distribution functions (PDFs) of the proton in a wide range of
momentum fractions $x$ and energy scales $Q$~\cite{DeRoeck:2011na} is a crucial ingredient 
for precision phenomenology at the Large Hadron Collider (LHC)~\cite{Watt:2011kp,Watt:2012fj}.  
Among all parton distributions, the gluon density $g(x,Q^2)$ is one of the least constrained PDFs since 
it does not couple directly to the photon in deep-inelastic scattering (DIS) measurements of the proton 
$F_2$ structure function~\cite{HERAPDF10:2009wt}. A precise determination of $g(x,Q^2)$ is of paramount importance 
for the LHC physics program both in standard model (SM) as well as in new physics searches. Indeed, gluons 
drive a significant fraction of the scattering processes at the LHC and gluon-gluon fusion is the dominant
channel for the production of the SM Higgs boson~\cite{Dittmaier:2011ti,Dittmaier:2012vm}, top-quark pairs 
or dijets, to mention a few.\\
%It is thus important for the LHC physics program to find new observables to further constrain the gluon PDF.

In global PDF analyses, the gluon density is directly constrained mostly by jet production and indirectly
constrained by scaling violations of $F_2(x,Q^2)$ in DIS and by the momentum sum rule. 
It is thus important to find new collider observables that provide independent information on $g(x,Q^2)$  
and help improve the accuracy of its determination and reduce its associated uncertainties.
Prompt photon production in hadronic collisions, defined as the production of photons not issuing from the
electromagnetic decays of hadrons,
%proton-proton (\pp) collisions and to a lesser extent in proton-antiproton (\ppbar) collisions, 
appears as an excellent observable to determine the gluon PDF, since at leading order (LO)
it probes the gluon directly through the quark-gluon ``Compton'' process 
$qg\rightarrow \gamma q$~\cite{Owens:1986mp,Aurenche:1988vi,Vogelsang:1995bg}.
As a matter of fact, up to about 12 years ago, prompt photon data were used  to constrain $g(x,Q^2)$
in global PDF fits~\cite{Lai:1996mg,Martin:1999ww}, as well as in independent determinations of
the strong coupling constant~\cite{Werlen:1999ij}. However, a series of measurements carried out 
at centre-of-mass energies $\sqrts\approx$~20--40~GeV by the fixed--target E706 
experiment~\cite{Alverson:1993da,Apanasevich:1997hm,Apanasevich:2004dr} showed 
an enhanced $\gamma$ cross section compared to next-to-leading-order (NLO) perturbative quantum chromodynamics (pQCD)
predictions~\cite{Baer:1990ra,Aurenche:1992yc,Gordon:1994ut}. The theoretical cross section deficit was only
partially cured by the inclusion of extra soft-gluon resummation 
contributions~\cite{Laenen:1998qw,Catani:1998tm,Kidonakis:2000gi,Kidonakis:2003bh,deFlorian:2005wf}.
Such data--theory discrepancies at fixed-target energies, together with the availability of more precise 
jet measurements from the Tevatron, lead to abandon the use of inclusive photon data in PDF analyses.\\

\begin{figure}[!htbp]
\centering
\epsfig{width=0.46\textwidth,figure=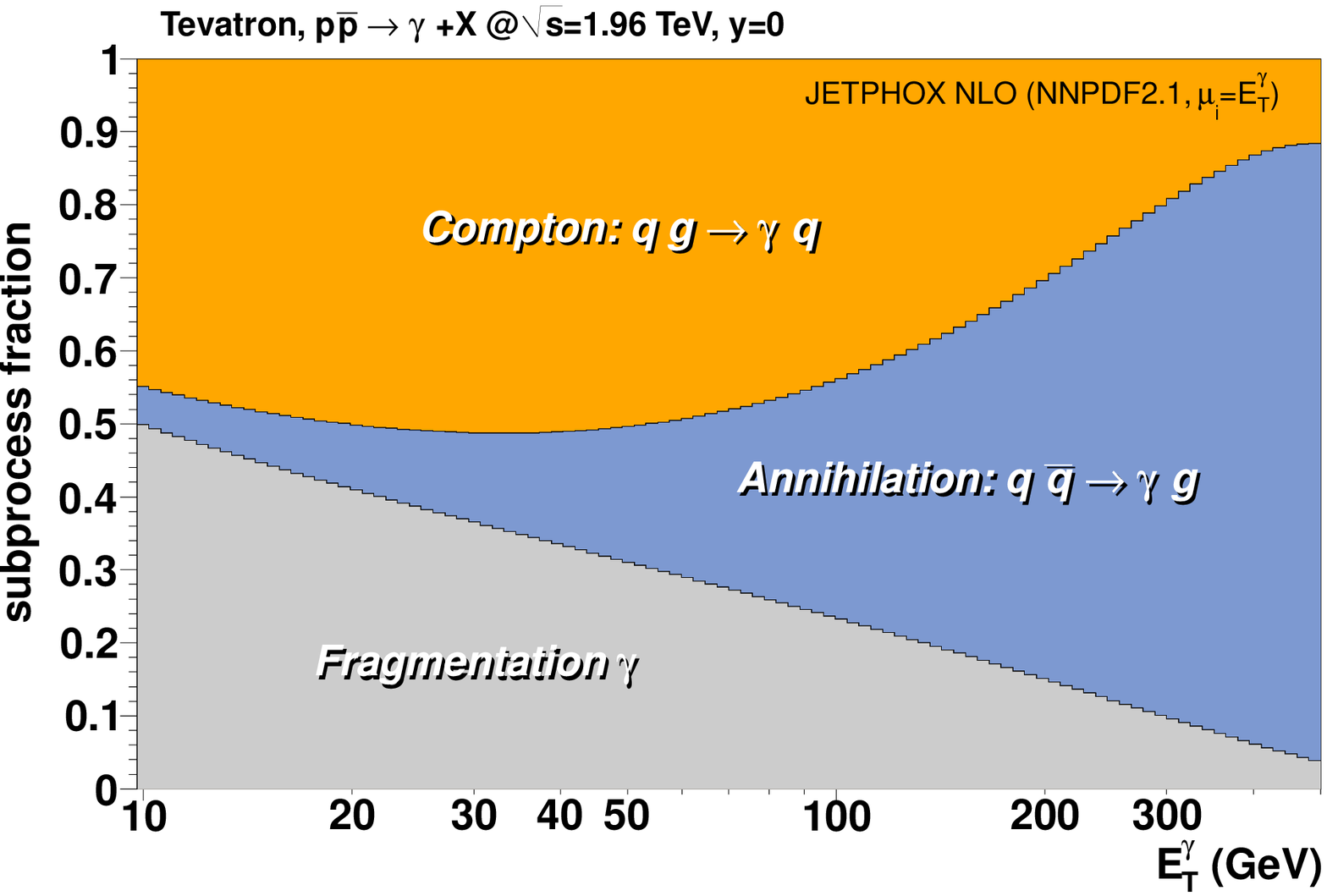}
\epsfig{width=0.46\textwidth,figure=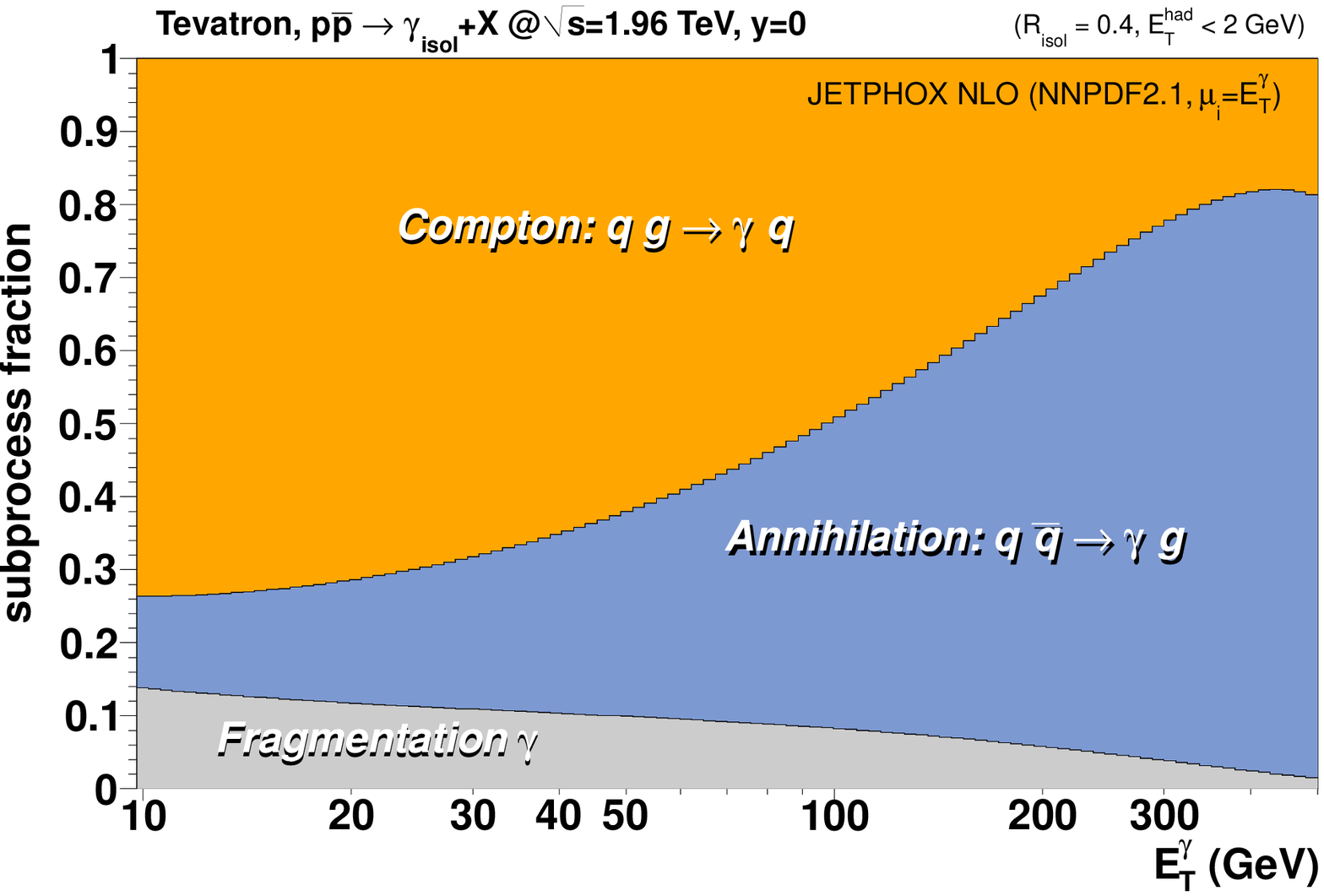}
\epsfig{width=0.46\textwidth,figure=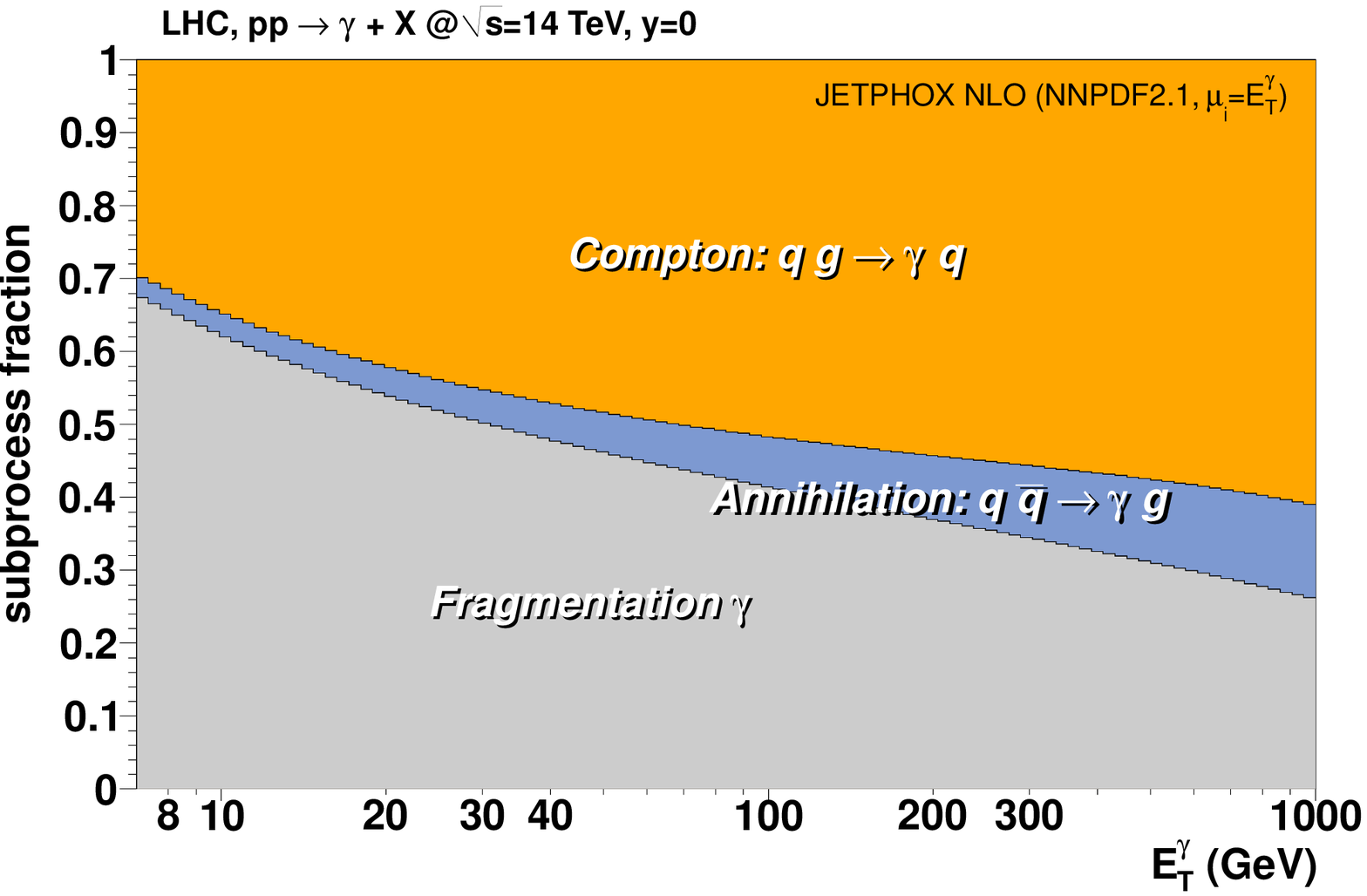}
\epsfig{width=0.46\textwidth,figure=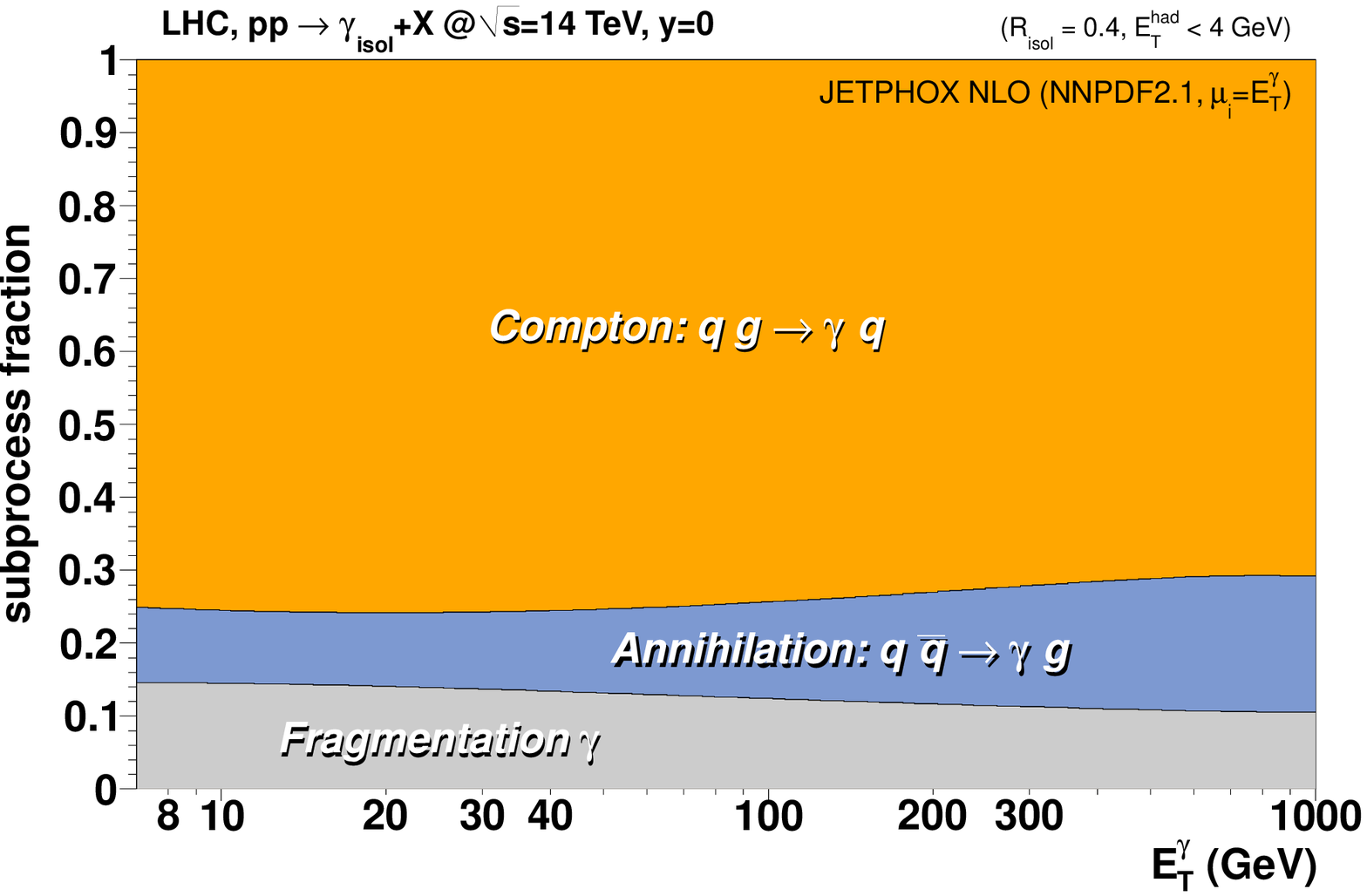}
\caption{\small Relative contributions of the $qg$-Compton, \qqbar-annihilation and fragmentation subprocesses in 
inclusive (left) and isolated (right) prompt-photon production in \ppbar\ (Tevatron at 1.96 TeV, top) and \pp\
(LHC at 14~TeV, bottom) collisions at midrapidity.
%increasing collider energies (top to bottom) and central rapidities. The left plots are for the total inclusive prompt-$\gamma$, 
%whereas the right ones include standard isolation cuts. %listed in Table~\ref{tab:compilation}.
Results obtained  at NLO with \jetphox~\protect\cite{jetphox} using NNPDF2.1~\protect\cite{Ball:2011mu}, 
theoretical scales set to $\mu$~=~$\ETg$, and BFG-II FFs~\protect\cite{Bourhis:1997yu} (see Sect.~\protect\ref{sec:setup}).
\label{fig:subproc}}
\end{figure}

The last PDF parametrisation that used the prompt photon data was MRST99~\cite{Martin:1999ww}. 
Since then, the issue of the compatibility of NLO pQCD with photon measurements has been discussed extensively.
For example, the authors of Ref.~\cite{Aurenche:2006vj} performed a systematic comparison of NLO calculations
with most of the available photon results by 2006 (see also Refs.~\cite{Vogelsang:1997cq,Aurenche:1998gv} for previous studies)
and showed that collider {\it isolated} photon data was in reasonable agreement with the theoretical predictions.
Indeed, as discussed in detail in~\cite{Ichou:2010wc}, by (i) increasing the $\sqrts$ from fixed-target to
collider energies, and by (ii) %removing photons that are surrounded by hadronic activity, 
requiring the photon to be isolated from any hadronic activity within a given distance around its
direction, one is left with an isolated-photon sample dominated by energy scales away from non-perturbative
effects (such as intrinsic-$k_T$ broadening~\cite{Apanasevich:1998ki}),
and where the uncertain contribution from photons issuing from the collinear 
fragmentation of final-state partons~\cite{GehrmannDeRidder:1998ba} are significantly reduced. 
Figure~\ref{fig:subproc} shows the relative contribution of the three leading diagrams -- direct Compton, direct
quark-antiquark annihilation, and fragmentation -- for inclusive (left) and for isolated (right)
photon production in proton-antiproton (\ppbar) at Tevatron (top) and in
proton-proton (\pp) collisions at the nominal LHC energy (bottom). %for the c.m. energies and kinematics considered in this paper. 
The fragmentation-$\gamma$ component, including the poorly known gluon--to--photon fragmentation 
function~\cite{Gluck:1992zx,Bourhis:1997yu} which dominates the low $\ETg$ part of the inclusive photon
spectra, is significantly reduced after application of isolation cuts and accounts only for 
less than 10--15\% of the cross sections at both energies. %Indeed, after application of isolation cuts, the 
The right panels of Fig.~\ref{fig:subproc} clearly indicate that the Compton component is the dominant contribution 
to the isolated-$\gamma$ spectra in wide ranges of the measured transverse energy $\ETg$, in
particular for \pp\ collisions, thus guaranteeing their increased sensitivity to $g(x,Q^2)$.\\

%Let us emphasize again only collider data are considered but not fixed 
%target data~\cite{Bonesini:1987bv,DeMarzo:1986vi,Ballocchi:1998au}. 
%There are many reasons to avoid fixed target
%photon data in comparisons with NLO pQCD: a larger dependence on the
%poorly known fragmentation functions, larger perturbative 
%corrections due to the lower scales ($\sqrts\le 40$~GeV) 
%as well as the possible 
%presence of non-perturbative effects like intrinsic $k_T$ broadening.
%We ignore as well the old ISR collider data~\cite{Angelis:1989zv}.

%%%%%%%%%%%%%%%%%%%%%%%%%%%%%%%%
%%%%%%%%%%%%%%%%%%%%%%%%%%%

In view of the various accurate measurements of isolated-photon spectra carried out recently
at the LHC~\cite{Aad:2010sp,arXiv:1108.0253,Chatrchyan:2011ue,Khachatryan:2010fm,Chatrchyan:2012vq}, 
which have added up about 130 new points to the existing world data, it is a timely moment to revisit %this issue and determine 
first if NLO pQCD provides a good description of all available isolated-$\gamma$ collider data, 
and if so, what quantitative constraints these new results impose on the gluon PDF.
However, the calculation of cross sections at NLO accuracy in hadronic collisions is usually a very
time-consuming process as it involves the Monte Carlo (MC) generation of a large number of event weights for the %to carry out the 
integration over the hard subprocess phase-space in order to cancel infrared divergences. 
%The repeated for any calculation with a different choice of parton densities within the proton. 
This fact makes the full-NLO calculation with different PDF choices, as in iterative fits 
of the parton densities, usually prohibitive in terms of the required processing time.
Computational alternatives exist, though, such as e.g. (i) the {\sc ApplGrid} framework~\cite{Carli:2010rw}
or the \textsc{FastNLO} software~\cite{Kluge:2006xs}, both of which make use of lookup tables with cross-section weights
which can be {\it a posteriori} fast-convoluted with any parton density set, and (ii) the NNPDF Bayesian
reweighting technique~\cite{Ball:2010gb,Ball:2011gg} which can be applied to any set of MC-based parton
densities and which relies on the computation of the $\chi^{2}$ distribution between the data and each MC
replica in order to determine the new (reweighted) PDFs.\\ 

In this paper we use the NNPDF reweighting method in order to quantify the impact of collider 
isolated-$\gamma$ data on $g(x,Q^2)$. Theoretical NLO pQCD predictions obtained with the 
\jetphox\ program~\cite{jetphox} are compared to all available experimental data in a very wide 
range of centre-of-mass (c.m.) energies ranging from 200~GeV up to 7~TeV. These comparisons are then 
used to quantitatively determine to which extent isolated-photons can be used to reduce the gluon 
PDF uncertainties.\\

The paper is organised as follows. In Sect.~\ref{sec:expdata} the experimental
isolated-photon results used in the analysis, more than 30 measurements with a
total of $\cO{400}$ data points, are presented.  
Section~\ref{sec:setup} describes the setup of the calculation based on the
the \jetphox\ code to provide NLO pQCD predictions and on the NNPDF-reweighting method 
of the parton densities. The main results of this analysis, the data--theory comparisons and the
quantitative impact of the photon data on the reweighted $g(x,Q^2)$, are discussed in Sect.~\ref{sec:results}. 
In Sect.~\ref{sec:higgs}, the predictions for Higgs cross sections
in the gluon-fusion production channel using the improved gluon distributions are presented. 
We conclude in Sect.~\ref{sec:summary} with a summary of the main results 
and an outline of possible future developments regarding the use of isolated-photon 
data in global PDF fits.

%%%%%%%%%%%%%%%%%%%%%%%%%%%%%%%%%%%%%%%%%%%%%%%%%%%%%%%%%%%%%%%%%%%%%%%%%%%%%%%%%%%%%%%%

\section{Experimental data}
\label{sec:expdata}

The world isolated-$\gamma$ data covers all prompt-photon measurements at collider machines in the last 30 years. 
Indeed, for c.m. energies above a hundred of~GeV, photon isolation is experimentally required 
in order to identify the high-$\ET$ prompt signal out of the overwhelming background of
photons from the decays of $\pi^0$ and $\eta$ mesons produced in the fragmentation of jets.
%In this section we review the experimental data used in the present analysis. First of all in 
Table~\ref{tab:compilation} summarises the details of the 35 existing experimental measurements. 
%considered in the present analysis. 
The table lists the characteristics of the more than 400 isolated-$\gamma$ data-points measured 
in \pp\ and \ppbar\ collisions from RHIC c.m. energies ($\sqrts$~=~200~GeV) up to the highest 
energies available so far at the LHC ($\sqrts$~=~7~TeV). 
The table updates the results collected in~\cite{HEP-EPS11} by including the latest
ATLAS~\cite{arXiv:1108.0253} and CMS~\cite{Chatrchyan:2012vq} measurements. 
For each system we provide the kinematical coverage in $\ETg$ and rapidity $y_\gamma$ 
of the measured photon, the corresponding range of parton fractional momentum $x$ 
covered by the measurement, %(obtained from the LO kinematics) 
as well as the applied photon isolation criteria used in each case 
(maximum hadronic energy E$_h$, or fraction of the photon energy $\varepsilon_h$, allowed within the
isolation cone of radius $R=\sqrt{(y_h-y_\gamma)^2+(\phi_h-\phi_\gamma)^2}$ around the photon direction).\\
%We avoid inclusive data since at collider energies the background
%from $\pi^0$ decays from jet fragmentation is overwhelming.

%%%%%%%%%%%%%%%%%%%%%
%\begin{sidewaystable}[htbp]
\begin{table}[htbp]
%\begin{small}
\begin{scriptsize}
%\begin{center} 
\centering
\begin{tabular}{c|c|c|c|c|c|c|c }\hline\hline
System  & Collab./Experiment & $\sqrts$ & $|y_\gamma|$ & $\ETg$ range &  $x$  & Data & Isolation \\
 & (collider) [Ref.]  & (TeV) & range  & (GeV) & range & points & radius, had. energy  \\
\hline
\pp & ATLAS (LHC) \cite{Aad:2010sp} & 7. & $<$0.6 & 15--100 & 5$\cdot 10^{-3}$--0.05 & 8  & $R$~=~0.4, E$_h<$~5 GeV\\ 
\pp & ATLAS (LHC) \cite{Aad:2010sp} & 7. & 0.6--1.37 & 15--100 & 3$\cdot 10^{-3}$--0.1 & 8  & $R$~=~0.4, E$_h<$~5 GeV\\ 
\pp & ATLAS (LHC) \cite{Aad:2010sp} & 7. & 1.52--1.81 & 15--100 & 2$\cdot 10^{-3}$--0.1 & 8  & $R$~=~0.4, E$_h<$~5 GeV\\ 

\pp & ATLAS (LHC) \cite{arXiv:1108.0253} & 7. & $<$0.6 & 45--400 & 5$\cdot 10^{-3}$--0.1 & 8  & $R$~=~0.4, E$_h<$~4 GeV\\ 
\pp & ATLAS (LHC) \cite{arXiv:1108.0253} & 7. & 0.6--1.37 & 45--400 & 5$\cdot 10^{-3}$--0.2 & 8  & $R$~=~0.4, E$_h<$~4 GeV\\ 
\pp & ATLAS (LHC) \cite{arXiv:1108.0253} & 7. & 1.52--1.81 & 45--400 & 2$\cdot 10^{-3}$--0.3 & 8  & $R$~=~0.4, E$_h<$~4 GeV\\ 
\pp & ATLAS (LHC) \cite{arXiv:1108.0253} & 7. & 1.81--2.37 & 45--400 & 2$\cdot 10^{-3}$--0.5 & 8  & $R$~=~0.4, E$_h<$~4 GeV\\ 

\pp & CMS (LHC) \cite{Khachatryan:2010fm} & 7. & $<$1.45 & 21--300 & 5$\cdot 10^{-3}$--0.1 & 11  & $R$~=~0.4, E$_h<$~5 GeV\\ 
\pp & CMS (LHC) \cite{Chatrchyan:2011ue} & 7. & $<$0.9  & 25--400 & 5$\cdot 10^{-3}$--0.2 & 15  & $R$~=~0.4, E$_h<$~5 GeV\\ 
\pp & CMS (LHC) \cite{Chatrchyan:2011ue} & 7. & 0.9--1.44 & 25--400 & 2$\cdot 10^{-3}$--0.3 & 15  & $R$~=~0.4, E$_h<$~5 GeV\\ 
\pp & CMS (LHC) \cite{Chatrchyan:2011ue} & 7. & 1.57--2.1 & 25--400 & $ 10^{-3}$--0.4 & 15  & $R$~=~0.4, E$_h<$~5 GeV\\ 
\pp & CMS (LHC) \cite{Chatrchyan:2011ue} & 7. & 2.1--2.5 & 25--400 & $10^{-3}$--0.5 & 15  & $R$~=~0.4, E$_h<$~5 GeV\\ 

\pp & CMS (LHC) \cite{Chatrchyan:2012vq}& 2.76 & $<$1.45 & 20--80 & $10^{-3}$--0.05 & 6 & $R$~=~0.4, E$_h<$~5 GeV\\ 

\hline

\ppbar & CDF (Tevatron) \cite{Aaltonen:2009ty} & 1.96 & $<$1.0 & 30--400 & 0.01--0.4 & 16  & $R$~=~0.4, $\varepsilon_h<0.1$\\
\ppbar & D0 (Tevatron) \cite{Abazov:2005wc} & 1.96 & $<$0.9 & 23--300 &  0.01--0.3 & 17  & $R$~=~0.4, E$_h<$~2 GeV\\ 

\ppbar & CDF (Tevatron) \cite{Acosta:2002ya} & 1.8 & $<$0.9 & 11--132 & 5$\cdot 10^{-3}$--0.2 & 17  & $R$~=~0.4, E$_h<$~4 GeV\\
\ppbar & CDF (Tevatron) \cite{Acosta:2004bg} & 1.8 & $<$0.9 & 10--65 & 5$\cdot 10^{-3}$--0.1 & 17  & $R$~=~0.4, E$_h<$~1 GeV\\
\ppbar & CDF (Tevatron) \cite{Abe:1994rra}  & 1.8 & $<$0.9 & 8--132 & 5$\cdot 10^{-3}$--0.2 & 16 & $R$~=~0.7, E$_h<$~2 GeV\\
\ppbar & D0 (Tevatron) \cite{Abbott:1999kd} & 1.8 & $<$0.9 & 10--140 & 5$\cdot 10^{-3}$--0.2 & 9  & $R$~=~0.4, E$_h<$~2 GeV\\ 
\ppbar & D0 (Tevatron) \cite{Abbott:1999kd} & 1.8 & 1.6--2.5 & 10--140 & $ 10^{-3}$--0.4 & 9  & $R$~=~0.4, E$_h<$~2 GeV\\
\ppbar & D0 (Tevatron) \cite{Abachi:1996qz} & 1.8 & $<$0.9 & 9--126  & 5$\cdot 10^{-3}$--0.2 & 23 &   $R$~=~0.4, E$_h<$~2 GeV\\ 
\ppbar & D0 (Tevatron) \cite{Abachi:1996qz} & 1.8 & 1.6 -- 2.5 & 9--126 & $ 10^{-3}$--0.4 & 23 &   $R$~=~0.4 E$_h<$~2 GeV\\ 

%\hdashline

\ppbar & CDF (Tevatron) \cite{Acosta:2002ya}& 0.63 & $<$0.9 & 8--38 & 0.01--0.2 & 7  & $R$~=~0.4, E$_h<$~4 GeV\\
\ppbar & D0 (Tevatron) \cite{Abazov:2001af}& 0.63  & $<$0.9 & 7--50 & 0.01--0.3 & 7 & $R$~=~0.4, E$_h<$~2 GeV\\ 
\ppbar & D0 (Tevatron) \cite{Abazov:2001af}& 0.63  & 1.6--2.5 & 7--50 & $ 10^{-3}$--0.4 & 7 & $R$~=~0.4, E$_h<$~2 GeV\\ 
\hline

\ppbar & UA1 (\spps) \cite{Albajar:1988im} & 0.63 & $<$0.8 & 16--100 & 0.03--0.3 & 16  & $R$~=~0.7, E$_h<$~2 GeV\\ 
\ppbar & UA1 (\spps) \cite{Albajar:1988im} & 0.63 & 0.8--1.4 & 16--70 & 0.01--0.4 & 10  & $R$~=~0.7, E$_h <$~2 GeV\\ 
\ppbar & UA1 (\spps) \cite{Albajar:1988im} & 0.63 & 1.6--3.0 & 16--70 & 0.01--0.5 & 13  & $R$~=~0.7, E$_h <$~2 GeV\\ 

\ppbar & UA2 (\spps) \cite{Alitti:1992hn} & 0.63 & $<$0.76 & 14--92 & 0.03--0.3 & 13 & $R$~=~0.265, $\varepsilon_h <$~0.25 \\
\ppbar & UA2 (\spps) \cite{Ansari:1988te} & 0.63 & $<$0.76 & 12--83 & 0.03--0.3 & 14 &    $R$~=~0.25, E$_h<$~0.1 GeV \\
\ppbar & UA2 (\spps) \cite{Ansari:1988te} & 0.63 & 1.0--1.8 & 12--51 & 0.01--0.4 & 8 &    $R$~=~0.53, E$_h<$~2 GeV \\

%\hdashline

\ppbar & UA1 (\spps) \cite{Albajar:1988im} & 0.546 & $<$0.8 & 16--51 & 0.03--0.2 & 6  & $R$~=~0.7, E$_h<$~2 GeV\\ 
\ppbar & UA1 (\spps) \cite{Albajar:1988im} & 0.546 & 0.8--1.4  & 16--46 & 0.02--0.4 & 5  & $R$~=~0.7, E$_h<$~2 GeV\\ 
\ppbar & UA1 (\spps) \cite{Albajar:1988im} & 0.546 & 1.6--3.0 & 16--38 & 0.01--0.5 & 5  & $R$~=~0.7, E$_h<$~2 GeV\\ 

\hline

\pp & PHENIX (RHIC) \cite{Adler:2006yt} & 0.2 & $<$0.35 & 3--16 & 0.03--0.2 & 17  & $R$~=~0.5, $\varepsilon_h <$~0.1\\ 

\hline 
\hline 
\end{tabular}\vspace{3mm} 
\caption{\small World systematics of isolated-photon data in \pp\ and \ppbar\ collisions. 
For each system, we quote (i) the experiment, collider and bibliographical reference, 
(ii) centre-of-mass energy $\sqrts$, the measured (iii) rapidity $|y_\gamma|$ and (iv) $\ETg$ ranges, 
(v) the parton fractional momenta $x$ probed, (vi) number of data points, and the (v) isolation criteria used.}
\label{tab:compilation} 
%\end{center} 
%\end{small}
\end{scriptsize}
%\end{sidewaystable}
\end{table}
%%%%%%%%%%%%%%%%%%%%%%%%%%%%%%%%%%%%%%%

%Let us discuss each dataset in more detail. In ascending order of collision energies,
Since we have decided not to consider the existing prompt-$\gamma$ measurements below $\sqrts \approx$~65~GeV,
both at fixed-target energies %~\cite{Bonesini:1987bv,DeMarzo:1986vi,Ballocchi:1998au}
%There are many reasons to avoid fixed target
%photon data in comparisons with NLO pQCD: a larger dependence on the
%poorly known fragmentation functions, larger perturbative 
%corrections due to the lower scales ($\sqrts\le 40$~GeV) as well as the possible 
%presence of non-perturbative effects like intrinsic $k_T$ broadening.
and at the CERN-ISR collider %~\cite{Angelis:1989zv} 
(see the compilation~\cite{Vogelsang:1997cq}),
as they deal with {\it inclusive}-$\gamma$ cross sections
with large dependence on poorly known parton-to-photon FFs and 
on non-perturbative corrections due to the smaller scales involved, 
the lowest-energy isolated-$\gamma$ measurement is that of the PHENIX
experiment in \pp\ at $\sqrts$~=~200~GeV at RHIC~\cite{Adler:2006yt}. 
Next in ascending order of collision energies are the oldest %isolated-$\gamma$ 
measurements by the CERN \spps\ UA1~\cite{Albajar:1988im} and UA2~\cite{Alitti:1992hn,Ansari:1988te} 
collaborations at $\sqrts$~=~546 and 630~GeV, which amount to a total of 90 data-points. 
The Tevatron datasets are available for $\sqrts$~=~0.63~TeV~\cite{Acosta:2002ya,Abazov:2001af}, 
1.8~TeV~\cite{Acosta:2002ya,Acosta:2004bg,Abbott:1999kd,Abachi:1996qz,Abe:1994rra} and 
1.96~TeV~\cite{Aaltonen:2009ty,Abazov:2005wc}, the most precise data being the latter.
The total number of data-points is 21 at 0.63~TeV, 114 at 1.8 TeV, and 33 at 1.96~TeV. 
Finally, the recent LHC measurements from ATLAS and CMS amounting to 133 data 
points~\cite{Aad:2010sp,arXiv:1108.0253,Khachatryan:2010fm,Chatrchyan:2011ue,Chatrchyan:2012vq}, 
cover the wider range in rapidity %, where forward production is covered 
(up to $|y_{\gamma}|$=~2.5) and have smaller systematical uncertainties
($\pm$(7~--20)\% depending on $\ETg$ and $y_\gamma$) 
than all other previous measurements. 
For ATLAS we consider both the 880~nb$^{-1}$~\cite{Aad:2010sp} and 36~pb$^{-1}$~\cite{arXiv:1108.0253} 
measurements, while for CMS we include the 7~TeV 2.9~pb$^{-1}$~\cite{Khachatryan:2010fm}
and 36~pb$^{-1}$ analyses~\cite{Chatrchyan:2011ue}, as well as the 
latest \pp\ measurement carried out at 2.76~TeV~\cite{Chatrchyan:2012vq} 
for reference heavy-ion collisions studies.\\

%Note that in Table~\ref{tab:compilation} we have separated measurements
%into different rapidity ranges. The motivation for this is to study
%separately both the compatibility with NLO pQCD and the impact
%on the gluon separately in different rapidity regions, as we have
%seen in Fig.~\ref{fig:correlations}, central production is more
%sensitive to photons that forward production and thus we expect
%better constraints for the central region data.

%To summarize the available data, in
% Figure~\ref{fig:allspectra} we show the cross sections
%from all available isolated-photon collider
%data first as a function of $\ETg$ and then as a function
%of $\xT$ with the cross section scaling by $1/\sqrts^n$ subtracted,
%with $n\sim 4.5$. The latter shows the universality of the QCD production
% mechanism that is valid 
%for all collider centre-of-mass energies, since all data satisfy the same 
%scaling with $\sqrts$ and collapse into a single curve.

\begin{figure}[htbp]
\centering
\epsfig{width=0.47\textwidth,height=8.cm,figure=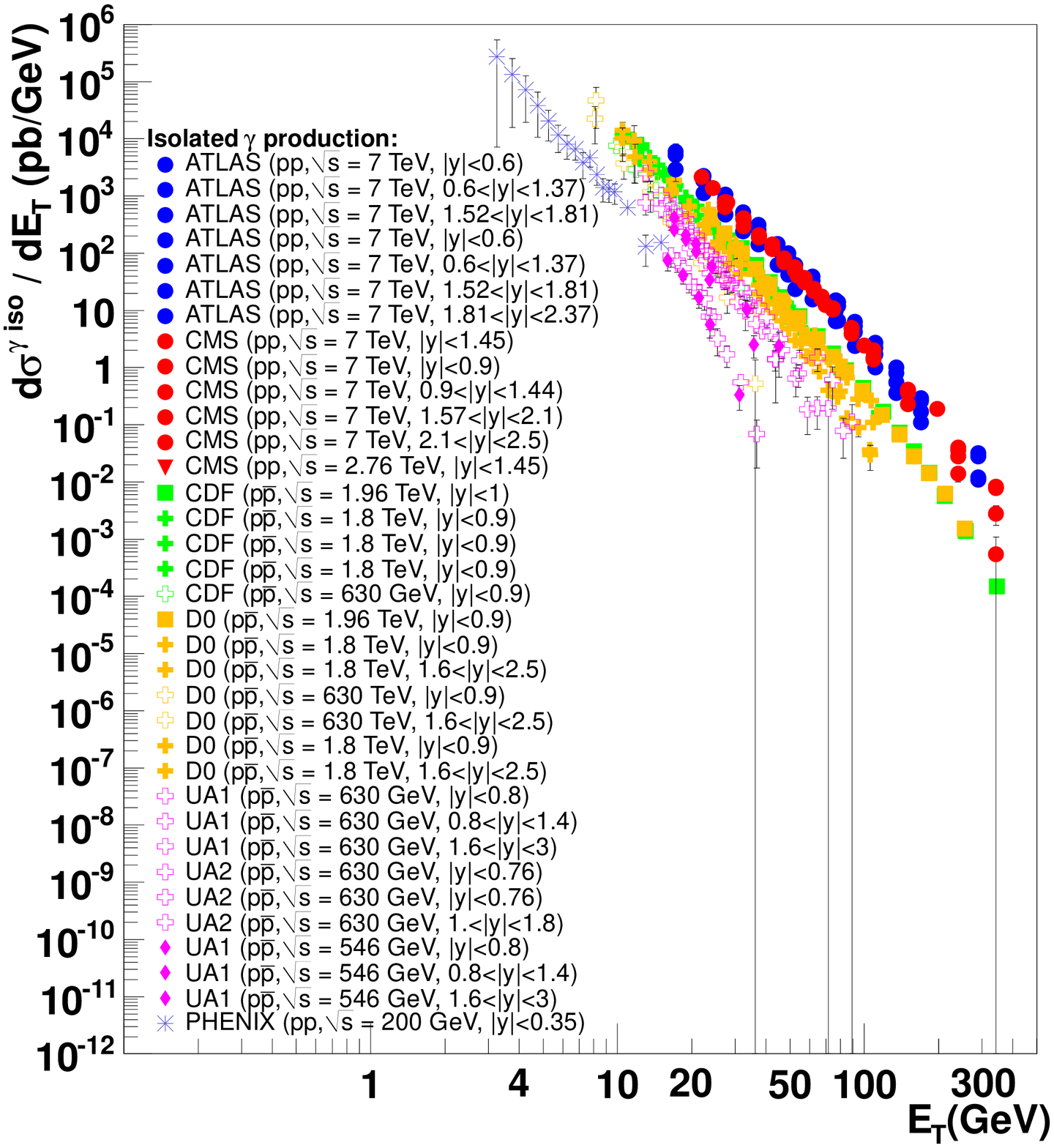}
\epsfig{width=0.52\textwidth,height=8.cm,figure=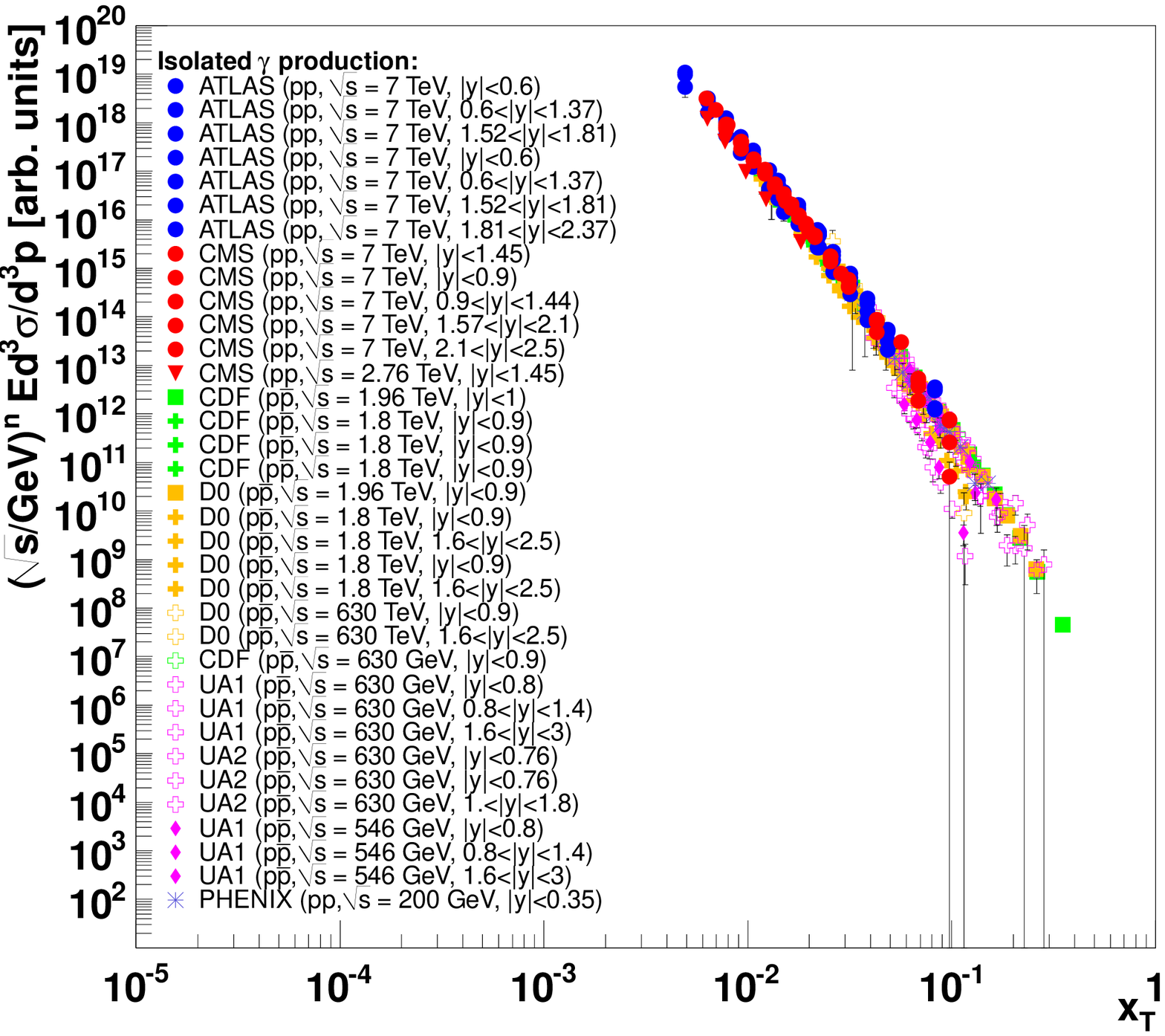}
\caption{\small World systematics of isolated-photon spectra measured 
in \pp\ and \ppbar\ collisions at collider energies (Table~\protect\ref{tab:compilation})
as a function of $\ETg$ (left) and $\xT$ (right) where the invariant cross sections 
have been scaled by $\sqrts^n$ with $n$~=~4.5.
 \label{fig:allspectra}}
\end{figure}

The isolated-photon data of Table~\ref{tab:compilation} are plotted in Fig.~\ref{fig:allspectra} as a
function of $\ETg$ (left panel) and as function of $\xT=2\ETg/\sqrts$ with the cross sections 
scaled by $\sqrts^{\,n}$ (right panel). All spectra in Fig.~\ref{fig:allspectra} (left) follow clear power-law
dependencies from 3 to 400~GeV spanning 9 orders of magnitude in the cross sections. 
The $\xT$ spectra (Fig.~\ref{fig:allspectra}, right) coalesce over a single curve when the 
cross sections are normalised by $\sqrts^n$ with exponent $n~\approx$~4.5. Such a behaviour is
very close to the $1/\pt^{n=4}$ dependence expected for partonic 2~$\to$~2 scattering
cross sections in the conformal QCD limit, disregarding scaling violations from PDF and running of $\as$~\cite{Arleo:2009ch}. 
A few deviations are visible, in particular for measurements that are either away from midrapidity for which the %tacit 
assumption $\xT = 2\ETg/\sqrts$ does not exactly hold and/or for which the highest photon energies are mis-reconstructed 
(see Sect.~\ref{sec:data-theory}). The fact that all data satisfy the same scaling with $\sqrts$ and collapse
into a single curve indicates the universality of the underlying partonic production mechanism.
The observed $\ETg$ power-laws and universal $\sqrts$-dependence of the cross sections 
are tell-tale of a perturbative origin for the production of isolated photons and confirm
the validity of pQCD calculations to study their yields.\\

%%%%%%%%%%%%%%%
%%%%%%%%%%%%%%%%

\begin{figure}[!htbp]
\centering
\epsfig{width=0.6\textwidth,figure=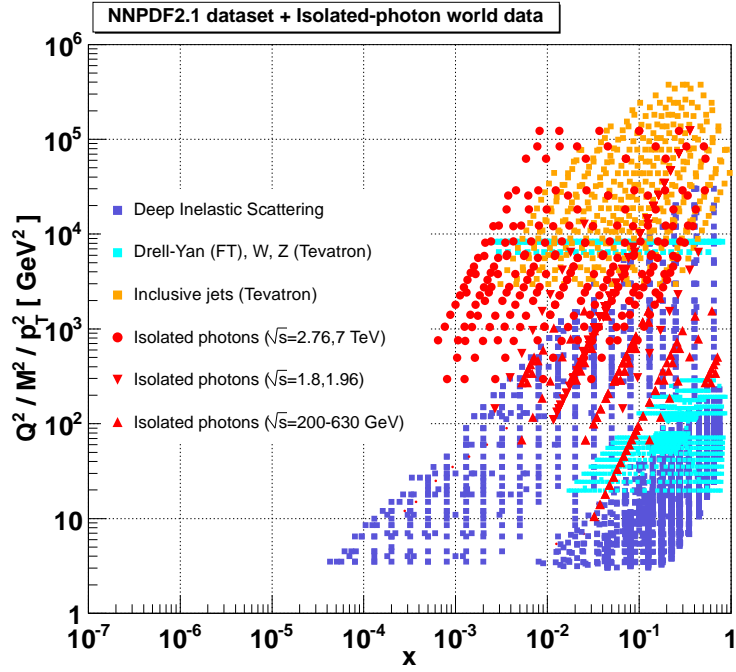}
\caption{\small Kinematical region in the $x-Q^2$ plane probed by experimental isolated-$\gamma$ data at collider energies
(red circles and triangles) which enter into this analysis (Table~\ref{tab:compilation}) 
compared to the coverage of DIS, Drell-Yan and jet datasets (squares) used in the NNPDF2.1 global fits.
\label{fig:x_Q2_map}}
\end{figure}

In Fig.~\ref{fig:x_Q2_map} we show a scatter plot of the world isolated-photon measurements listed
in Table~\ref{tab:compilation} together with the DIS, Drell-Yan and jet data (about 3500 points) 
used in the NNPDF2.1 global fits. For each experimental data-point we plot two values of
parton fractional momentum $x_{\pm} = \xT \cdot e^{\pm y_\gamma}/\sqrts$, assuming leading-order partonic
kinematics. The plot shows that the photon LHC data extend the kinematic coverage in particular over the region 
$x\approx 10^{-3} - 10^{-2}$ at moderately large energy scales ($\ETgSq \approx 10^{3} - 10^{5}$ GeV$^2$) 
not directly covered by the other experimental datasets.

%%%%%%%%%%%%%%%%%%%%%%%%%%%%%%%%
%%%%%%%%%%%%%%%%%%%%%%%%%%%

%%%%%%%%%%%%%%%%%%%%%%%%%%%%%%%%%%%%%%%%%%%%%%%%%%%%%%%%%%%%%%%%%%%%%%%%%%%%%%%%%%%%%%%%
%\clearpage

\section{Theoretical setup}
\label{sec:setup}

In this section the basic ingredients of the \jetphox\ program used to
compute the isolated-photon cross sections are discussed, and
the NNPDF reweighting technique employed to quantify the impact of new 
data on the proton PDFs is briefly recalled.

\subsection{Isolated-photon cross sections}
\label{sec:jetphox}

Two types of processes contribute at leading order to prompt photon production %cross section 
in \pp\ and \ppbar\ collisions: the `direct' contribution, where the photon is emitted directly 
from a pointlike coupling to the hard parton-parton vertex, and the `fragmentation' 
(called also `anomalous' in the past) contribution, in which the photon originates from the 
collinear fragmentation of a final-state parton. Schematically, the differential photon 
cross section as a function of transverse energy $\ETg$ and rapidity $y_\gamma$ can be written as %~\cite{Aurenche:2006vj}: 
%\begin{equation} 
%d\sigma^{\gamma}  =d\sigma^{{dir}}(\mu_{_R},\mu_{_F},\mu_{_{ff}}) +\sum_{k=q,\bar{q},g}d\sigma^{{frag}}_{k}(\mu_{_R},\mu_{_F},\mu_{_{ff}}) \otimes D_{\gamma/k}(\mu_{_{ff}}) 
%\label{eq:1}
%\end{equation}
\begin{eqnarray}
\label{eq:dsigma_nlo}
\lefteqn{d\sigma  \equiv d\sigma_{_{\rm dir}}+d\sigma_{_{\rm frag}} = \sum_{a,b=q,\bar{q},g}\int dx_a dx_b \; f_a(x_a;\mu_{_{\rm F}}^2) f_b(x_b;\mu_{_F}^2) \; \times } \\
& & \left[ d\hat{\sigma}_{ab}^{\gamma}(p_{\gamma}, x_a,x_b; \mu_{_{\rm R}},\mu_{_{\rm F}},\mu_{_{\rm ff}})+
\! \sum_{c=q,\bar{q},g}\int^1_{z_{min}}\! \frac{dz}{z^2} d\hat{\sigma}_{ab}^c (p_{\gamma},x_a,x_b,z;\mu_{_{\rm R}},\mu_{_{\rm F}},\mu_{_{\rm ff}}) D^{\gamma}_c (z;\mu_{_{\rm ff}}^2)\right] \nonumber
\end{eqnarray}
where $f_{a}(x_{a};\mu_{_{\rm F}}^2)$ is the parton distribution function of  parton
species $a$ inside the incoming protons at momentum fraction $x_a$; $d\hat{\sigma}_{ab}$ are the parton-parton
subprocess differential cross sections; and $D_{\gamma/k}(z;\mu_{_{\rm ff}}^2)$ is the fragmentation function of parton 
$k$ to a photon carrying a fraction $z$ of the parent parton energy,
integrated from $z_{min}=\xT \cosh y_\gamma $ to 1. %, with $\xT\equiv  2\ETg/\sqrts$, to 1. 
The scaled momentum $\xT$ is a good representative for the typical
$x$ values probed in the PDFs at central rapidities, while
at forward rapidities one is probing $x\sim \xT \exp(-y_\gamma)$.
%and factorisation scale $\mu_{_F}$; $\alpha_s(\mu_{_R})$ is the strong coupling defined
%in the $\overline{\mbox{MS}}$ renormalisation scheme at the  renomalisation scale $\mu_{_R}$.  ,
%where $d\sigma^{{frag}}_{k}$ describes the production of a parton $k$ in a
%hard collision 
The arbitrary parameters $\mu_{_{\rm R}}$, $\mu_{_{\rm F}}$ and $\mu_{_{\rm ff}}$
are respectively the renormalisation, initial-state factorisation, and fragmentation scales which %, loosely speaking, 
encode any residual dependence of the cross sections to higher-order 
contributions which are missing in the calculation.\\
%The probability that a high $E_{_T}$ parton of species $k$ (quark or gluon) ends 
%up splitting into a photon is encoded into a fragmentation function of parton 
%$k$ to a photon, $D_{\gamma/k}(z,\mu_{_{\rm ff}},)$, obtained from fits of the experimental
%$e^+e- \to q \bar q \to \gamma\,X$ data~\cite{ffs}, %defined in some arbitrary fragmentation scheme, 
%at some arbitrary fragmentation scale $\mu_{_{\rm ff}}$.\\

The study provided in this article relies on the \jetphox\ calculation of both $d\sigma_{_{\rm dir}}$ and
$d\sigma_{_{\rm frag}}$ at NLO accuracy~\cite{jetphox} in the strong coupling 
$\as(\mu_{_{\rm R}})$, i.e. all diagrams up to the order $\mathcal{O}(\alpha\as^2)$ 
are included, in the $\overline{\mbox{MS}}$ renormalisation scheme.
The results of the NLO calculation of $d\sigma_{_{\rm dir}}$ have been known for a long time~\cite{ABDFS}. 
The calculation of the NLO corrections to $d\sigma_{_{\rm frag}}$ became also
available later~\cite{Aversa:1988vb,Gordon:1994ut,jetphox}. 
We note that the distinction between $d\sigma_{_{\rm dir}}$ and $d\sigma_{_{\rm frag}}$ 
is arbitrary and only its sum is physically observable, e.g. bremsstrahlung from 
a quark leg can be considered as ``fragmentation'' or as ``NLO direct'' depending 
on the value of the fragmentation scale considered.\\
%The knowledge of  $\LMS$, e.g. from deep-inelastic scattering
%experiments, completely specifies  the NLO expression of the running coupling
%$\alpha_s(\mu_{_R})$. The NLO correction terms to $d\sigma_{_{\rm dir}}$ and {frag},
%$\kd_{ij}$~\cite{ABDFS,GV} and $\kf_{ij,k}$~\cite{ACGG} respectively, are known
%and their expressions in the $\overline{\mbox{MS}}$ scheme will be used. 
%The dependence of these functions on the kinematical variables $x_1, x_2, z, \sqrts, E_{_T}$ and $y$ has not been explicitly displayed. 
%The structure and fragmentation functions have been determined at the required level of accuracy by NLO fits to the data.
%%%% All the quantities entering the above
%%%% equations have been either calculated ($\kd_{ij}$ and $\kf_{ij,k}$) or

A few recent works exist that have computed various beyond-NLO corrections
to the inclusive production of prompt photons: %, but which are not relevant for this analysis:
\begin{itemize}
\item The small-$x$ high-energy corrections~\cite{Diana:2009xv} were computed in Ref.~\cite{Diana:2010ef} 
focusing on the low-$\ETg$ region of the spectrum where power terms of the type $\alpha^k_s\ln^p(x)$ are enhanced
when the scaling variable $x$ becomes small. It was found that such type of next-to-NLO (NNLO)
corrections are negligible in the kinematical range of the available collider data and, thus, ignored
in the following.
\item Resummation at next-to-leading-logarithmic (NLL)~\cite{Laenen:1998qw,Catani:1998tm,Kidonakis:2000gi,Kidonakis:2003bh,deFlorian:2005wf}
or even next-to-next-to-leading (NNLL)~\cite{Becher:2009th} accuracy of threshold and recoil contributions 
due to soft gluon emissions which are large close to the phase space boundary when $\ETg$ is about 
half of the c.m. energy ($\xT \to 1)$ have been
%obtained, first in $d\sigma_{_{\rm dir}}$~\cite{los,cmn,cmnov,ko,sv}, and more recently in 
%$d\sigma_{_{\rm frag}}$ as well \cite{deFlorian:2005wf}. This resummation is performed 
obtained for $d\sigma_{_{\rm dir}}$ and $d\sigma_{_{\rm frag}}$. 
The effect of this resummation is important at very large photon $\ET$ corresponding to values
$\xT \gtrsim$~0.2, %Interestingly, the N(N)LL results 
and provide a much reduced %$\mu_{_R}$ and $\mu_{_F}$ 
scale dependence than the NLO approximation. 
%but, in general, they are consistent with the values of the 
%cross sections obtained at NLO when the scales are reduced somewhat compared
%to the typical $\mu=\ETg$ value~\cite{Aurenche:2006vj}. 
In the current collider data the typical ranges involved ($x\approx$~0.001--0.2, see Fig.~\ref{fig:x_Q2_map}) 
lead to small threshold corrections which we do not consider either.
\item At very large photon transverse energies $\ETg\approx$~1--2~TeV, not yet measured at the LHC, 
one should also consider additional corrections due to electroweak boson exchanges 
which decrease the photon yields by about 10--20\%~\cite{Kuhn:2005gv}. 
%Such contributions will need to be taken into account in
\end{itemize}

In this paper, we use the \jetphox\ (version 1.3.0) Monte Carlo program~\cite{jetphox} to compute the NLO pQCD
predictions for isolated-photon production. We take into account five active quark flavours. The gluon box diagram 
$g\,g\to g\,\gamma$ is included in the calculations although its contribution to the single inclusive spectrum is found
to be just of a few percent. The NNPDF2.1 parton densities (with 100 replicas per system)~\cite{Ball:2011mu} 
were interfaced to \jetphox\ via the \lhapdf\ (version 5.8.5) package~\cite{Bourilkov:2006cj}. 
The value of $\as$~=~0.119 used in the calculations is that provided by the NNPDF2.1 parametrisation itself.
The renormalisation, factorisation and fragmentation scales are all set equal to the photon transverse energy,
$\mu_{_{\rm F}}=\mu_{_{\rm R}}=\mu_{_{\rm ff}}=\ETg$. Such $\mu$ scales are found to result in spectra that agree well with the 
central values of the experimental data. At LHC and Tevatron energies, the sensitivity to changes in the 
arbitrary theoretical scales was discussed in~\cite{Ichou:2010wc} and found to be of about $\pm$(10--15)\% 
in the measured $\ETg$ ranges.
The parton--to--photon fragmentation functions used are the BFG-II (``large gluon'') set~\cite{Bourhis:1997yu}. 
The isolated-$\gamma$ spectrum obtained with the alternative BFG-I (``small gluon'') set is just a
few percent smaller than the one obtained with BFG-II in the lowest $\ETg$ range~\cite{Ichou:2010wc} since a
significant  fraction (up to 80\%, see Fig.~\ref{fig:subproc}) of the fragmentation photons are removed by the
application of isolation cuts. \\

%The theoretical uncertainties affecting the isolated-photon cross section in hadronic collisions from scale
%variations and parton distributions were extensively discussed in Ref.~\cite{Ichou:2010wc}.

We run \jetphox\ for all the systems in Table~\ref{tab:compilation}  
within the ($\ETg,y_\gamma$) kinematics ranges of every measurement.
The MC photon isolation criteria are also matched as closely as possible to each one 
of the experimental cuts. %\footnote{The ``smooth cone'' Frixione criterion~\cite{Frixione:1998jh} 
%would allow one to completely remove the fragmentation component 
%from the isolated cross section, but it has not yet been used in the experimental analyses.}.
We histogram the partonic configurations generated in $\ETg$-bins of the same size as  
for the experimental data in order to be able to compute the $\chi^2$ bin-by-bin and  
avoid potential problems with steeply-falling spectra such as those discussed in
Ref.~\cite{Lafferty:1994cj}. Running for each one of the 100 NNPDF2.1 replicas with acceptable MC statistics
(statistical uncertainty below 1\%) for all $\ETg$ bins is a very CPU-demanding task. 
The \jetphox\ MC production for 100 PDF replicas takes about 7 (resp. 10) hours per each million direct 
(resp. fragmentation) events in a $\cO{\rm 2~GHz}$ CPU core, which means that about one week 
of computer-time is needed to obtain the 100 spectra for each one of the 35 systems considered.

%%%%%%%%%%%%%%%%%%%%%%%%%%%%%%%%%%%%%%%%%%%%%%%%%%%%%%%%%%%%%%%%%%%%%%%%%%%%%%%%%%%%%%%%

\subsection{PDF reweighting}
\label{sec:rw}

In order to quantify the impact of the collider isolated-photon
data on the proton PDFs we use the Bayesian reweighting method described
in Refs.~\cite{Ball:2010gb,Ball:2011gg}. This technique can be applied
to any PDF-set that estimates PDF uncertainties based on a MC method,
such as the NNPDF family~\cite{DelDebbio:2004qj,DelDebbio:2007ee,Ball:2008by,Ball:2009mk,Ball:2010de,Ball:2011mu,Ball:2011gg,Ball:2011uy}
(but also MSTW~\cite{watttalk2} and HERAPDF~\cite{glazov} have produced MC sets). 
The method  allows one to straightforwardly determine the impact of a new dataset on PDFs
by means of Bayesian inference. The only ingredient needed are the data--theory ``goodness-of-fit''
$\chi^{2}_k$ for each $k$-th MC replica (where $k$~=~1,...,$N_{\rm rep}$ runs over the full set of replicas: 
$N_{\rm rep}$~=~100 or 1000 in the NNPDF2.1 set). 
This technique reduces the problem of the slowness of \jetphox\ 
(or in general any NLO hadronic computation) which forbids its direct use 
within a PDF-fitting program, since %using reweighting 
the theoretical predictions need to be computed only once for each PDF replica.\\

The $\chi^{2}_k$ for each replica is defined as follows:
\be
\chi^{2}_k = \frac{1}{N_{\rm dat}}\sum_{i=1}^{N_{\rm dat}}
\frac{\lp \sigma_i^{{\rm (th)},(k)}-\sigma_i^{{\rm (exp)}}\rp^2}{\Delta_{\rm tot}^2} \ ,
\label{eq:chi2} 
\ee
where $\sigma_i^{{\rm (th)},(k)}$ is the NLO theoretical prediction for the
isolated-photon cross section, Eq.~(\ref{eq:dsigma_nlo}), obtained with the $f_k$ PDF replica, 
$\sigma_i^{{\rm (exp)}}$ is the corresponding experimental measurement
and $\Delta_{\rm tot}$ accounts for the %stands for the sum in quadrature of the statistical and systematic 
experimental uncertainties. $N_{\rm dat}$ is the number
of data points of each particular measurement.
We note that the experimental covariance matrix is not available for any of the isolated-photon datasets and
thus we are forced to add in quadrature the statistical and systematic uncertainties into $\Delta_{\rm tot}$.
Likewise, absolute normalisation uncertainties (e.g. from the integrated luminosity) should be in principle included
in the experimental covariance matrix using the $t_0$ method (as consistently done within 
the NNPDF analysis)~\cite{Ball:2009qv}, but they are typically smaller than other experimental 
uncertainties and also added in quadrature here.\\

Once the $\chi^{2}_k$ for each replica has been computed,
the new weight  of the replica is given by
\begin{equation}
w_k = 
\frac{(\chi^{2}_k)^{{1\over 2}(n-1)} 
e^{-\frac{1}{2}\chi^{2}_k}}
{\smallfrac{1}{N_{\rm rep}}\sum_{k=1}^{N}(\chi^{2}_k)^{{1\over 2}(n-1)}
e^{-\frac{1}{2}\chi^{2}_k}}\;.
\label{eq:weights}
\end{equation}
The weights $w_k$, when divided by the number of MC replicas of the prior PDF set ($N_{\rm rep}$), 
are then simply the probabilities of the replicas $f_k$, given the $\chi_{k}^2$ to the new added experimental results.
If the new data constrains the PDFs, reweighting will be less
efficient that refitting because of the discarded replicas
with low weight. One can quantify this efficiency loss  by using the Shannon 
entropy to compute the effective number of replicas left after reweighting:
\begin{equation}
N_{\rm eff}\equiv \exp \left\{\smallfrac{1}{N_{\rm rep}}\sum_{k=1}^{N_{\rm rep}} w_k \ln (N_{\rm rep}/w_k)\right\}.
\label{eq:effective}
\end{equation}
Clearly $0<N_{\rm eff}<N_{\rm rep}$ and the reweighted fit has the same accuracy as a 
refit with $N_{\rm eff}$ replicas. This effective number of replicas
$N_{\rm eff}$ is a useful estimator of the impact of each individual
dataset. The smaller $N_{\rm eff}$ is, the more the new dataset constrains the PDFs.\\

It is often the situation that experimental uncertainties
are under or over estimated.
It is then possible to rescale the uncertainties of the data by a factor 
$\alpha$, and then use inverse probability to calculate the probability density 
for the rescaling parameter $\alpha$
\begin{equation}
\mathcal{P}(\alpha)\propto \smallfrac{1}{\alpha}\sum_{k=1}^{N_{\rm rep}} w_k(\alpha).
\label{eq:rescaling}
\end{equation}
Here $w_k(\alpha)$ are the weights Eq.~(\ref{eq:weights}) evaluated by replacing 
$\chi^2_k$ with $\chi^2_k/\alpha^2$, and are thus proportional to 
the probability of $f_k$ given the new data with rescaled errors.
Averaging $w_k(\alpha)$ in the reweighted fit thus gives the probability density 
for $\alpha$. If this probability density peaks close to unity, 
the new data are consistent with the pre-reweighting data, 
while if it peaks far above (below) one then it is 
likely that the errors in the data have been under (over) estimated.
%In the following we normalize the distribution to unity.
The distribution of the rescaling variable $\alpha$, normalised to unity,
is used in the next section to investigate
if the bad agreement between NLO pQCD and a few datasets could be
due to possibly underestimated experimental uncertainties.

%%%%%%%%%%%%%%%%%%%%%%%%%%%%%%%%%%%%%%%%%%%%%%%%%%%%%%%%%%%%%%%%%%%%%%%%%%%%%%%%%%%%%%%%

\section{Results}
\label{sec:results}

%Now we discuss the results of our analysis. 
The comparison between NLO pQCD and the experimental data for all systems listed in
Table~\ref{tab:compilation} is presented in the next subsection. With this information
we compute for each case the data--theory $\chi^2_k$ distributions Eq.~(\ref{eq:chi2}), 
the weights Eq.~(\ref{eq:weights}), the effective number of replicas Eq.~(\ref{eq:effective}), 
the sensitivity to individual PDF flavours, and finally the associated constraints on the gluon distribution.

\subsection{Comparison between data and NLO pQCD}
\label{sec:data-theory}

\begin{figure}[htbp!]
\centering
\epsfig{width=0.45\textwidth,figure=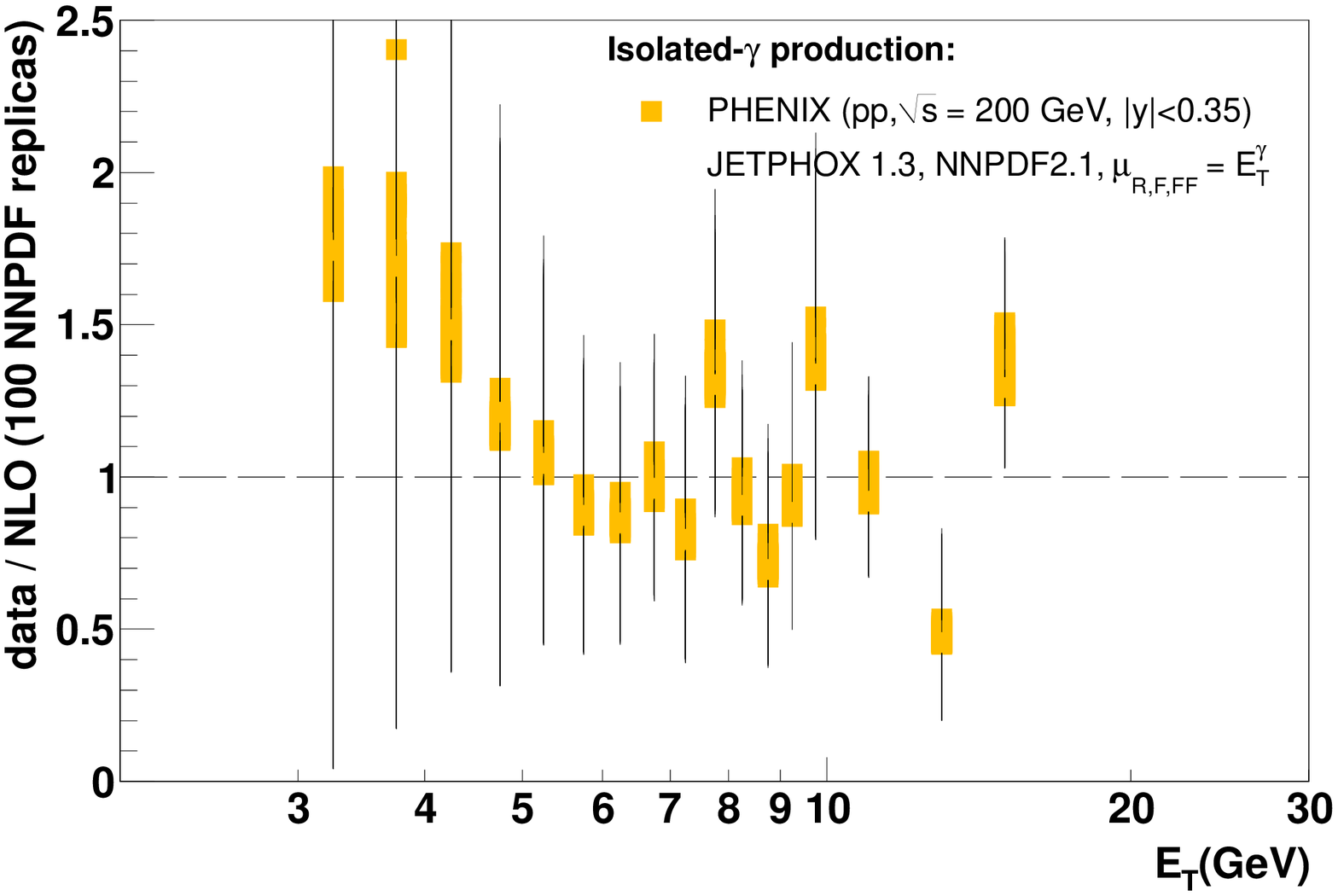}
\epsfig{width=0.45\textwidth,figure=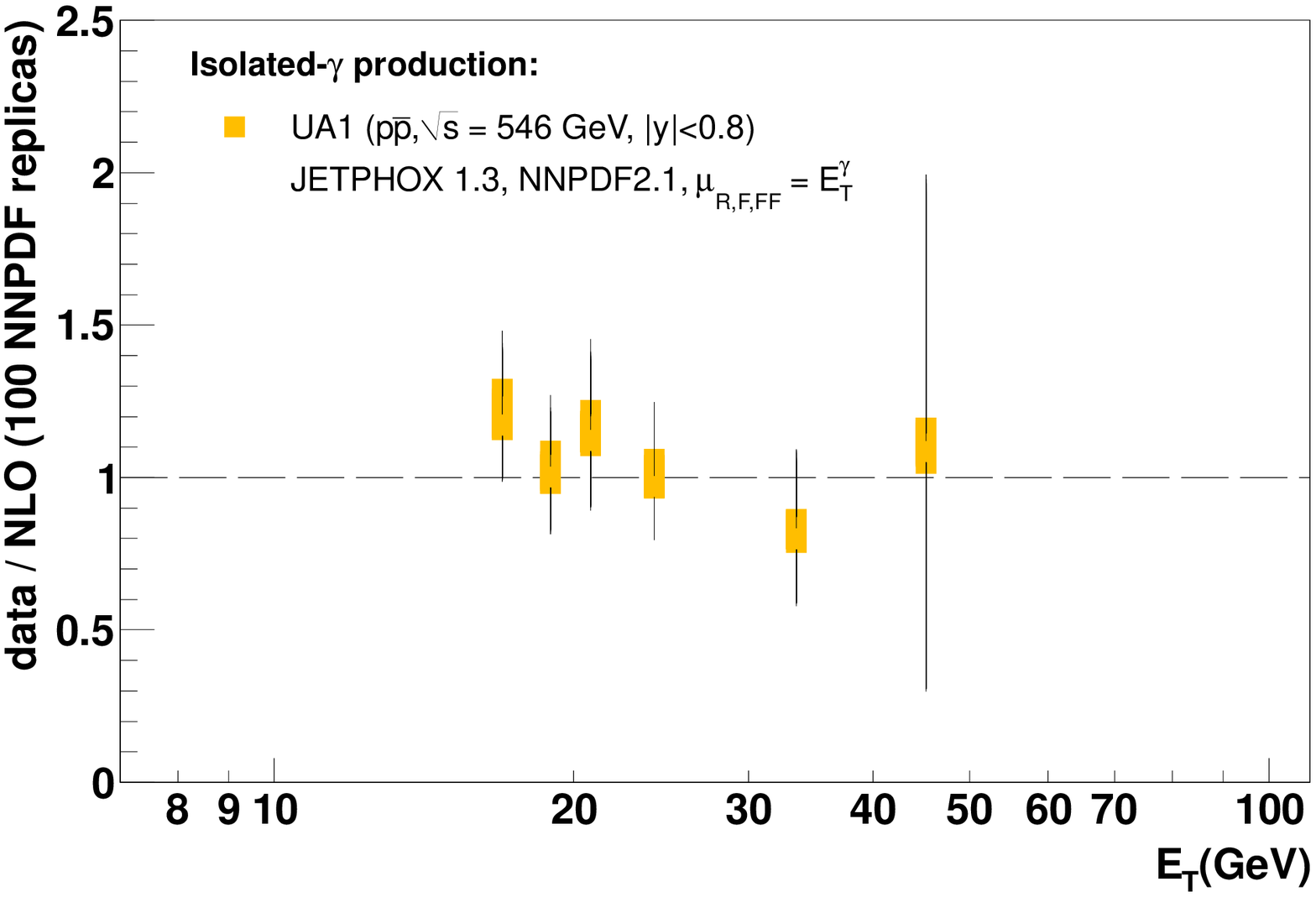}\\
\epsfig{width=0.45\textwidth,figure=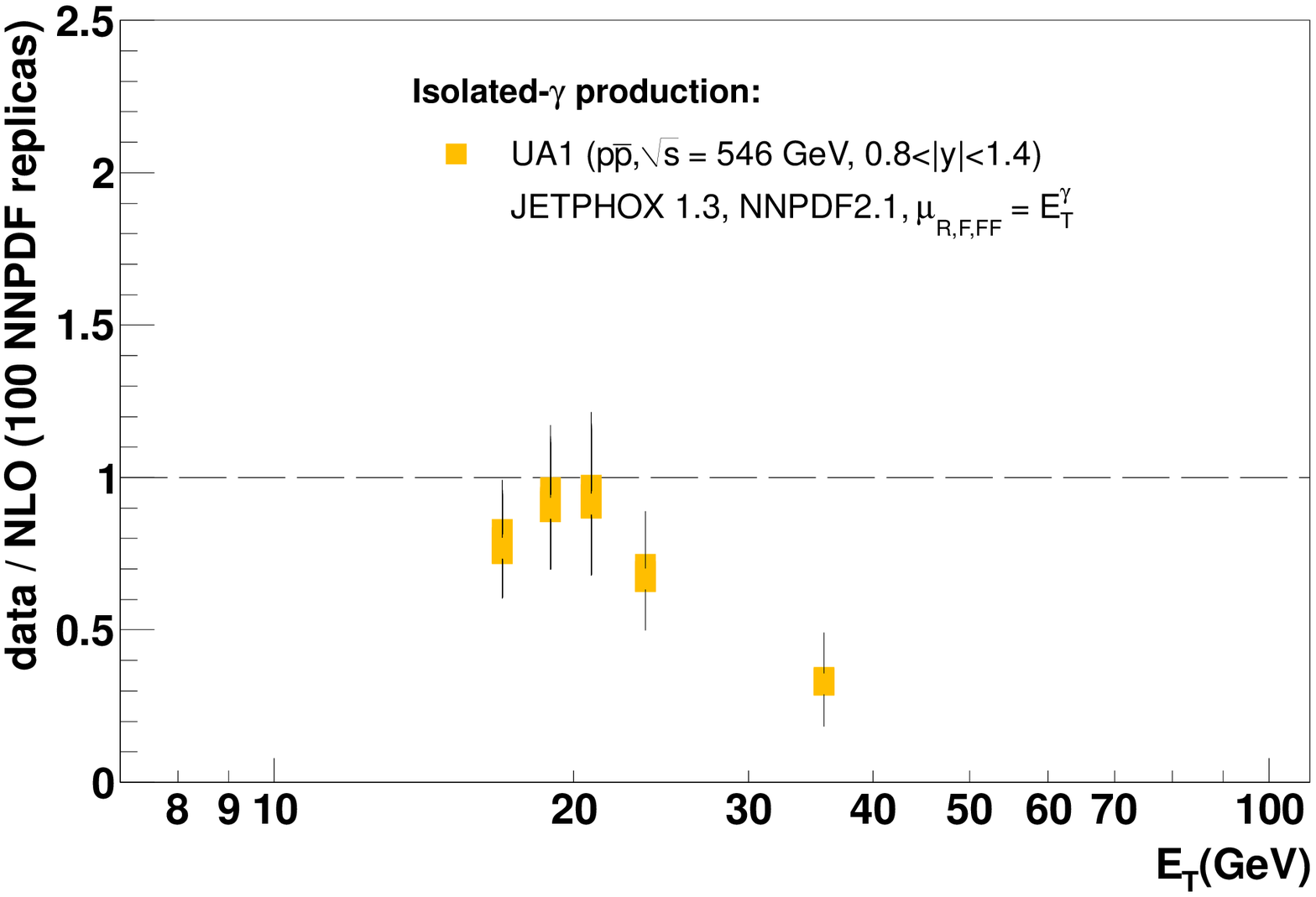}
\epsfig{width=0.45\textwidth,figure=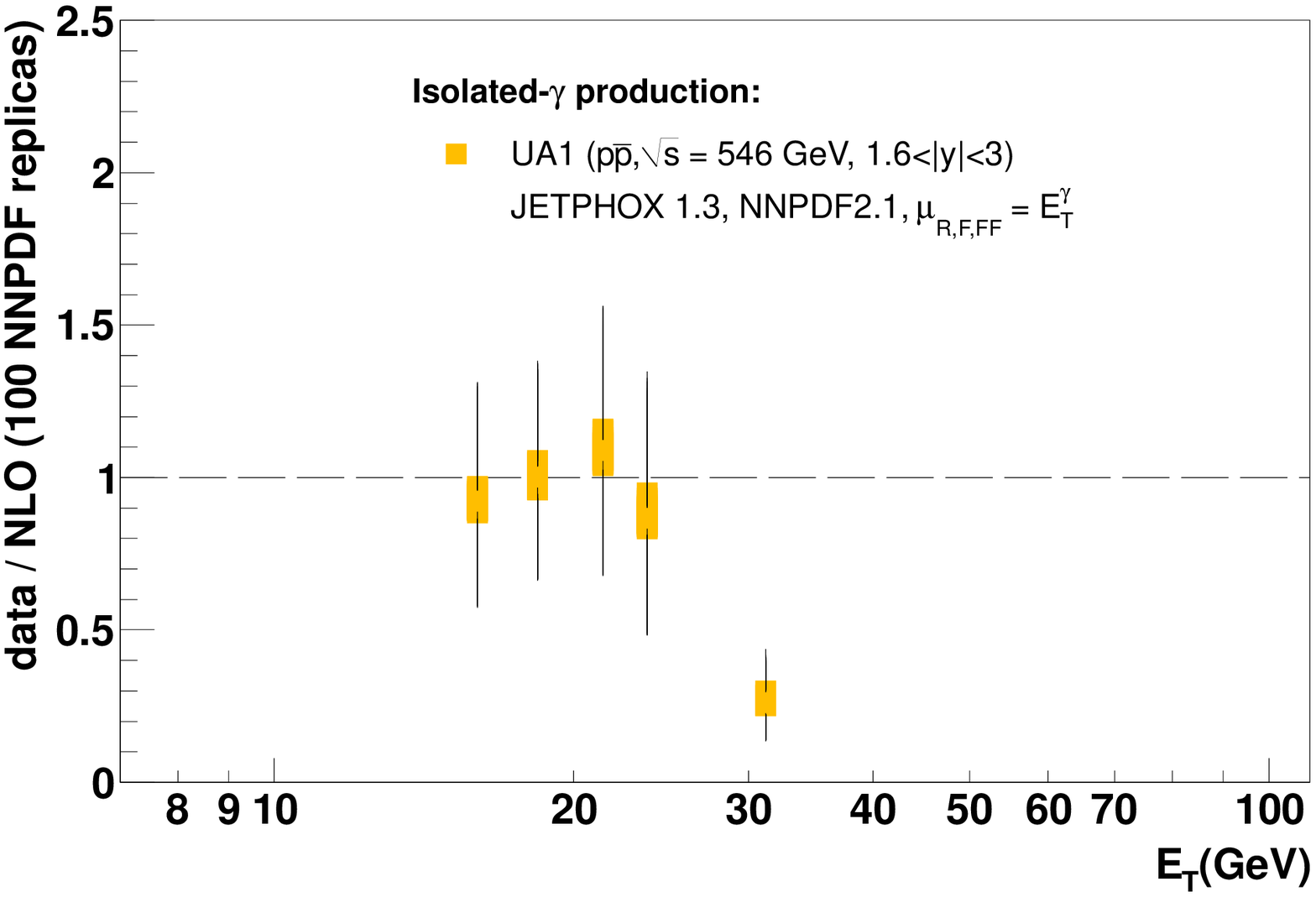}
\caption{\small Ratio of isolated-photon data and NLO pQCD predictions for \pp\ collisions at
$\sqrts$~=~200~GeV (PHENIX) and \ppbar\ collisions at $\sqrts$~=~546~GeV (UA1) and various photon rapidities. 
The (yellow) band indicates the range of predictions for each one of the 100 NNPDF2.1 replicas, 
and the bars show the total experimental uncertainty. 
\label{fig:datatheo_200_546}}
\end{figure}

\begin{figure}[htbp!]
\centering
\epsfig{width=0.32\textwidth,figure=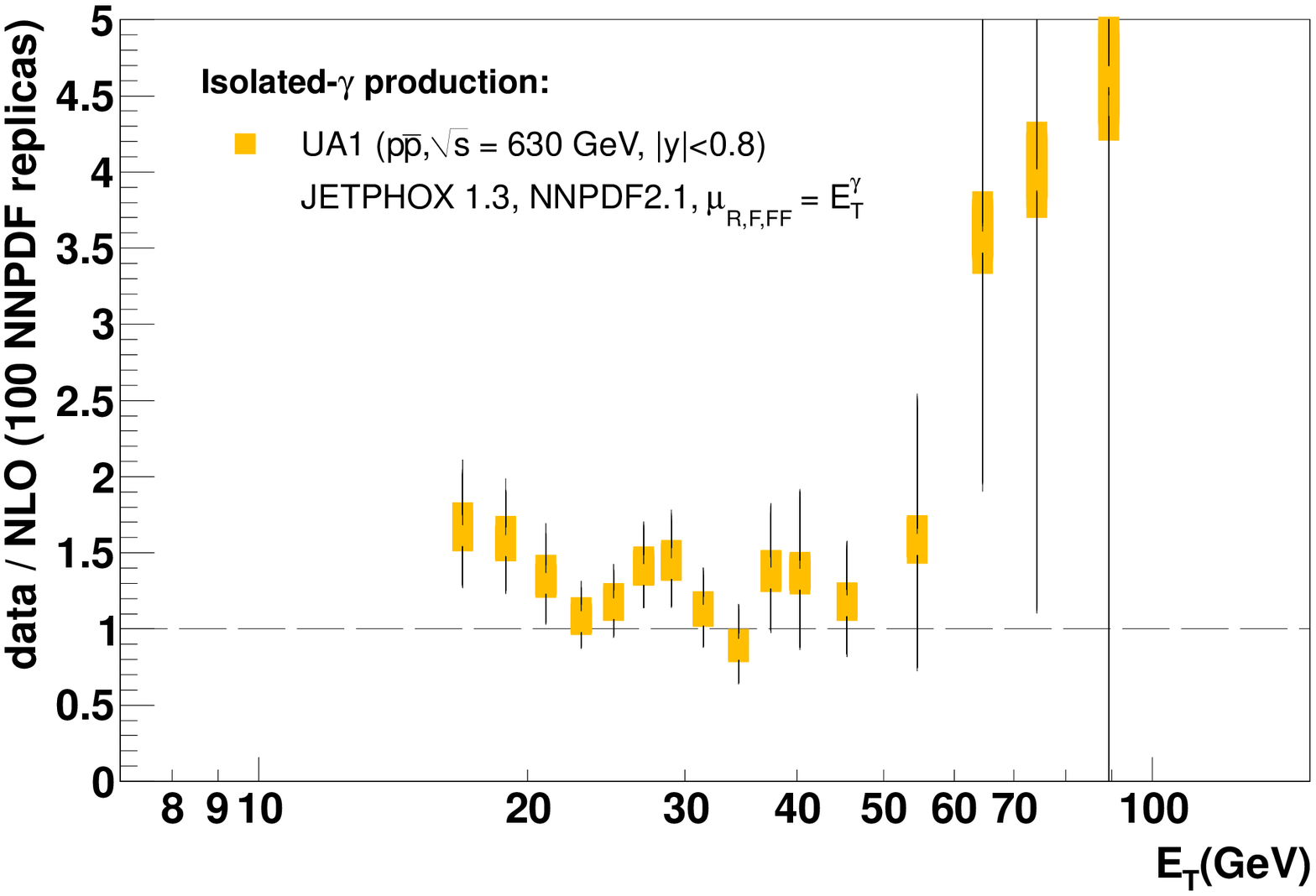}
\epsfig{width=0.32\textwidth,figure=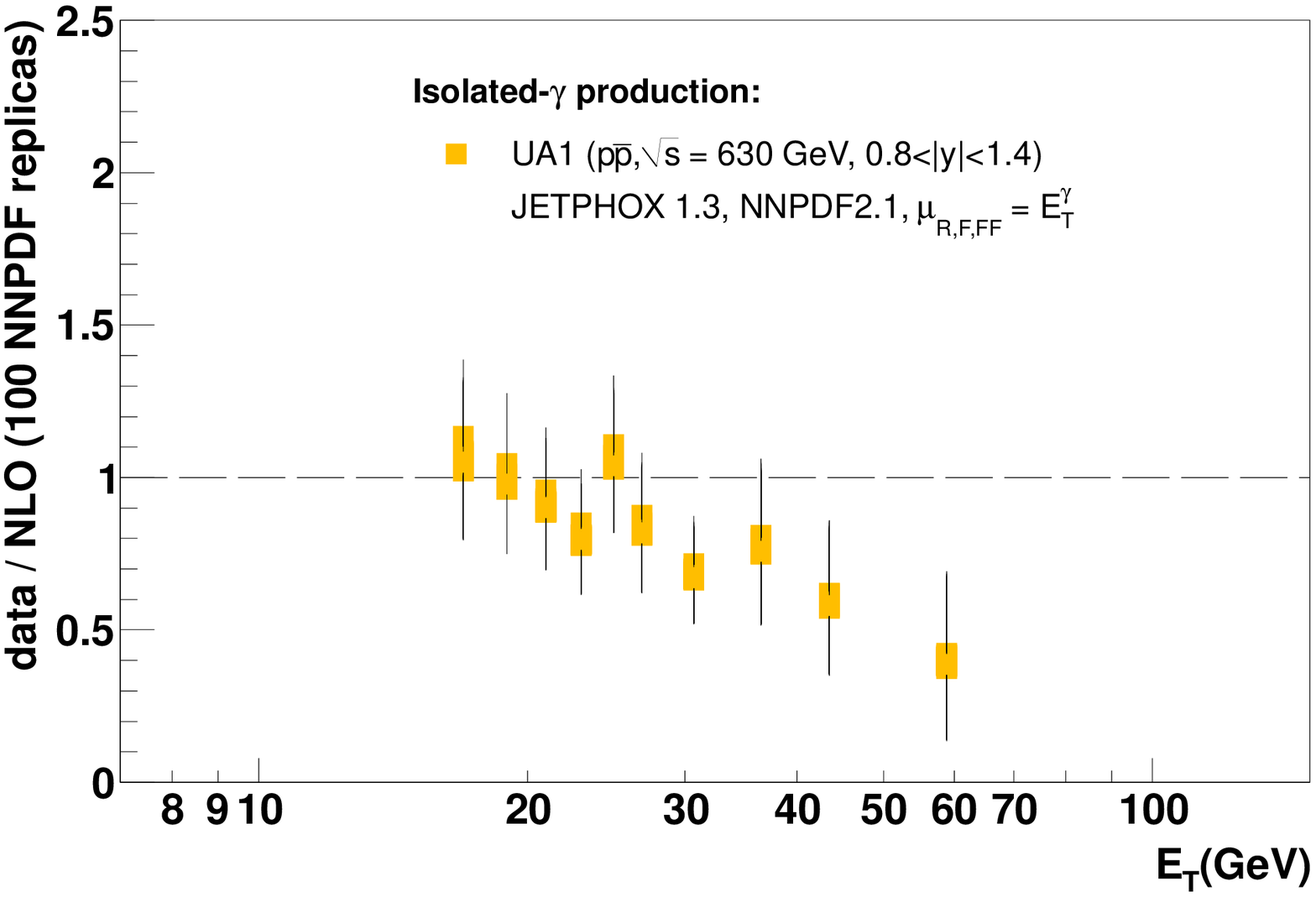}
\epsfig{width=0.32\textwidth,figure=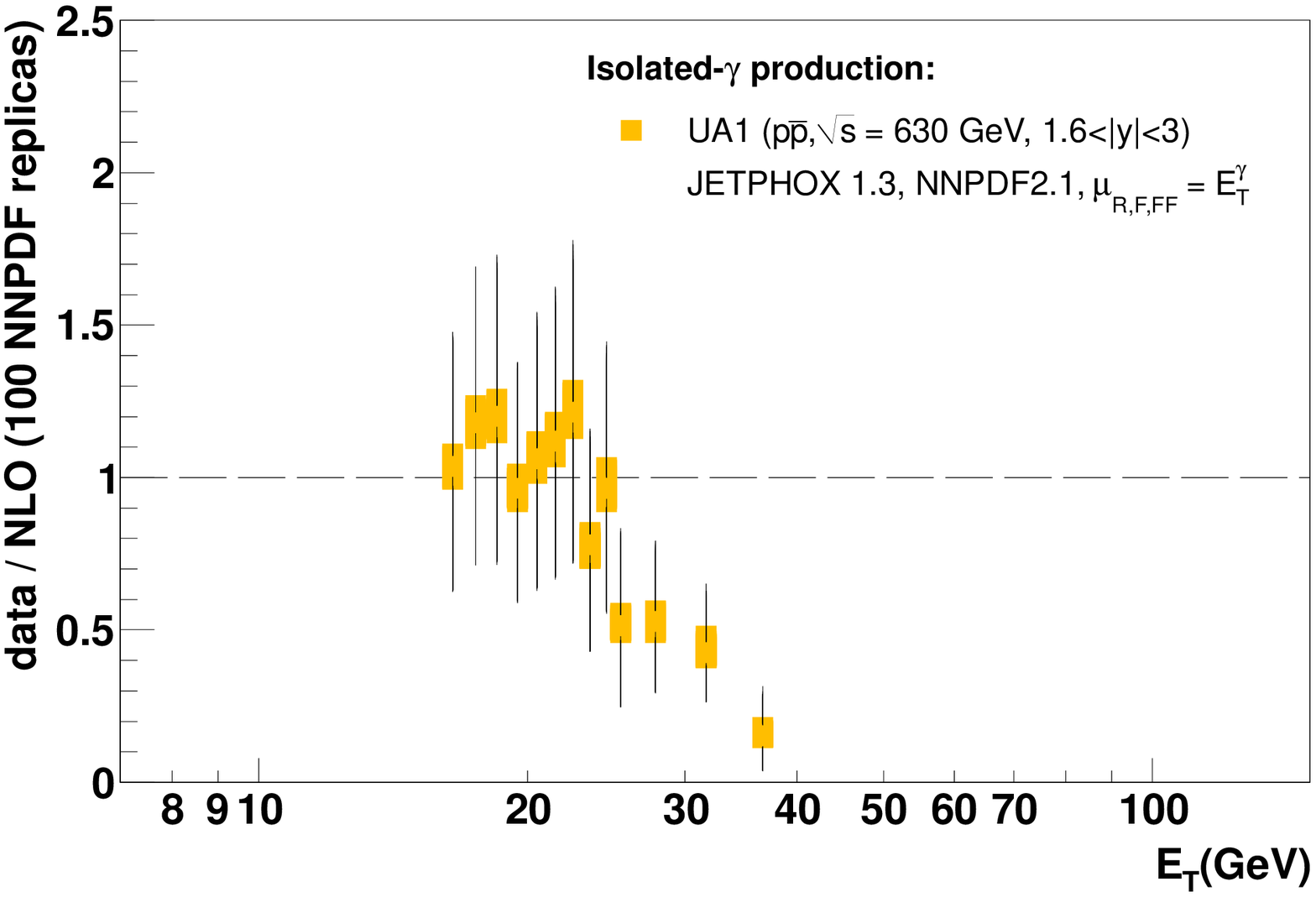}\\
\epsfig{width=0.32\textwidth,figure=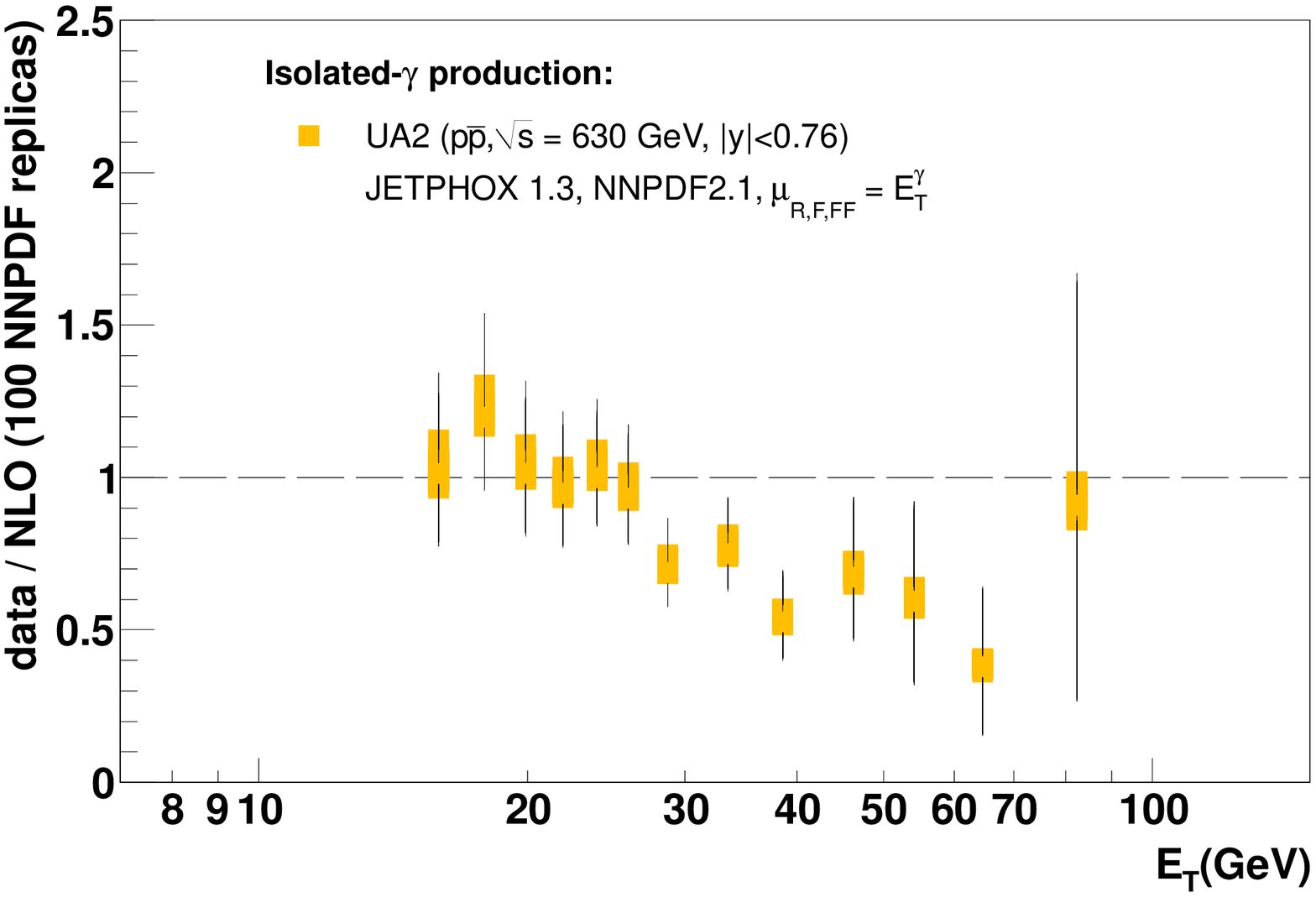}
\epsfig{width=0.32\textwidth,figure=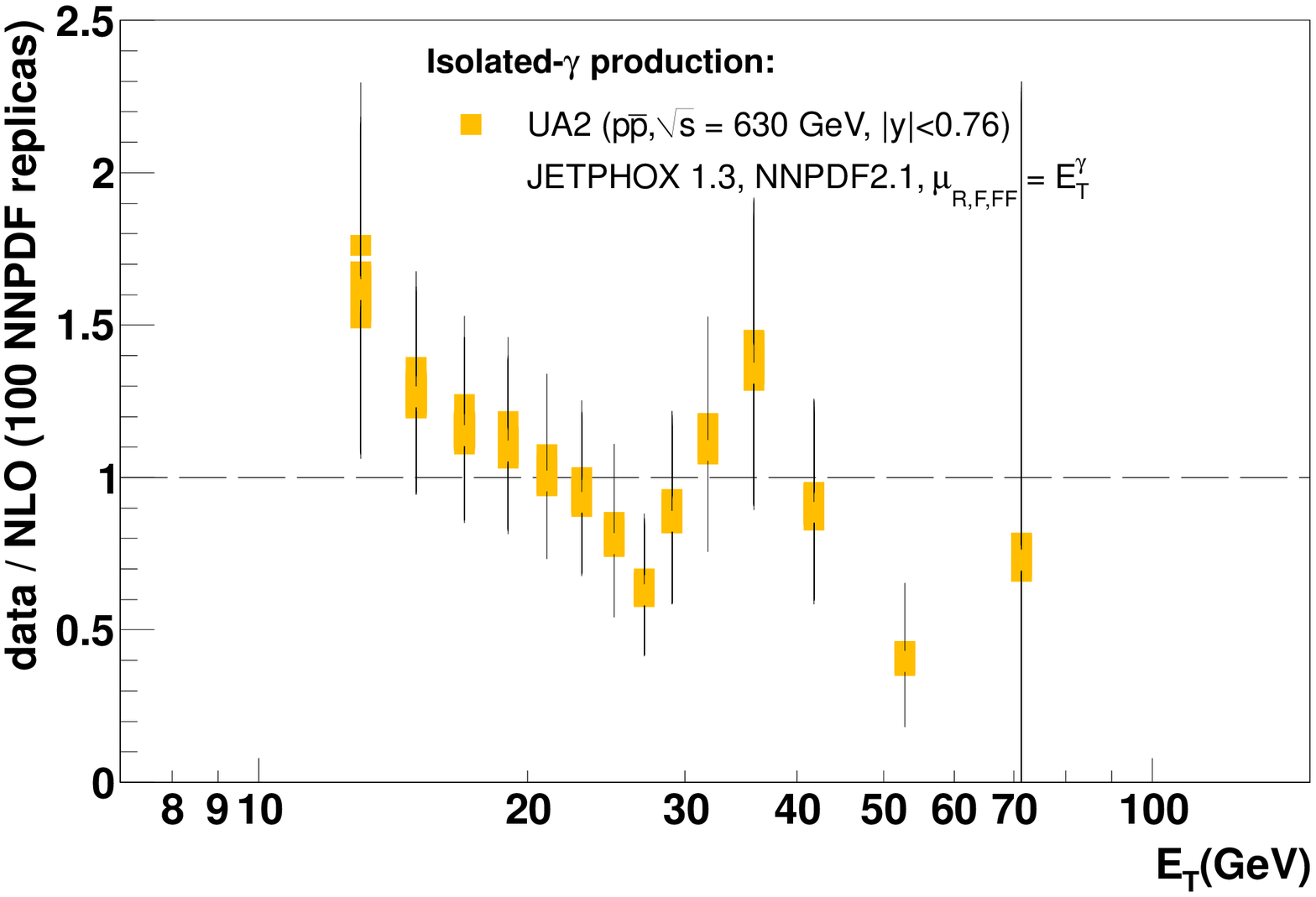}
\epsfig{width=0.32\textwidth,figure=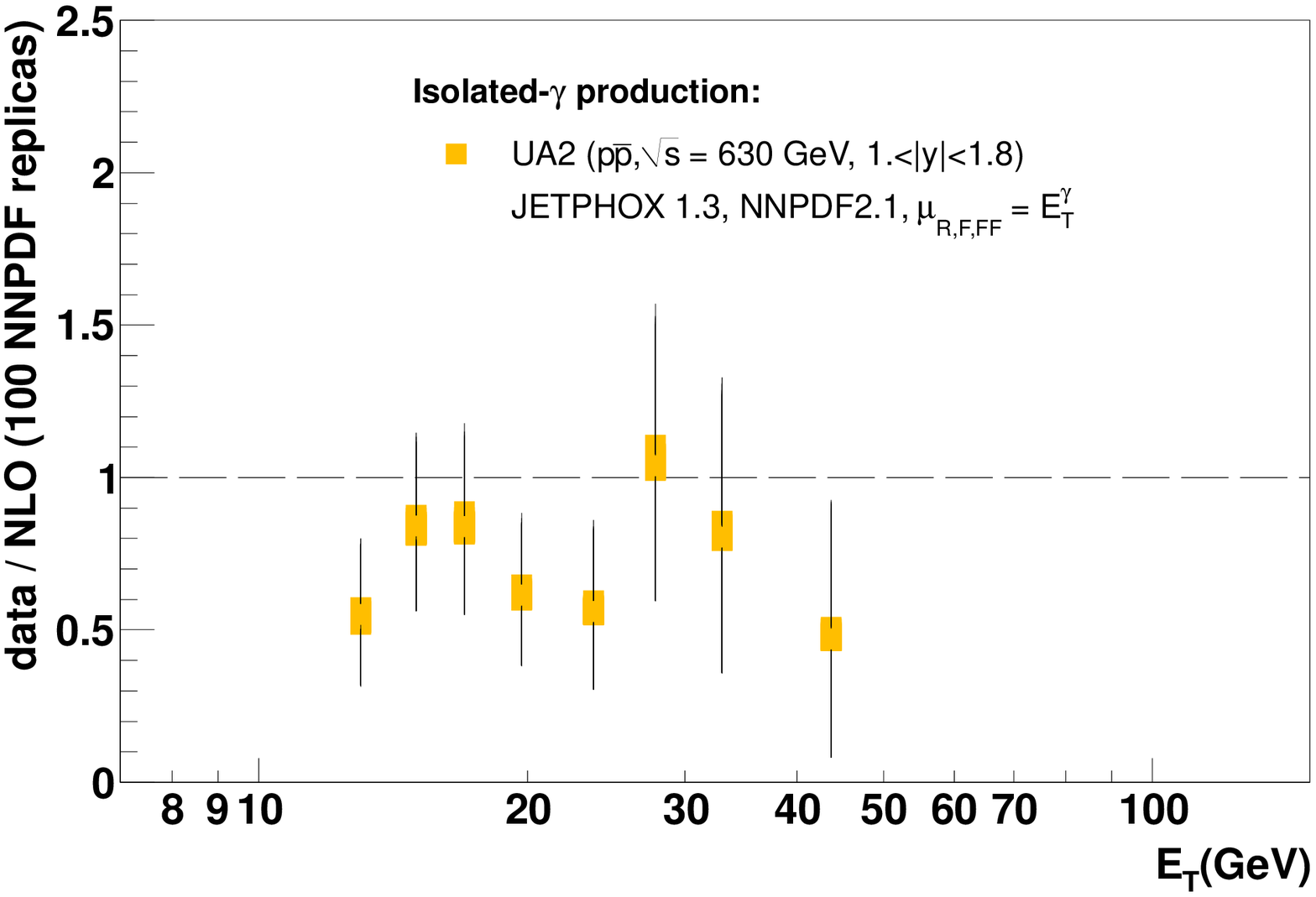}\\
\epsfig{width=0.32\textwidth,figure=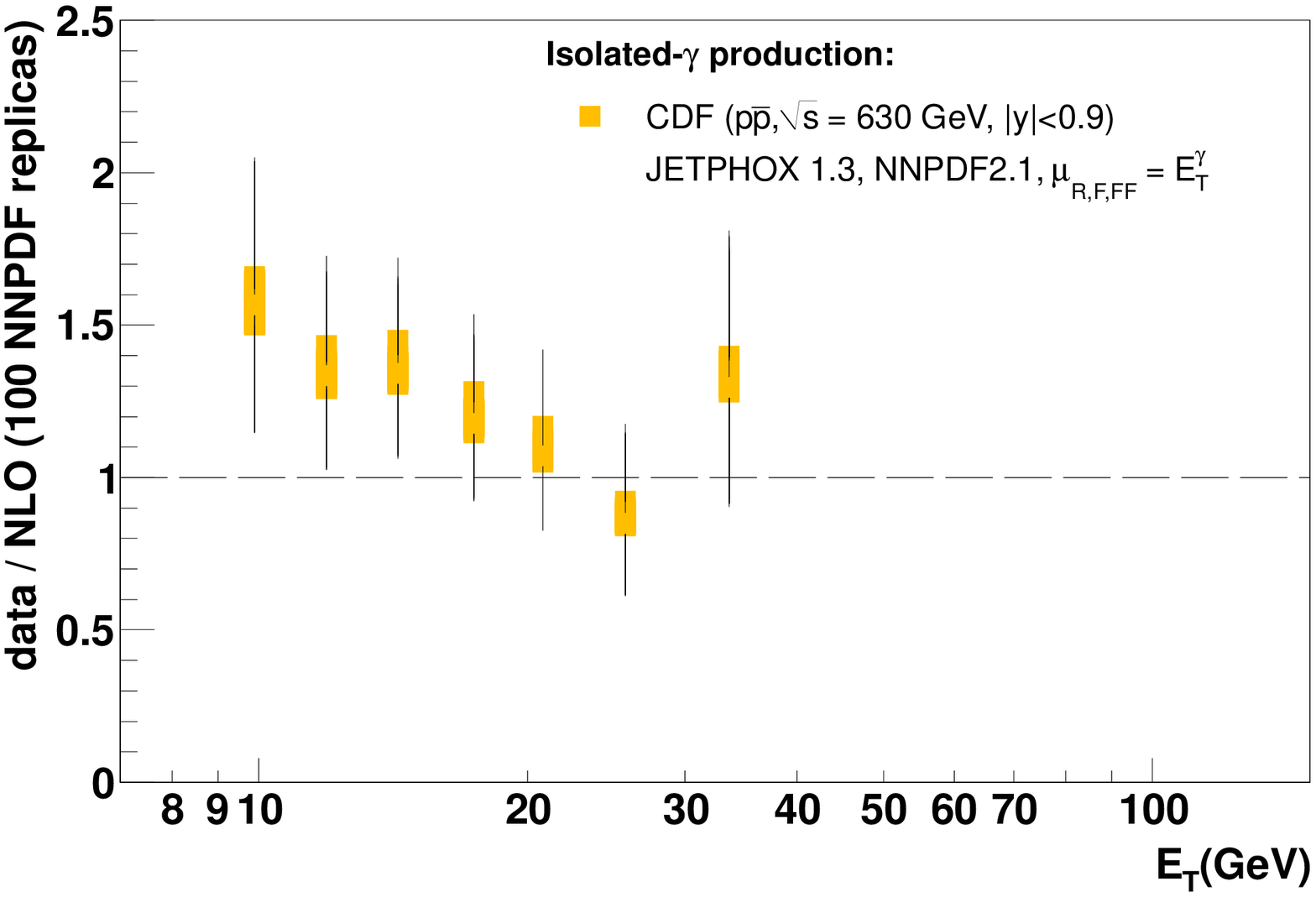}
\epsfig{width=0.32\textwidth,figure=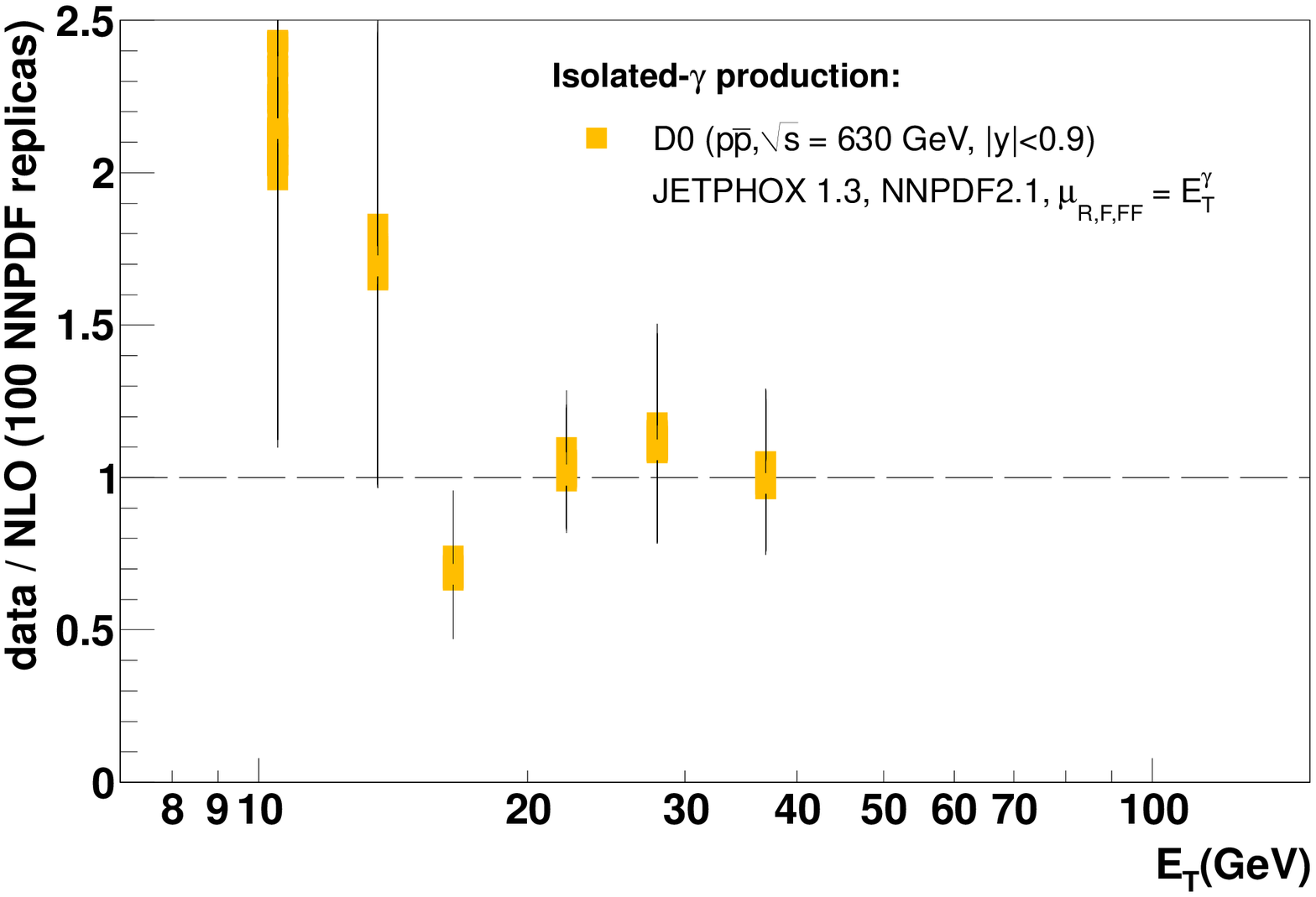}
\epsfig{width=0.32\textwidth,figure=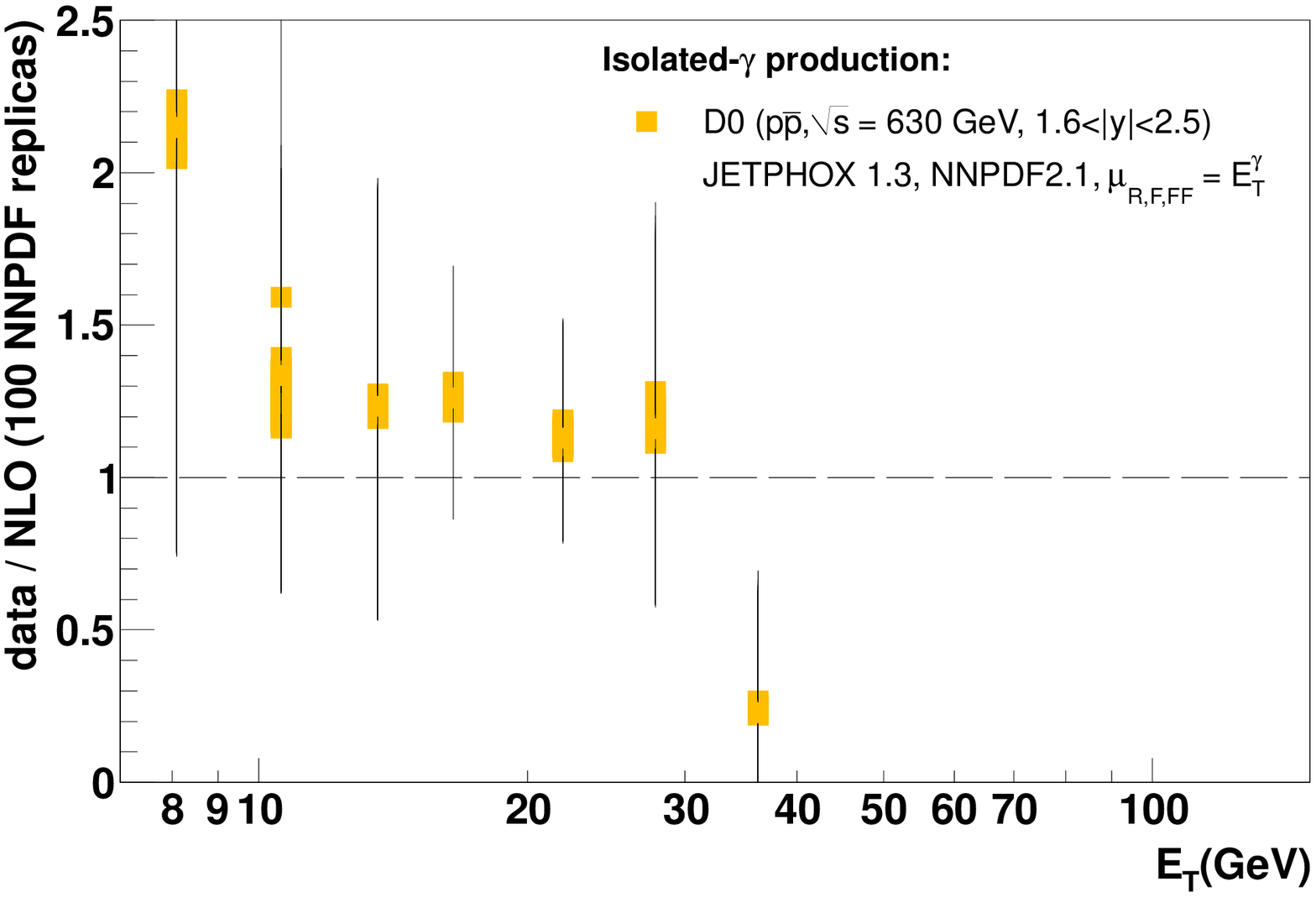}
\caption{\small Ratio of isolated-photon data  (UA1 and UA2 in the 3 top and middle plots, and CDF and D0 in the 3
  bottom panels) and NLO pQCD predictions for \ppbar\ collisions at $\sqrts$~=~630~GeV at various photon rapidities. 
The (yellow) band indicates the range of predictions for each one of the 100 NNPDF2.1 replicas, 
and the bars show the total experimental uncertainty.
\label{fig:datatheo_630}}
\end{figure}

\begin{figure}[htbp!]
\centering
\epsfig{width=0.43\textwidth,figure=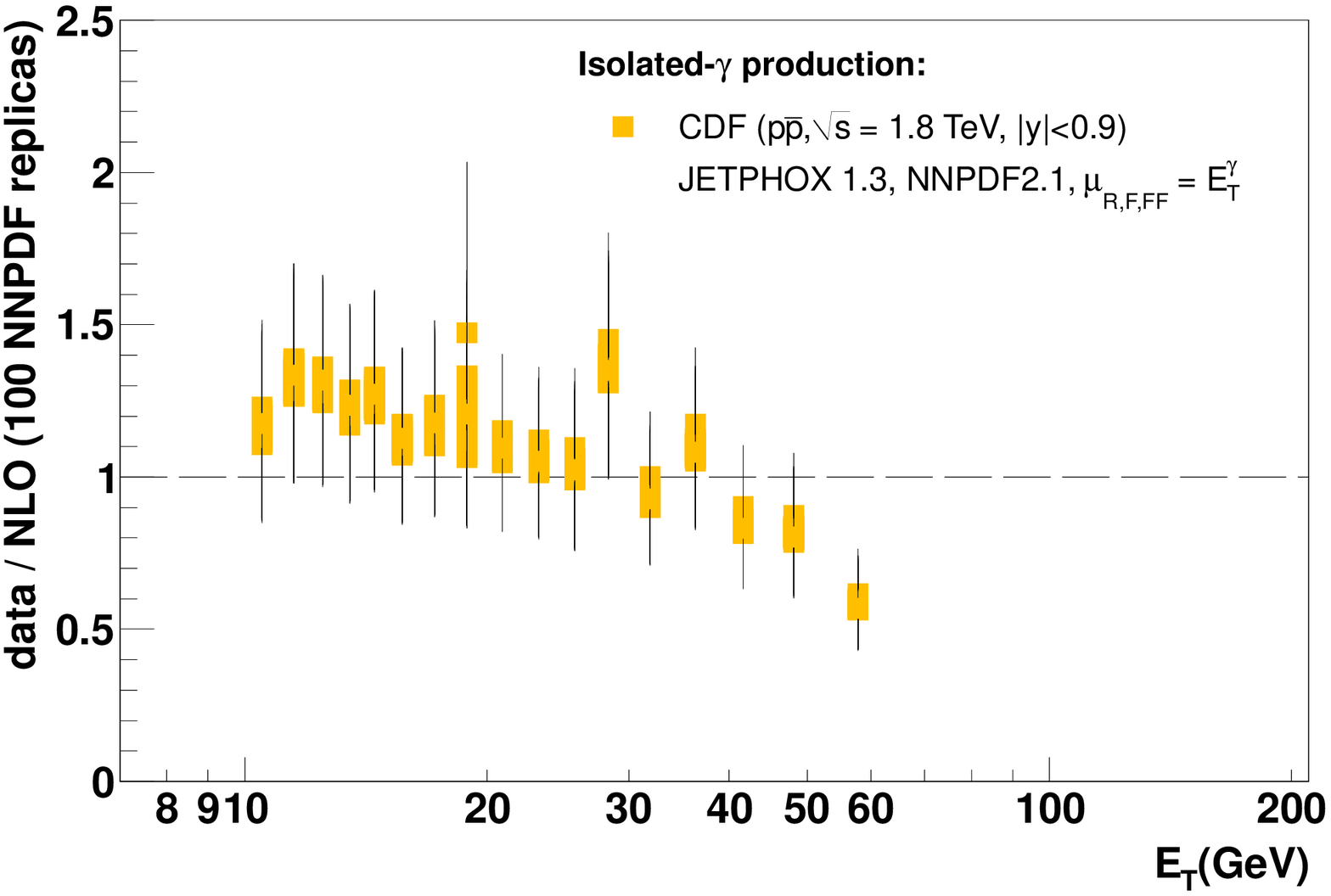}
\epsfig{width=0.43\textwidth,figure=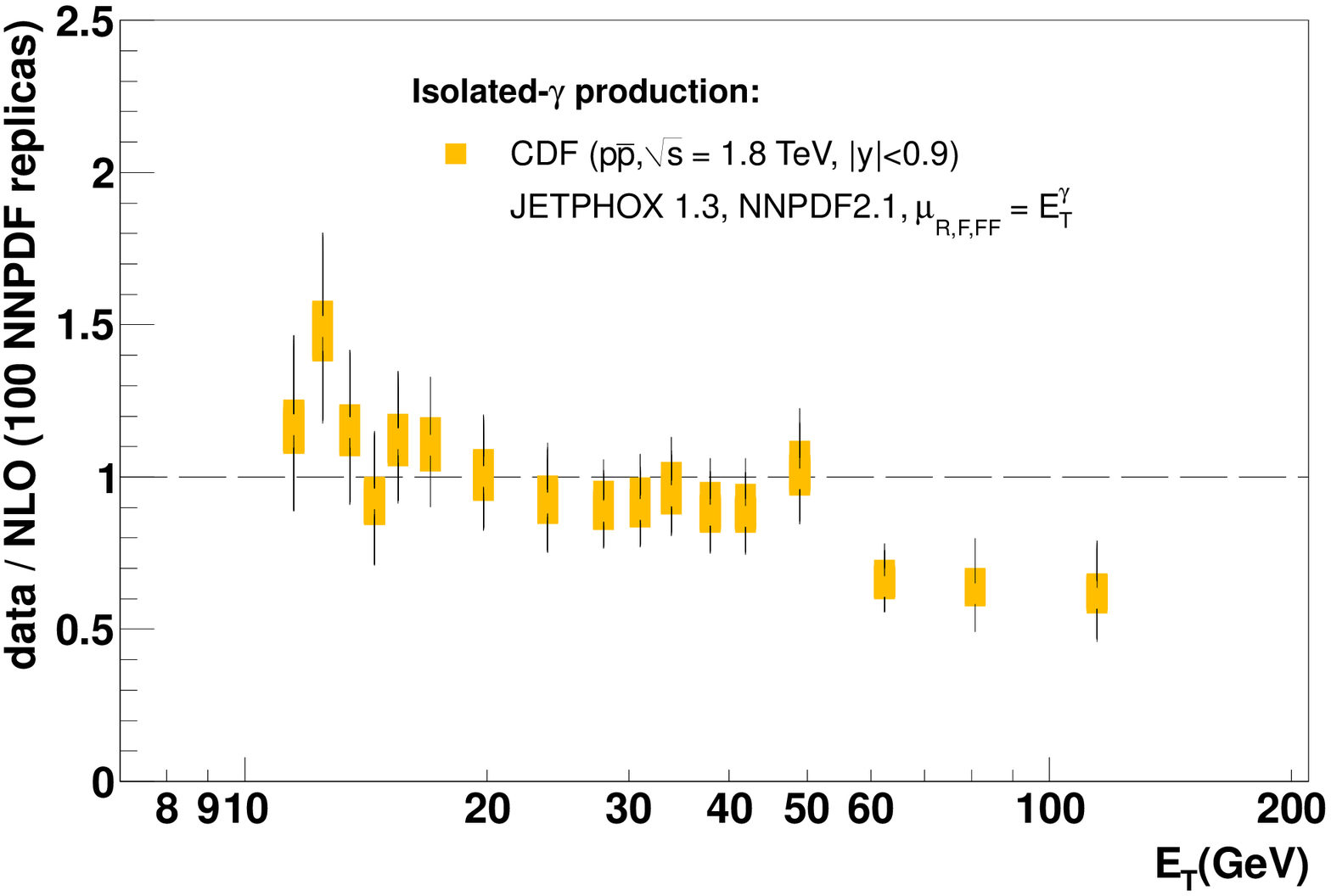}\\
\epsfig{width=0.43\textwidth,figure=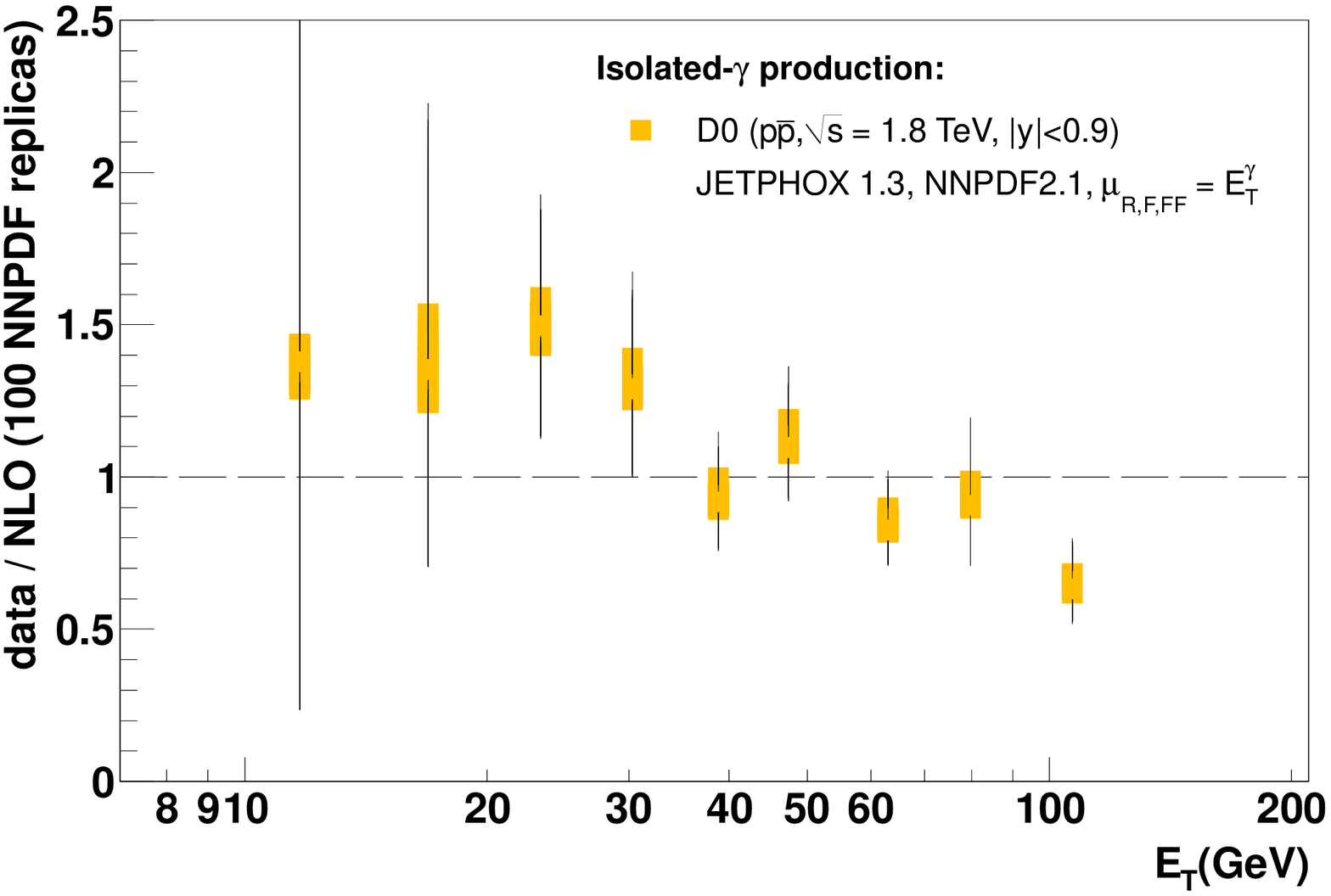}
\epsfig{width=0.43\textwidth,figure=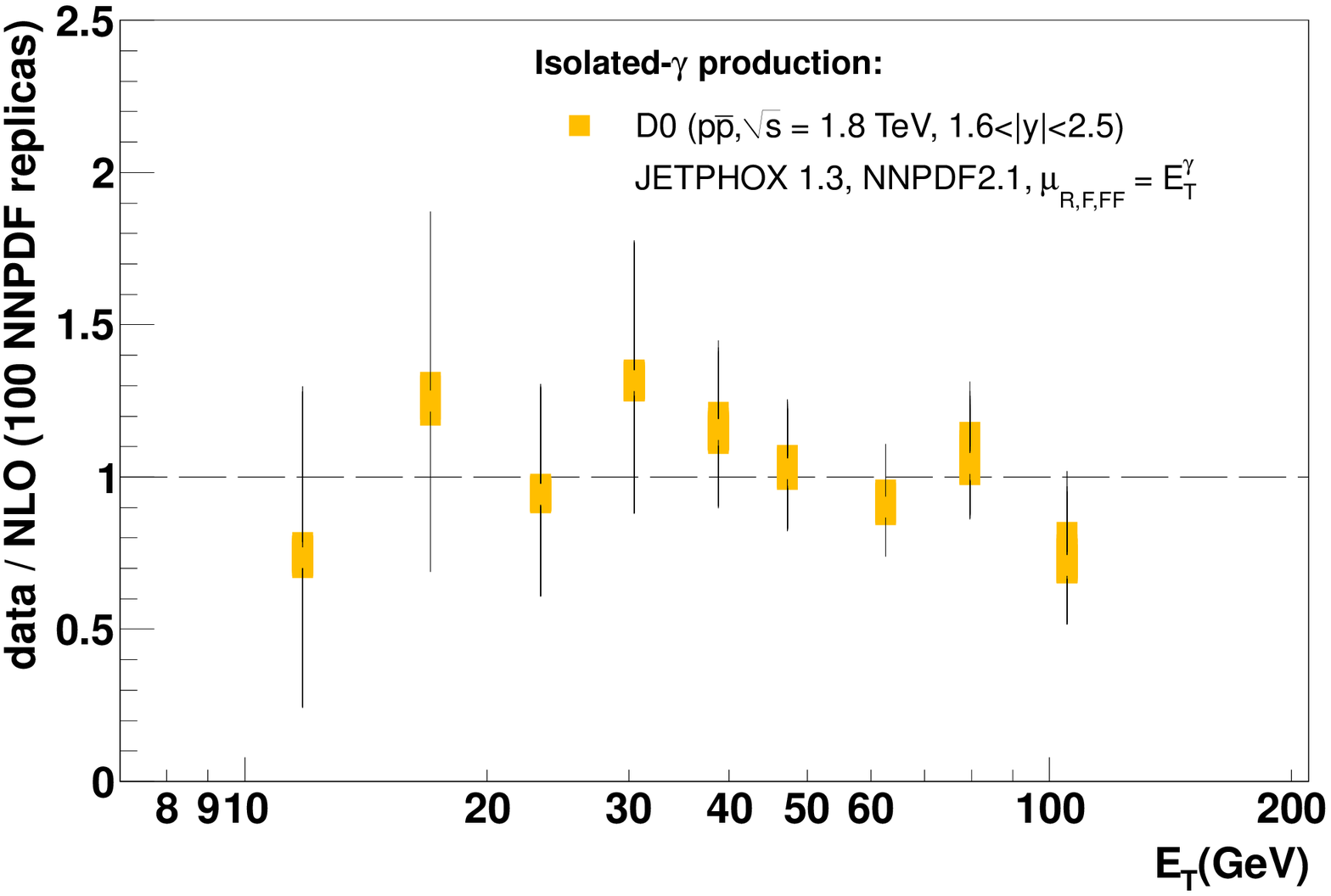}
\caption{\small Ratio of isolated-photon data (CDF and D0) and NLO pQCD predictions for \ppbar\ collisions at
$\sqrts$~=~1.8~TeV at various photon rapidities. 
The (yellow) band indicates the range of predictions for each one of the 100 NNPDF2.1 replicas, 
and the bars show the total experimental uncertainty.
\label{fig:datatheo_tev1}}
\end{figure}

\begin{figure}[htbp!]
\centering
\epsfig{width=0.49\textwidth,figure=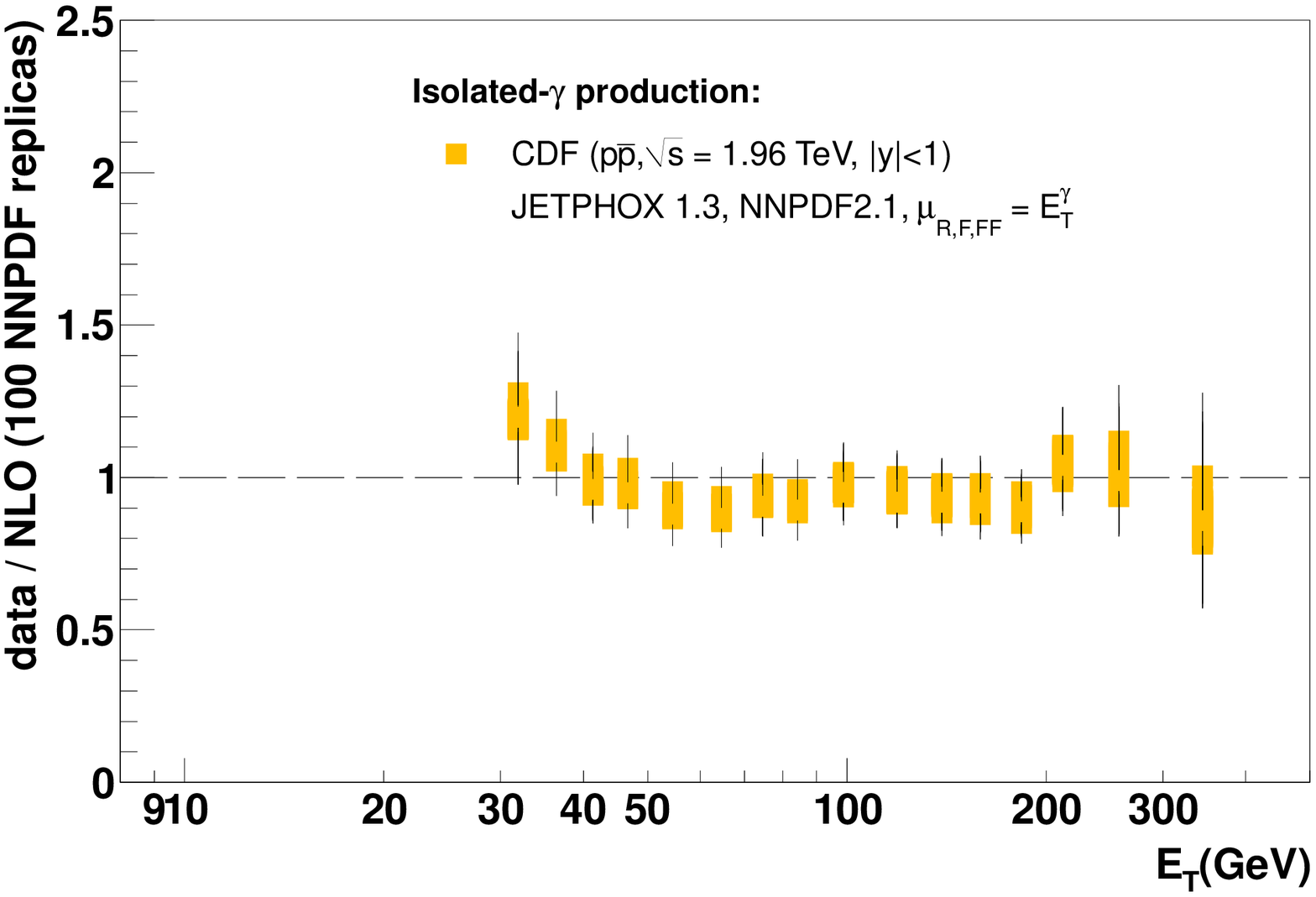}
\epsfig{width=0.49\textwidth,figure=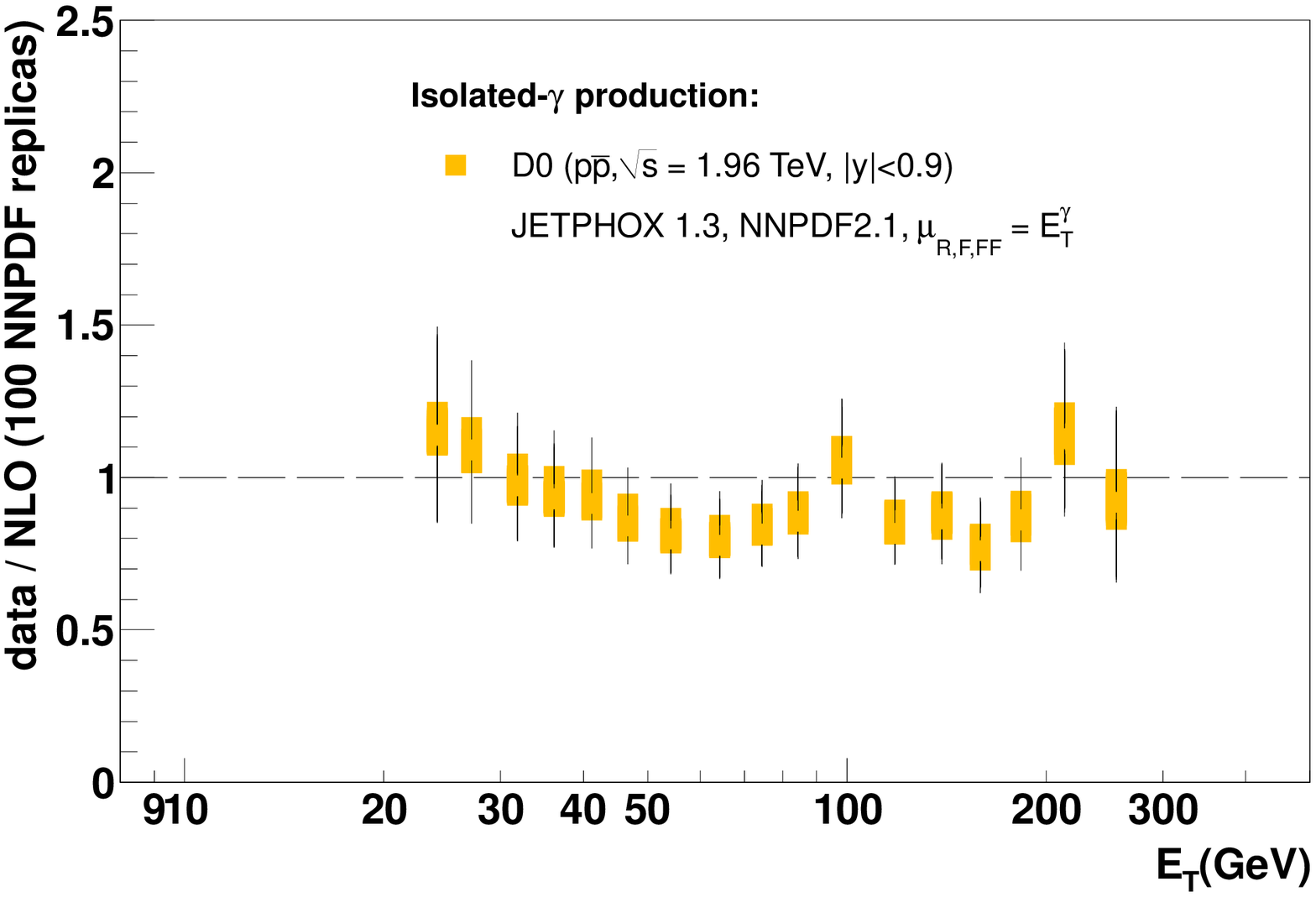}
\caption{\small Ratio of isolated-photon data (CDF and D0) and NLO pQCD predictions for \ppbar\ collisions at
$\sqrts$~=~1.96~TeV at central rapidities. 
The (yellow) band indicates the range of predictions for each one of the 100 NNPDF2.1 replicas, 
and the bars show the total experimental uncertainty.
\label{fig:datatheo_tev2}}
\end{figure}

\begin{figure}[htbp!]
\centering
\epsfig{width=0.32\textwidth,figure=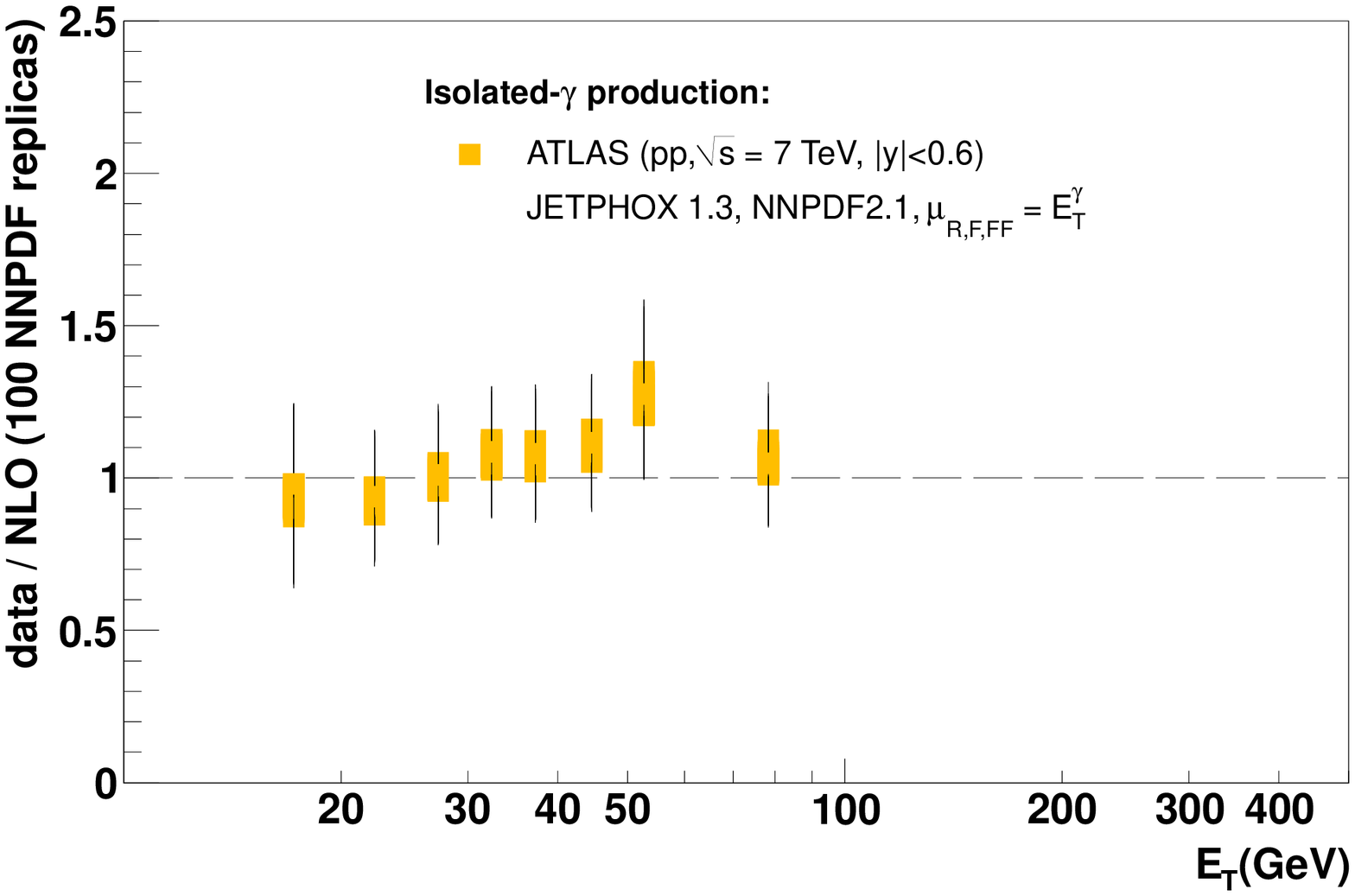}
\epsfig{width=0.32\textwidth,figure=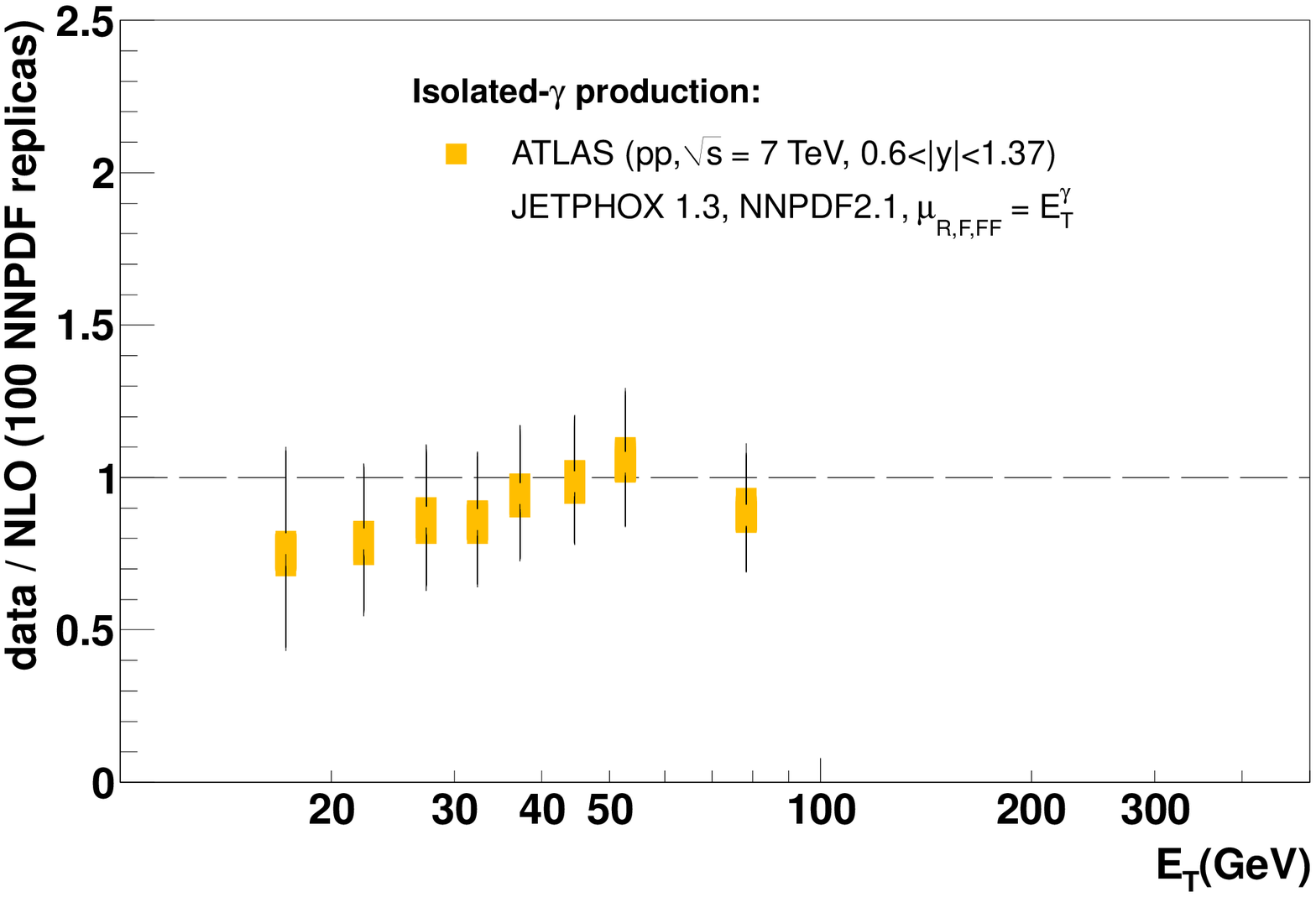}
\epsfig{width=0.32\textwidth,figure=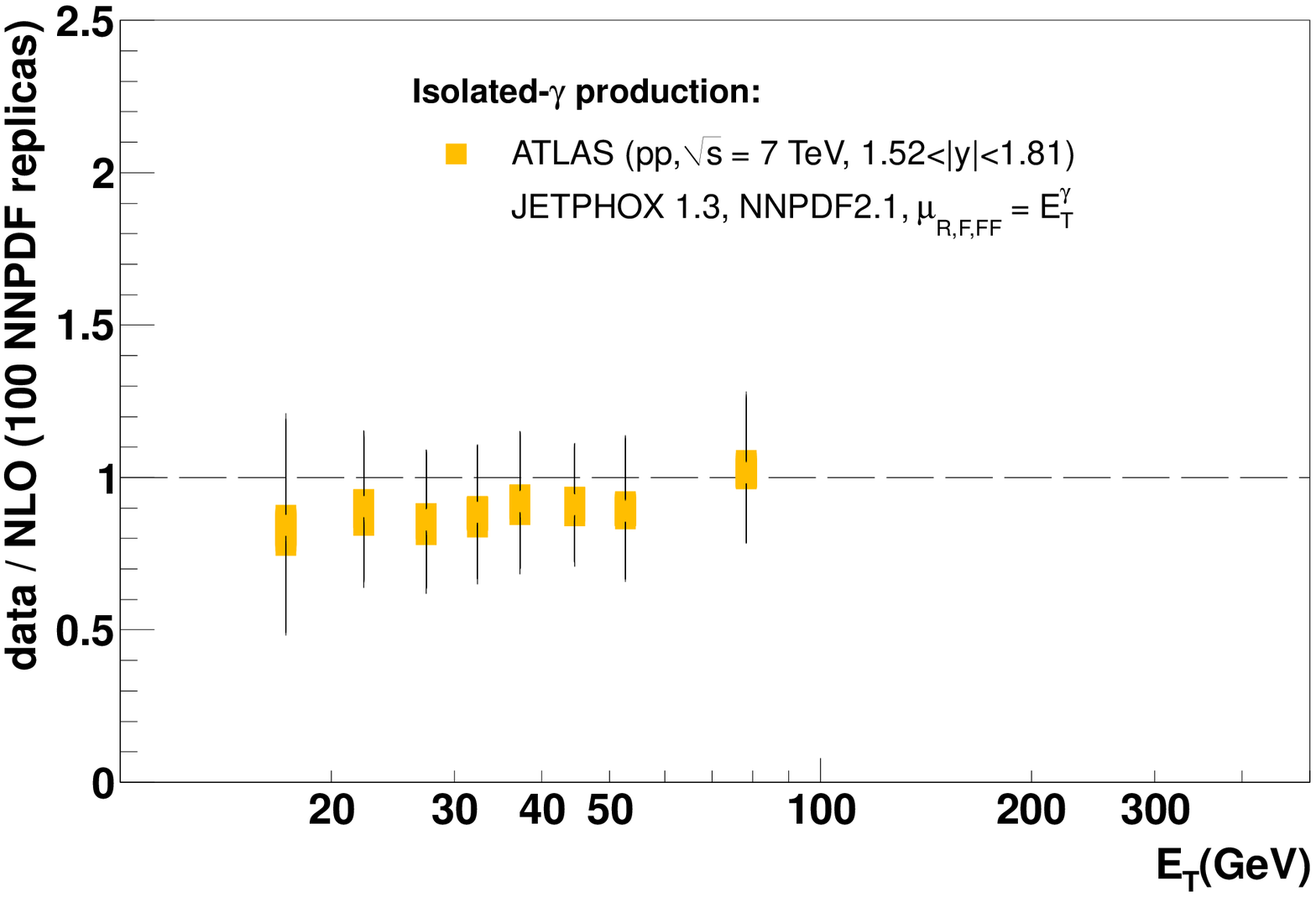}
\epsfig{width=0.32\textwidth,figure=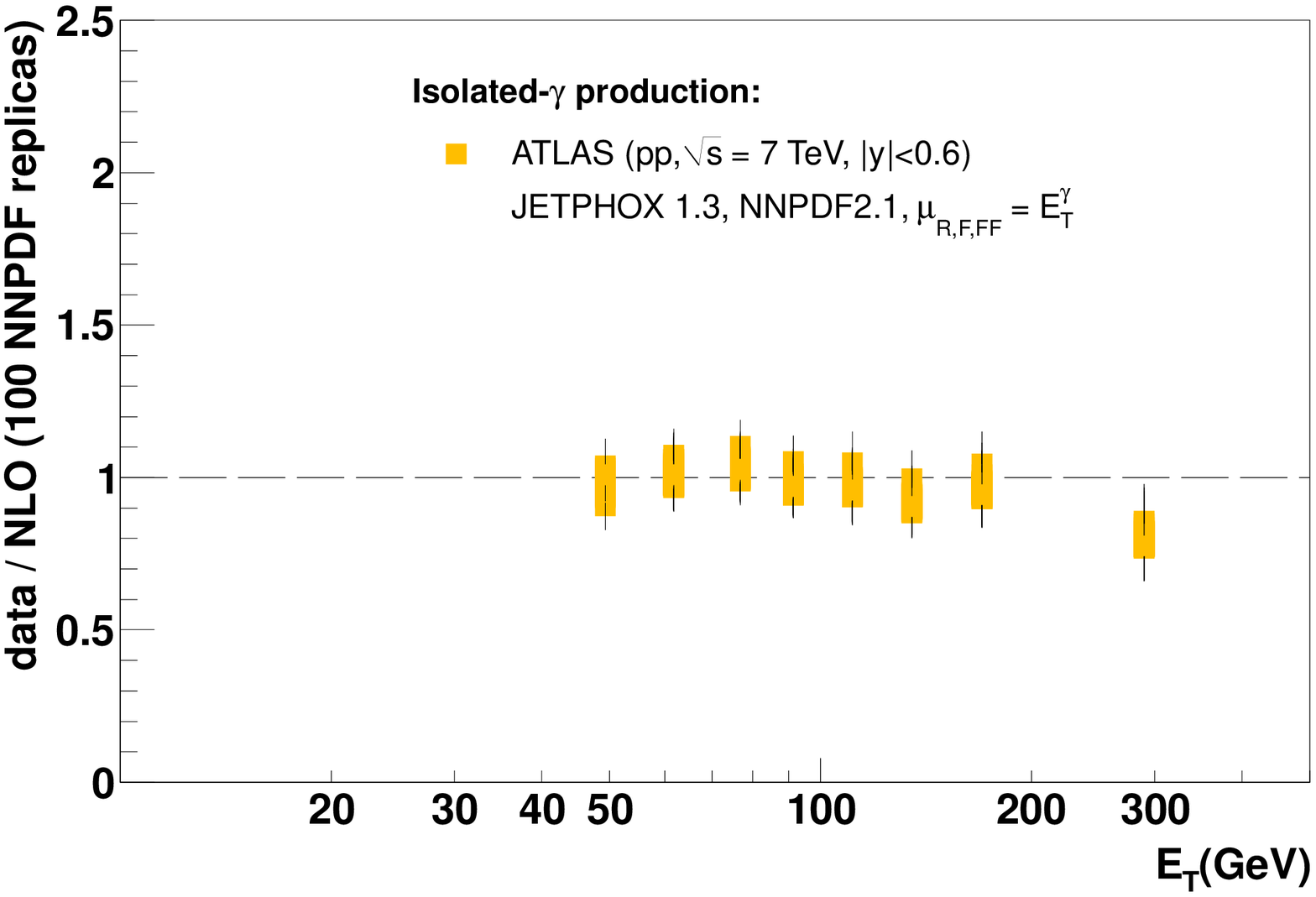}
\epsfig{width=0.32\textwidth,figure=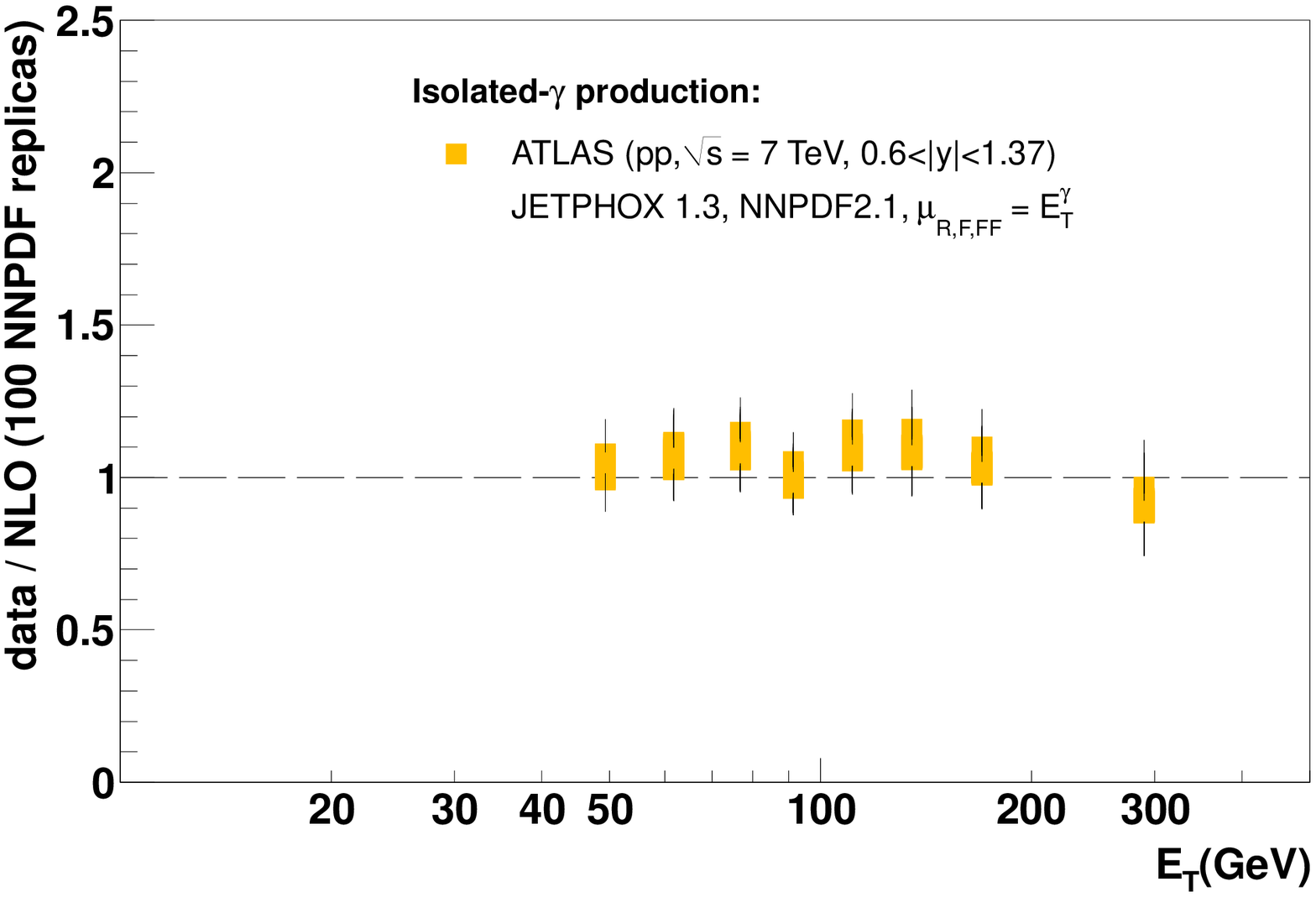}
\epsfig{width=0.32\textwidth,figure=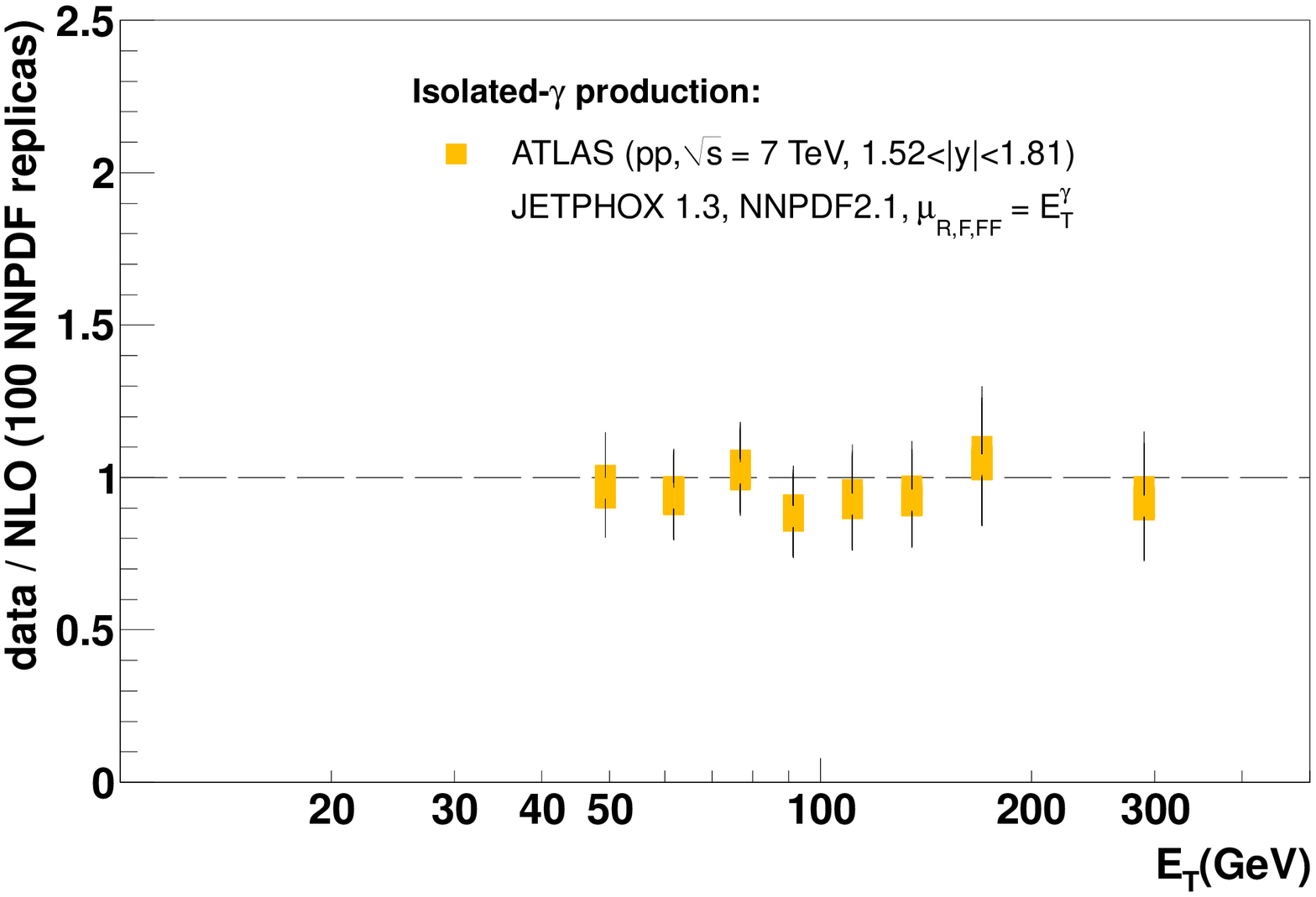}
\caption{\small Ratio of ATLAS isolated-photon datasets (880~nb$^{-1}$, top, and 36~pb$^{-1}$, bottom) 
and NLO pQCD predictions in \pp\ collisions at $\sqrts$~=~7~TeV at various photon rapidities. 
The (yellow) band indicates the range of predictions for each one of the 100 NNPDF2.1 replicas, 
and the bars show the total experimental uncertainty.
\label{fig:datatheo_atlas}}
\end{figure}

\begin{figure}[htbp!]
\centering
\epsfig{width=0.32\textwidth,figure=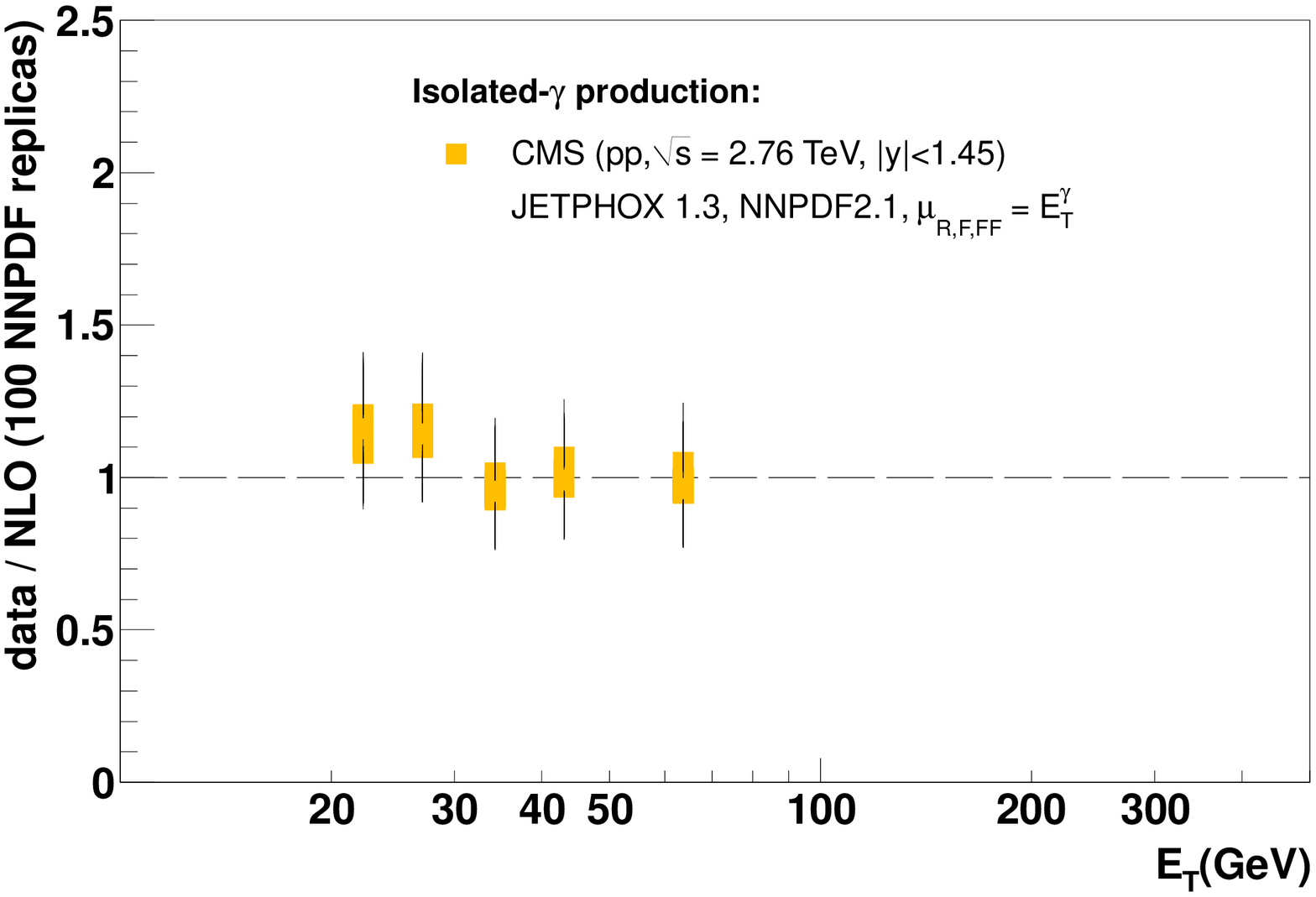}
\epsfig{width=0.32\textwidth,figure=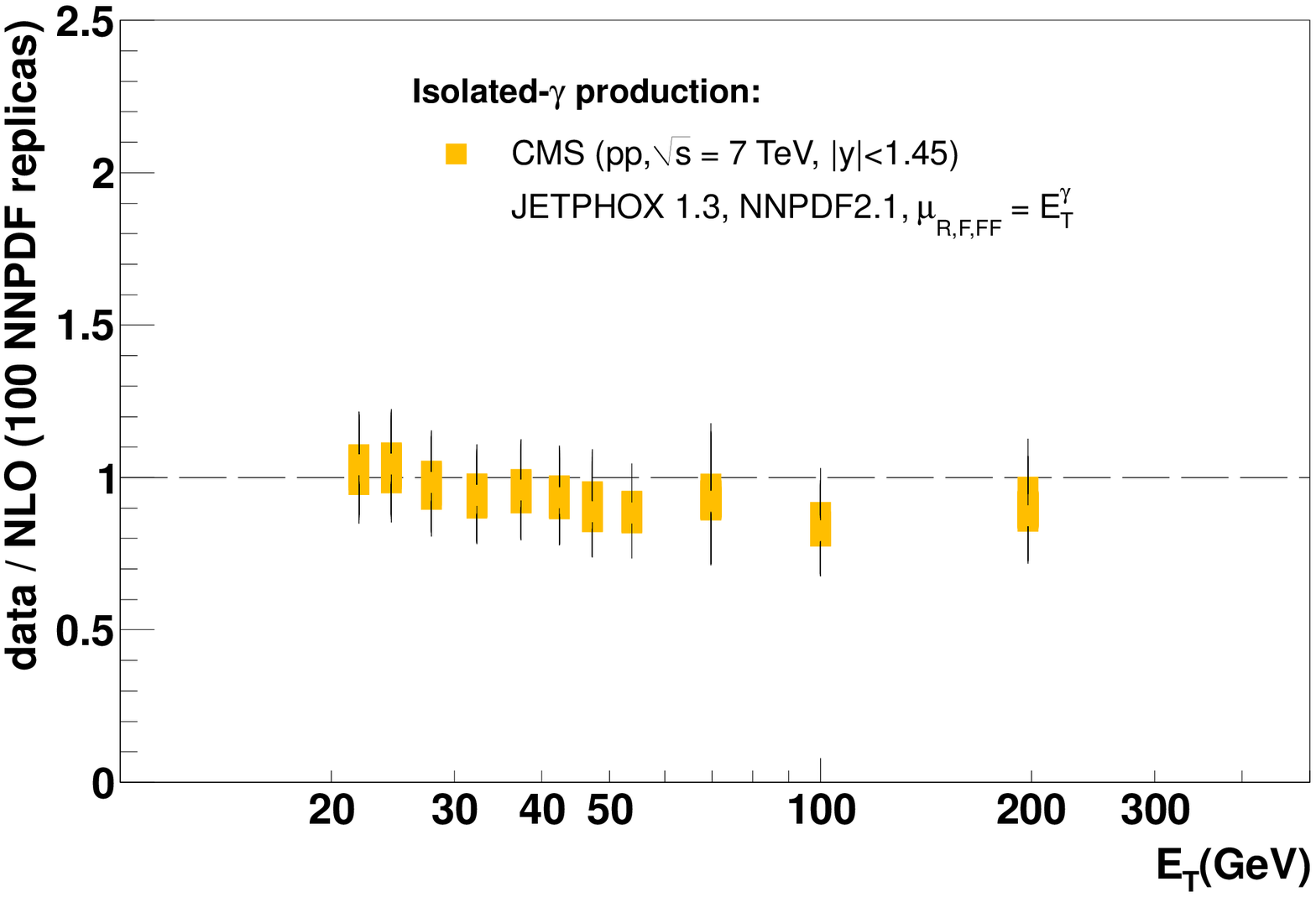}
\epsfig{width=0.32\textwidth,figure=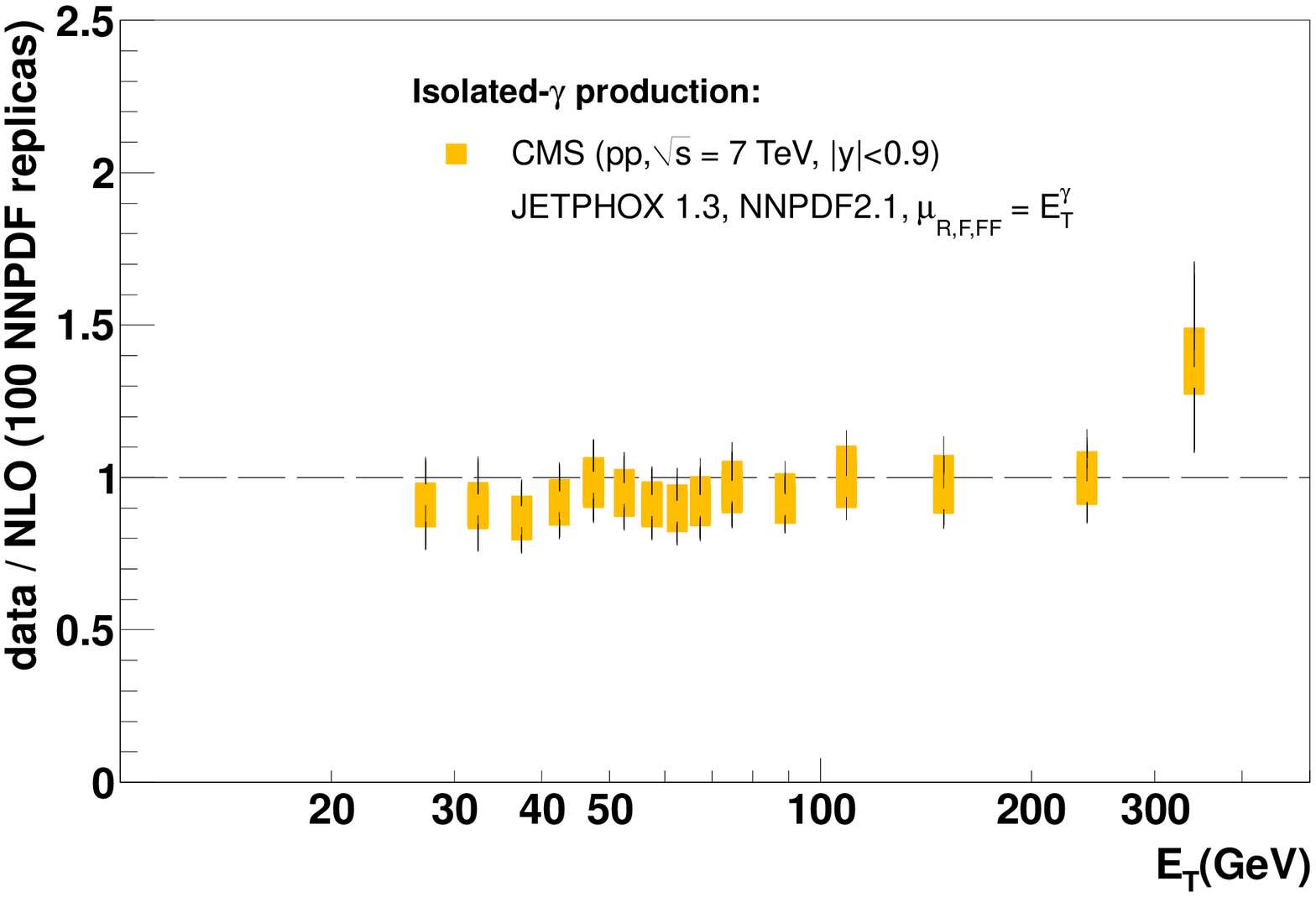}\\
\epsfig{width=0.32\textwidth,figure=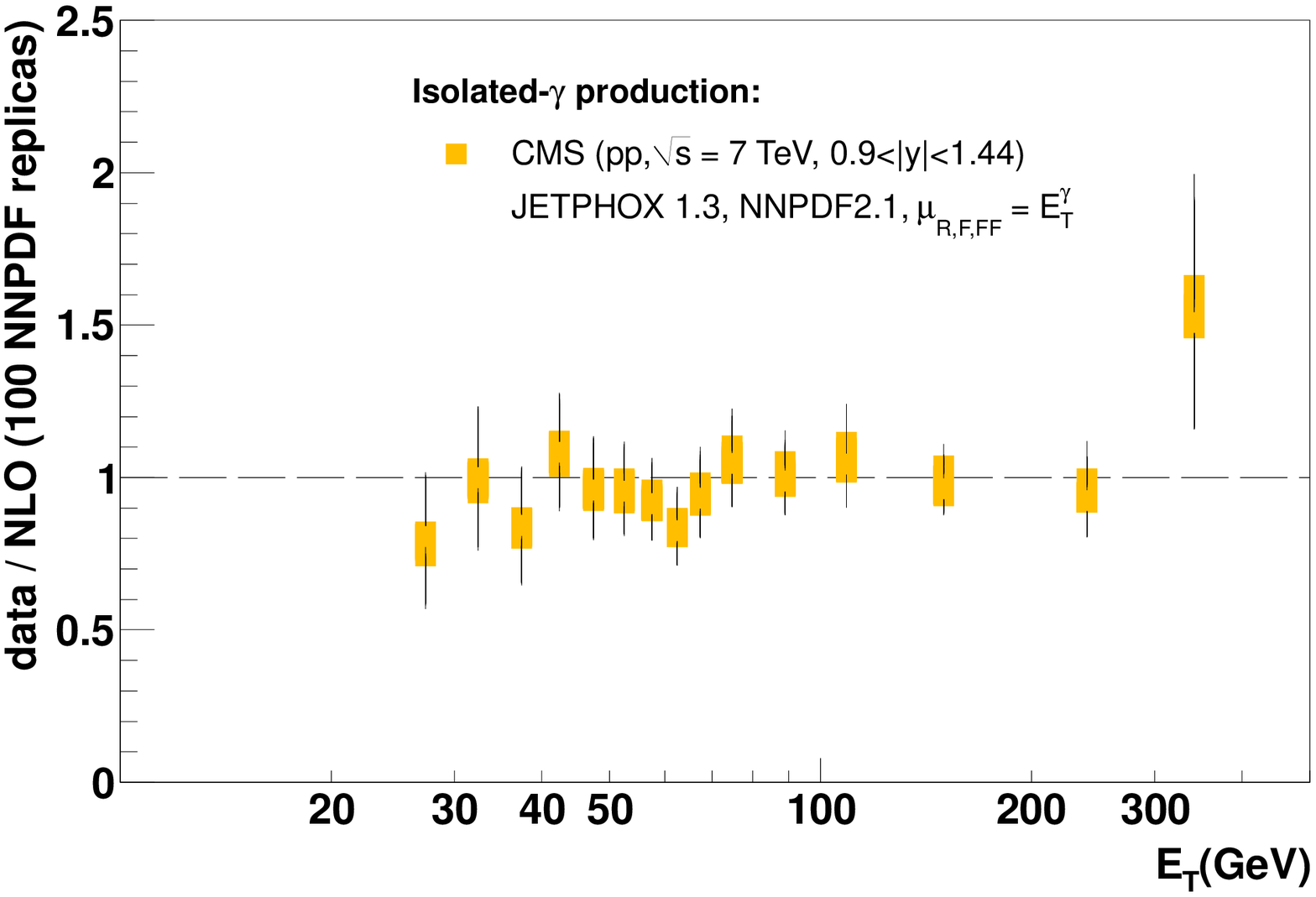}
\epsfig{width=0.32\textwidth,figure=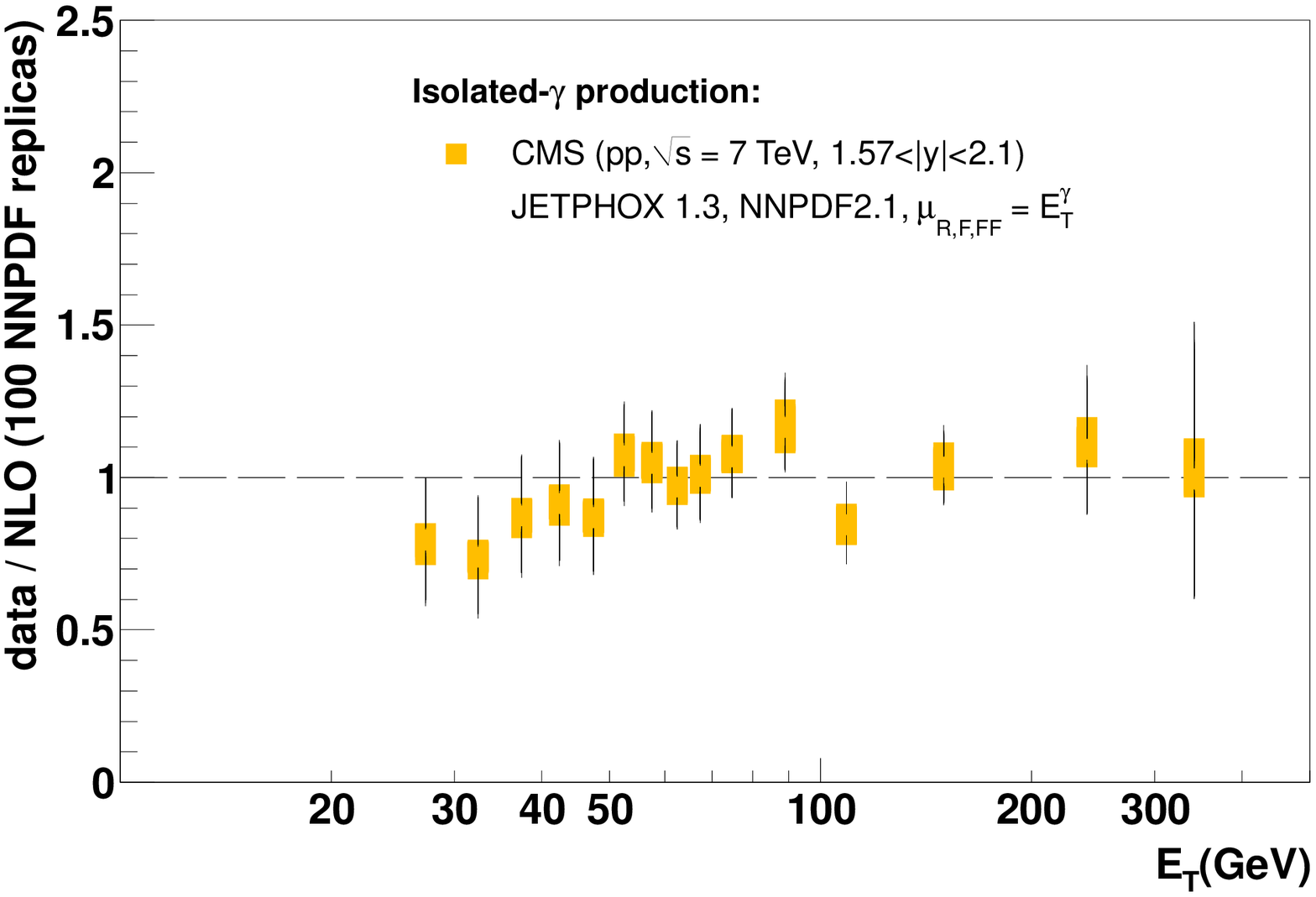}
\epsfig{width=0.32\textwidth,figure=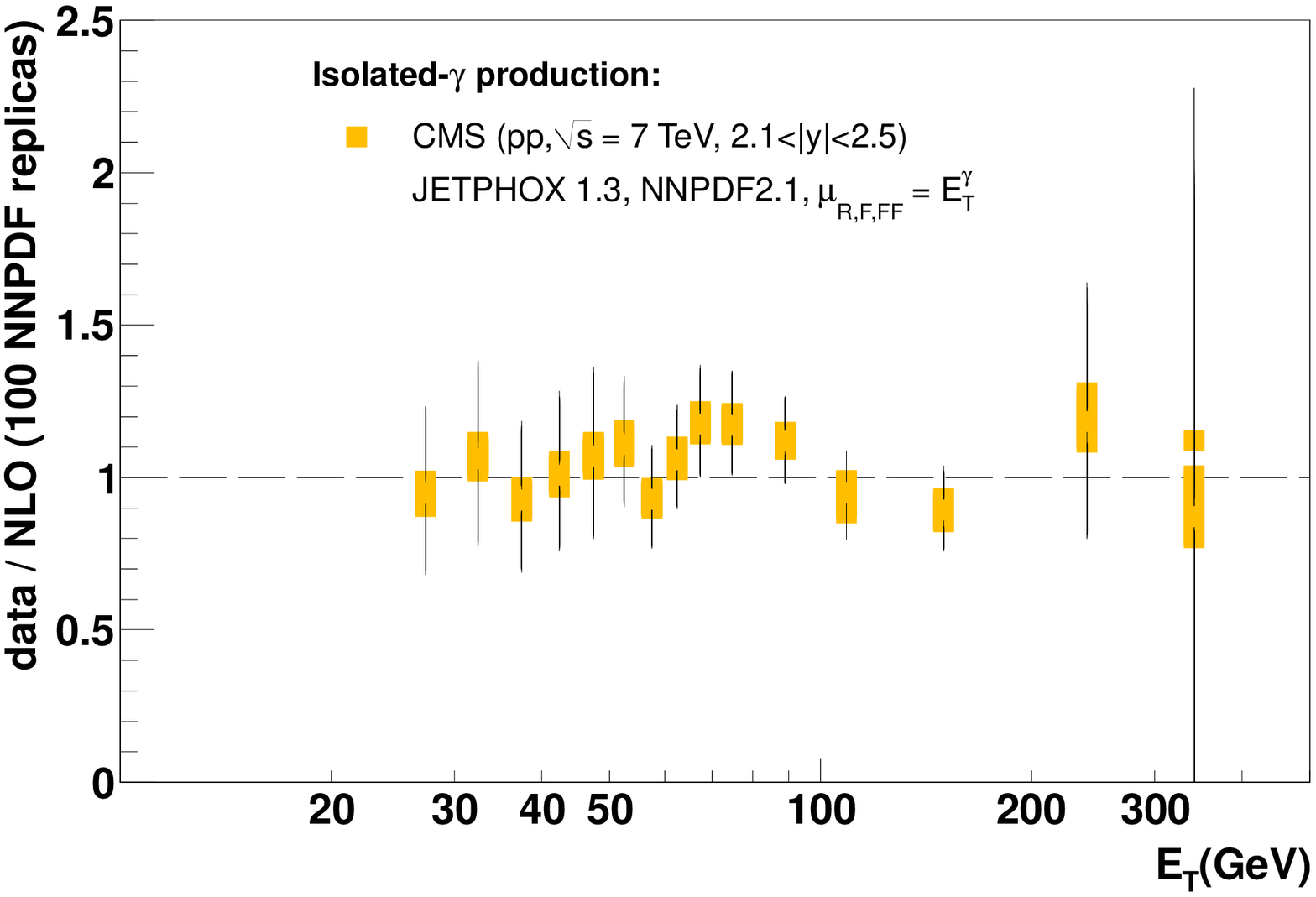}
\caption{\small Ratio of CMS isolated-photon data and NLO pQCD predictions in \pp\ collisions at
$\sqrts$~=~2.76~TeV (top left) and $\sqrts$~=~7~TeV (2.9~nb$^{-1}$, top-middle, and 36~pb$^{-1}$, top-right
and bottom, datasets) at various photon rapidities. 
The (yellow) band indicates the range of predictions for each one of the 100 NNPDF2.1 replicas, 
and the bars show the total experimental uncertainty.
% CMS~\cite{Khachatryan:2010fm,Chatrchyan:2011ue}. The top left
%plot correspond to the 2.9~nb$^{-1}$ dataset, while the other four plots show
%the comparison for the four rapidity bins of the 36~pb$^{-1}$ measurement.
\label{fig:datatheo_cms}}
\end{figure}

Using the \jetphox\ program with the setup discussed in Sect.~\ref{sec:jetphox}
we have computed the $\ETg$-differential cross sections using the $N_{\rm rep}=100$ replicas of the NNPDF2.1
NLO parton distributions for all the systems listed in Table~\ref{tab:compilation}.
The ratios between data and theory are shown in 
Figs.~\ref{fig:datatheo_200_546}--\ref{fig:datatheo_cms},
where the (yellow) band corresponds to the distribution
of the predictions obtained which each one of the 100 replicas, while the
outer error-bars cover the sum in quadrature of the statistical and systematic
uncertainties of the measurement.\\

Figure~\ref{fig:datatheo_200_546} shows the data/NLO ratios for the isolated-photons measured at the lowest
c.m. energies: $\sqrts$~=~200 GeV (RHIC) and 546~GeV (\spps). 
The PHENIX results agree well with NLO pQCD within the quite large experimental uncertainties.
For UA1 and UA2, the data/theory ratio is around unity below $\ETg\approx$~25~GeV,
but the highest transverse energy points are clearly overestimated by NLO in two cases, 
likely due to an experimental mis-reconstruction of the photon energy at the end of the spectra 
measured at the time (see below).\\

%%%%%%%%%%%%%%%
%%%%%%%%%%%%%%%%

The comparison of the theoretical predictions to the  UA1, UA2, CDF and D0 data at 630~GeV c.m. 
energies and various photon rapidities are shown in Fig.~\ref{fig:datatheo_630}. With a few exceptions, a general
trend appears indicating that the measured cross-sections are higher (respectively lower) than the NLO
calculations at the low (resp. high) end of the photon energy spectrum. The disagreement is larger for
increasingly forward rapidities. Yet, given the relatively large experimental uncertainties, 
theory and data are in general consistent and show $\chi^2$ not very far from one in most of the cases.\\
%In some cases, which correspond to systems with bad $\chi^2$,
%the data are systematically below the theory by more than a factor 2.

%%%%%%%%%%%%%%%
%%%%%%%%%%%%%%%%

The Tevatron Run-I measurements at $\sqrts$~=~1.8~TeV are compared to the NLO predictions in
Fig.~\ref{fig:datatheo_tev1}. The two oldest measurements~\cite{Abe:1994rra,Abachi:1996qz} (not plotted here) 
are systematically below not only the pQCD predictions but also the spectra obtained by the same 
experiments in a basically identical kinematic range a few years later~\cite{Acosta:2002ya,Acosta:2004bg,Abbott:1999kd}. 
By discarding those superseded datasets from further analysis, we find a generally good consistency of the
1.8-TeV data with NLO pQCD.\\ 
%Although a couple of measurements (in particular at high photon energy) 
%, again data and theory are mostly consistent given the relatively large experimental uncertainties.\\
%We note that for the Run-I at the Tevatron 
%($\sqrts$~=~1.8~TeV) there exists other measurements~\cite{Abe:1994rra,Abachi:1996qz} 
%which we have discarded since they are superseded by updated datasets at the same collision
%energy already included in our study~\cite{Acosta:2002ya,Acosta:2004bg,Abbott:1999kd} (moreover, the oldest 
%measurements are systematically below the most recent ones and inconsistent with the latter).

%%%%%%%%%%%%%%%
%%%%%%%%%%%%%%%%

The comparison of NLO to the Tevatron CDF and D0 Run-II results  at $\sqrts$~=~1.96~TeV 
is shown in Fig.~\ref{fig:datatheo_tev2}.
While at the level of $\chi^2$ the data--theory agreement is good, in both cases the
spectral shape seems somewhat different at small photon $\ET$
where the data rise steeper as compared to the theory.
The origin of this different shape is still not well
understood, although Ref.~\cite{Aurenche:2006vj} argues that the
discrepancy decreases with a suitable scale choice.\\

%%%%%%%%%%%%%%%
%%%%%%%%%%%%%%%%

Data--theory comparisons for ATLAS and CMS measurements are shown in Fig.~\ref{fig:datatheo_atlas}
and~\ref{fig:datatheo_cms} respectively.
We observe a very good agreement for all rapidity ranges, except maybe in a few of the lowest-$\ETg$ bins where 
the central value of the data points tends to undershoot a bit the theoretical predictions.\\

%%%%%%%%%%%%%%%
%%%%%%%%%%%%%%%%

%%%%%%%%%%%%%%%
%%%%%%%%%%%%%%%%

\begin{figure}[ht!]
\centering
\epsfig{width=0.80\textwidth,figure=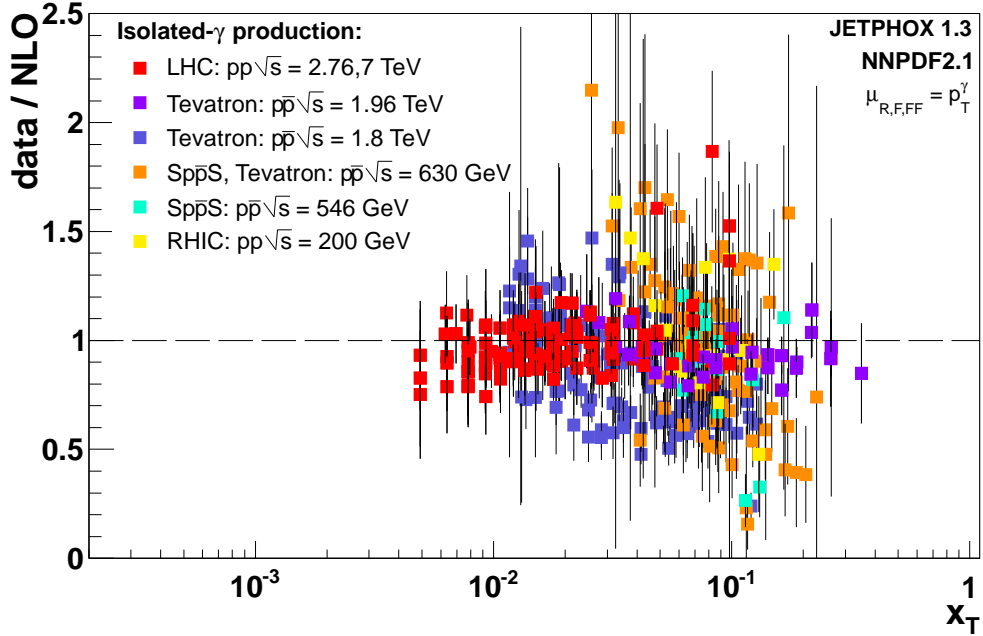}
\caption{\small Summary plot for the data/theory ratios for all
collider isolated-photon data considered in this analysis, as a
function of $\xT = 2\ETg/\sqrts$. For each system, the NLO prediction 
used is the one obtained with the central PDF replica of the NNPDF2.1 set.
The error bars indicate the total experimental uncertainty.
%The datasets have been collected into
%three groups: LHC data, Tevatron Run-I and Run-II data and
%collider experiments with $\sqrts$ between 200 and 630~GeV.
\label{fig:datatheo_summary}}
\end{figure}

Figure~\ref{fig:datatheo_summary} shows the overall comparison of the full collider photon dataset
considered in this analysis with the NLO predictions obtained using the central NNPDF2.1 replicas, 
as a function of $\xT$. Most of the data/theory ratios are around unity indicating an overall good agreement
between NLO pQCD and the experimental measurements. No systematic deviation is observed in a wide range of $\xT$
corresponding to a wide kinematical coverage of photon $\ET$, rapidity and collision energy.
This is to be contrasted with similar systematics studies~\cite{Aurenche:2006vj} that indicate clear data--NLO
deviations for the inclusive-$\gamma$ production in the E706 results at fixed-target energies.\\

%%%%%%%%%%%%%%%
%%%%%%%%%%%%%%%%

In Table~\ref{tab:compilation} we quote the average $\chi^2$ over all replicas between each one of
the datasets and the NLO calculations. As one can see, for a large majority of cases the agreement is quite
good, while in a few cases the $\chi^2$ obtained is rather poor ($\chi^2\gg$~1). The total initial $\chi^2$ of
all the systems considered is $\chi^2=1.3$, while after reweighting it decreases to $\chi^2_{\rm rw}=1.1$ 
for the whole dataset. This global result confirms at the quantitative level that there is a good agreement
between NLO pQCD and the experimental results measured at all collider energies.\\

%%%%%%%%%%%%%%%%%%%%%
%\begin{sidewaystable}[htbp]
\begin{table}[Hhtbp!]
\begin{small}
%\begin{footnotesize}
%\begin{center} 
\centering
\begin{tabular}{c|c|c|c|c|c|c|c}\hline\hline
System  & Collab./Experiment & $\sqrts$ & $|y|$ & $\ETg$ range &  
\hspace{0.2cm} $\chi^2$ \hspace{0.2cm} & \hspace{0.2cm} $\chi^2_{\rm rw}$ \hspace{0.2cm} & \hspace{0.2cm} $\la \alpha \ra$ \hspace{0.2cm} \\
\hline
\pp & ATLAS (LHC)  & 7. & $<$0.6 & 15--100 & 0.2 & 0.5  & 0.6 \\
\pp & ATLAS (LHC)  & 7. & 0.6--1.37 & 15--100 & 0.5 & 0.3 & 0.8 \\
\pp & ATLAS (LHC)  & 7. & 1.52--1.81 & 15--100 & 0.3  & 0.6 & 0.6 \\

\pp & ATLAS (LHC)  & 7. & $<$0.6 & 45--400 &  0.7 & 0.8 & 1.0 \\
\pp & ATLAS (LHC) & 7. & 0.6--1.37 & 45--400 &  0.3 & 0.4 & 0.7 \\
\pp & ATLAS (LHC)  & 7. & 1.52--1.81 & 45--400 & 0.3  & 0.3 & 0.7 \\
\pp & ATLAS (LHC)  & 7. & 1.81--2.37 & 45--400 &  1.6 & 1.7 & 1.6 \\

\pp & CMS (LHC)  & 7. & $<$1.45 & 21--300 &  0.6  & 0.6 & 0.8 \\
\pp & CMS (LHC)  & 7. & $<$0.9  & 25--400 &  1.3   & 1.1 & 1.2  \\
\pp & CMS (LHC) & 7. & 0.9--1.44 & 25--400 &  0.8  & 0.8 & 1.0 \\
\pp & CMS (LHC)  & 7. & 1.57--2.1 & 25--400 &  0.7  & 0.8 & 1.0 \\
\pp & CMS (LHC)  & 7. & 2.1--2.5 & 25--400 & 0.5 & 0.5 & 0.8 \\

\pp & CMS (LHC)& 2.76 & $<$1.45 & 20--80 &  0.2  & 0.2 & 0.7 \\

\hline

\ppbar & CDF (Tevatron)   & 1.96 & $<$1.0 & 30--400 & 0.7 & 0.8 & 1.3 \\
\ppbar & D0 (Tevatron)  & 1.96 & $<$0.9 & 23--300 & 1.4 & 1.3 &  0.9 \\

\ppbar & CDF (Tevatron)  & 1.8 & $<$0.9 & 11--132 &  3.6 &  2.9 & 2.1 \\
\ppbar & CDF (Tevatron)  & 1.8 & $<$0.9 & 10--65 & 1.0  & 1.1 & 1.1 \\
\ppbar & D0 (Tevatron)  & 1.8 & $<$0.9 & 10--140 & 3.1 & 2.9 &  2.1 \\
\ppbar & D0 (Tevatron) & 1.8 & 1.6--2.5 & 10--140 & 1.5  & 0.6 & 1.9 \\

%\hdashline

\ppbar & CDF (Tevatron) & 0.63 & $<$0.9 & 8--38 & 1.1 & 1.1 & 1.3 \\
\ppbar & D0 (Tevatron)  & 0.63  & $<$0.9 & 7--50 & 0.9 & 1.1 & 1.3 \\
\ppbar & D0 (Tevatron)  & 0.63  & 1.6--2.5 & 7--50 & 0.8 & 0.9 & 1.2 \\
\hline

\ppbar & UA1 (\spps)  & 0.63 & $<$0.8 & 16--100 & 1.5 & 1.5 & 1.3 \\ 
\ppbar & UA1 (\spps)   & 0.63 & 0.8--1.4 & 16--70 & 1.7 & 1.7 & 1.5 \\
\ppbar & UA1 (\spps)   & 0.63 & 1.6--3.0 & 16--70 & 5.9 & 4.2 & 2.7 \\

\ppbar & UA2 (\spps)   & 0.63 & $<$0.76 & 14--92 & 1.1 & 1.1 & 1.2 \\
\ppbar & UA2 (\spps)  & 0.63 & $<$0.76 & 12--83 & 3.0 & 2.7 & 1.9 \\
\ppbar & UA2 (\spps)  & 0.63 & 1.0--1.8 & 12--51 & 1.7 & 1.7 & 1.5 \\

%\hdashline

\ppbar & UA1 (\spps)  & 0.546 & $<$0.8 & 16--51 & 0.4 & 0.4 & 0.8 \\
\ppbar & UA1 (\spps)  & 0.546 & 0.8--1.4  & 16--46 & 6.1 & 5.7 & 3.3 \\
\ppbar & UA1 (\spps)  & 0.546 & 1.6--3.0 & 16--38 & 7.4 & 6.1 & 3.6 \\

\hline

\pp & PHENIX (RHIC)  & 0.2 & $<$0.35 & 3--16 & 0.6 & 0.6 & 0.8 \\

\hline 
\hline 
\end{tabular}\vspace{3mm} 
\caption{\small Summary of the $\chi^2$-analysis between NLO pQCD and the world
isolated-$\gamma$ data. For each system we list the initial data--theory $\chi^2$ (6th column),
the $\chi^2_{\rm rw}$ (7th column) obtained after including each corresponding dataset via PDF reweighting,
and $\la \alpha \ra $ (last column), the mean of the associated $\mathcal{P}(\alpha)$ distribution.}
\label{tab:chi2datasets} 
%\end{center} 
%\end{footnotesize}
\end{small}
%\end{sidewaystable}
\end{table}
%%%%%%%%%%%%%%%%%%%%%%%%%%%%%%%%%%%%%%%

\begin{figure}[ht!]
\centering
\epsfig{width=0.32\textwidth,figure=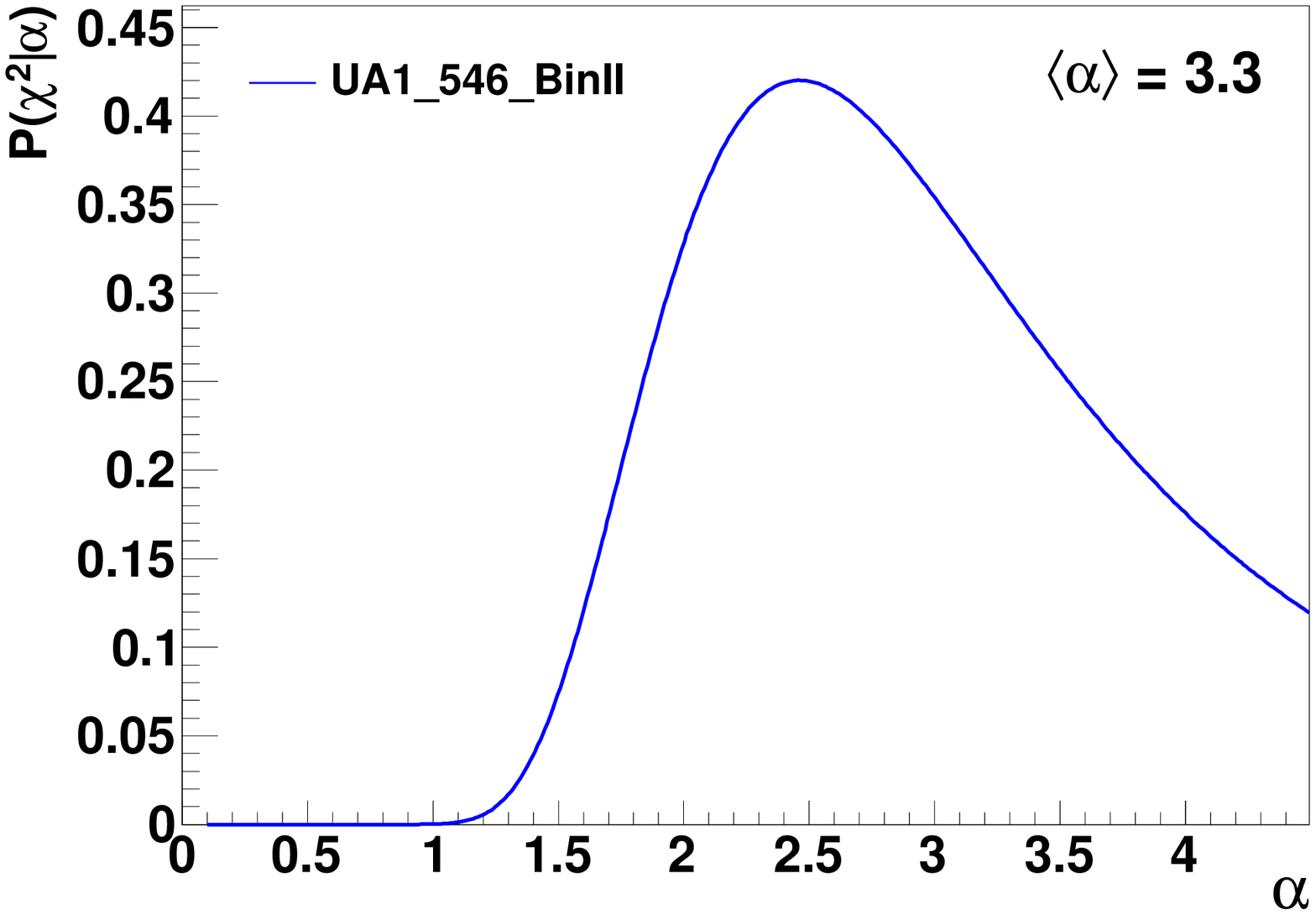}
\epsfig{width=0.32\textwidth,figure=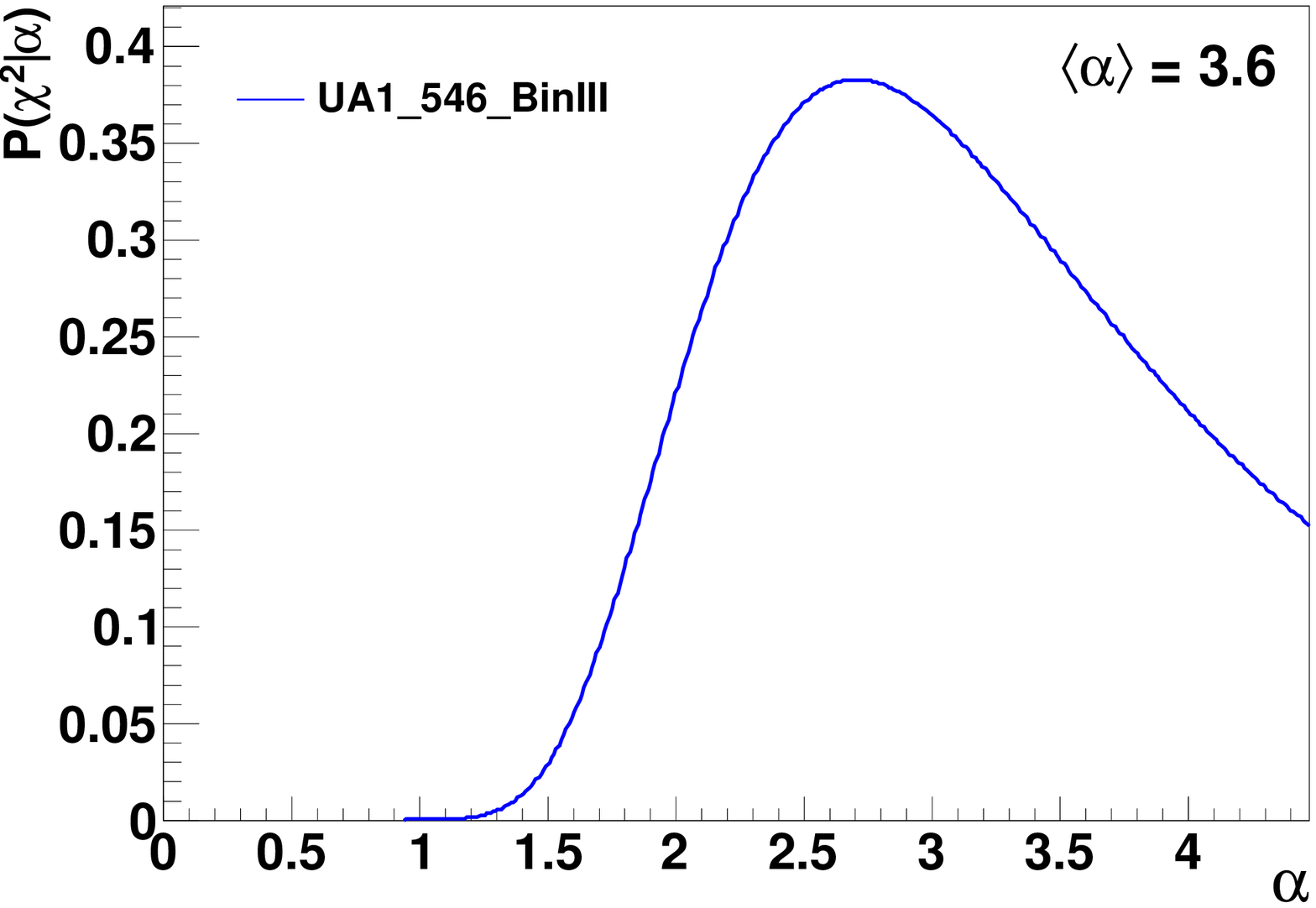}
\epsfig{width=0.32\textwidth,figure=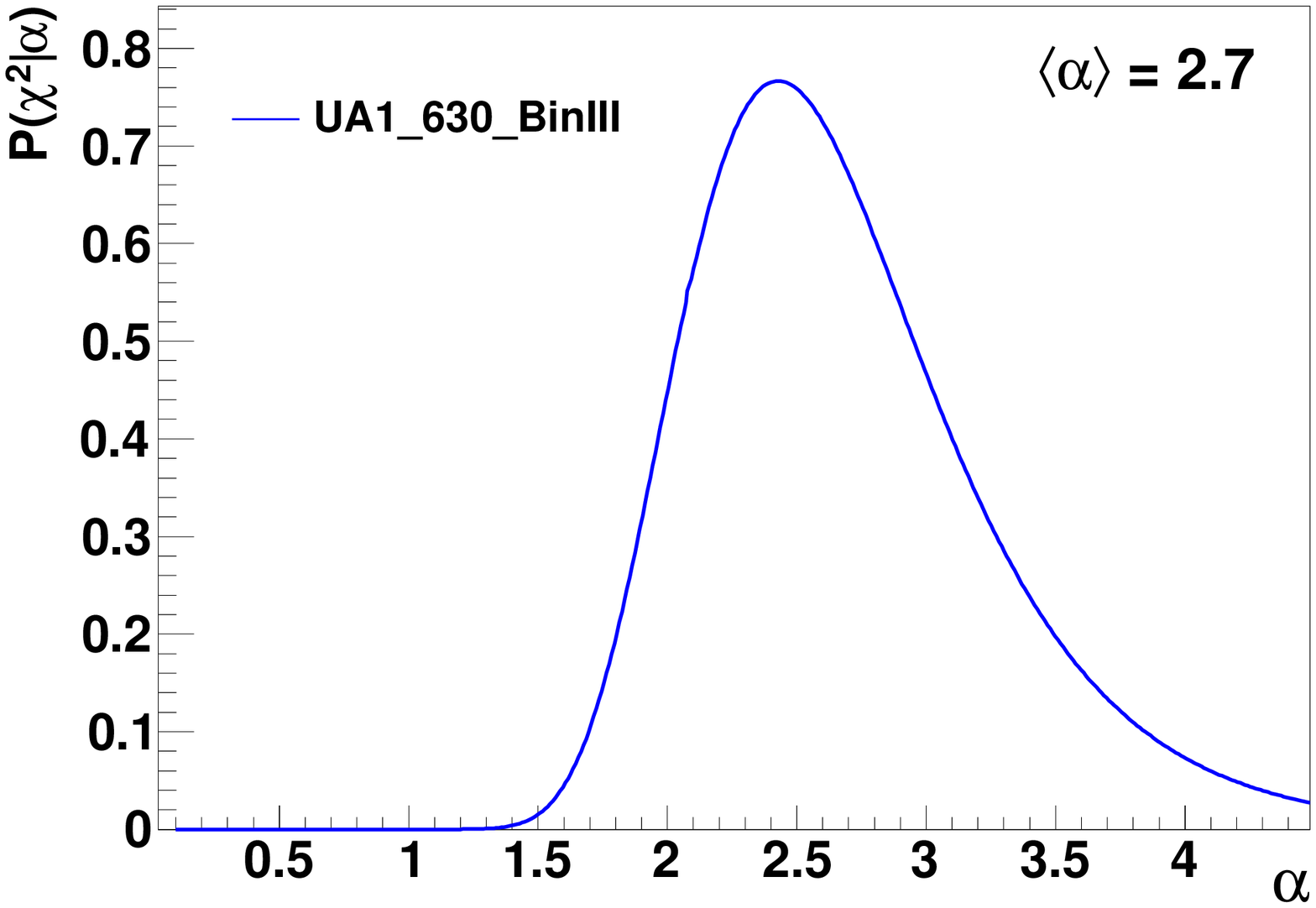}\\
\epsfig{width=0.32\textwidth,figure=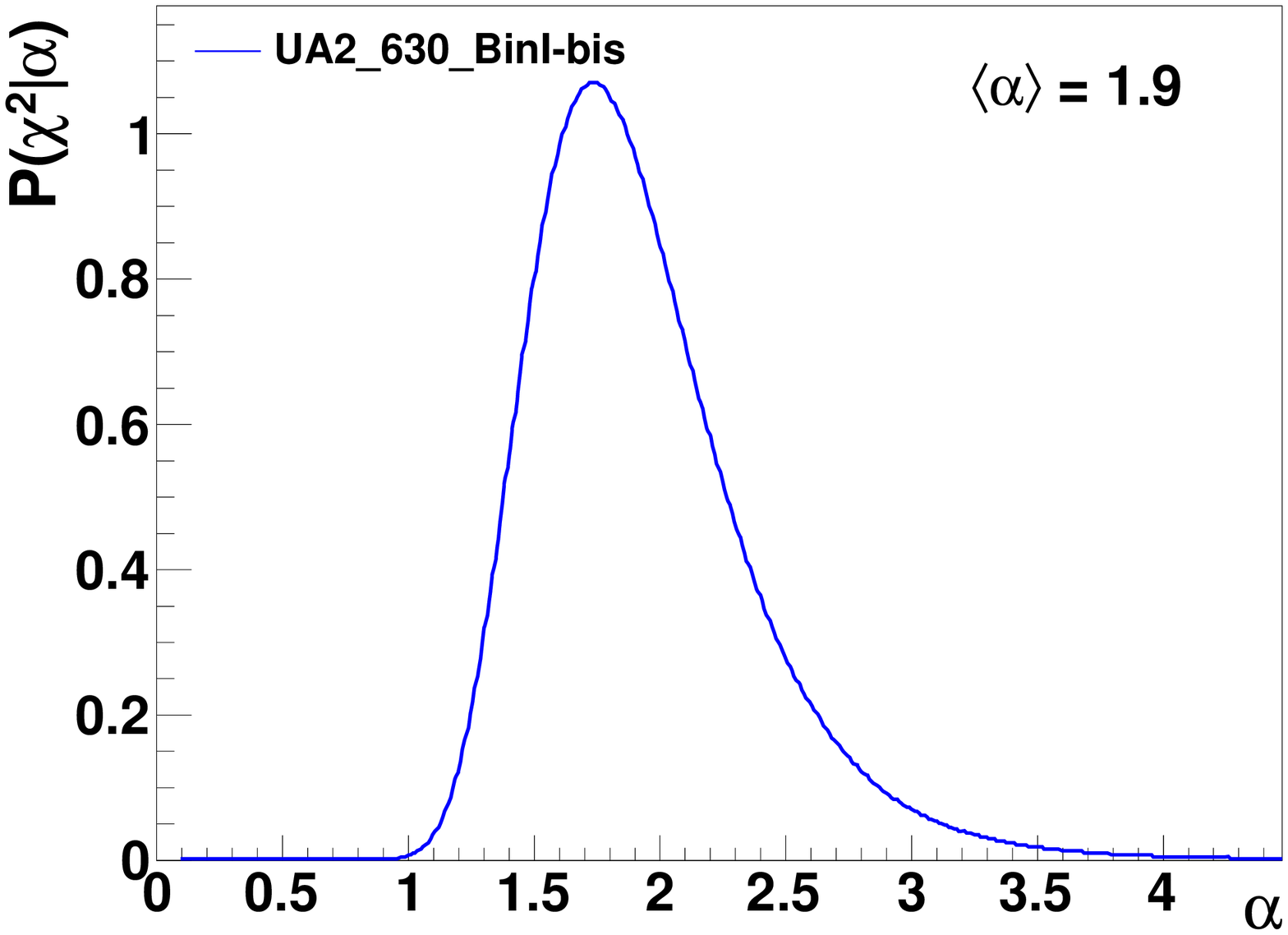}
\epsfig{width=0.32\textwidth,figure=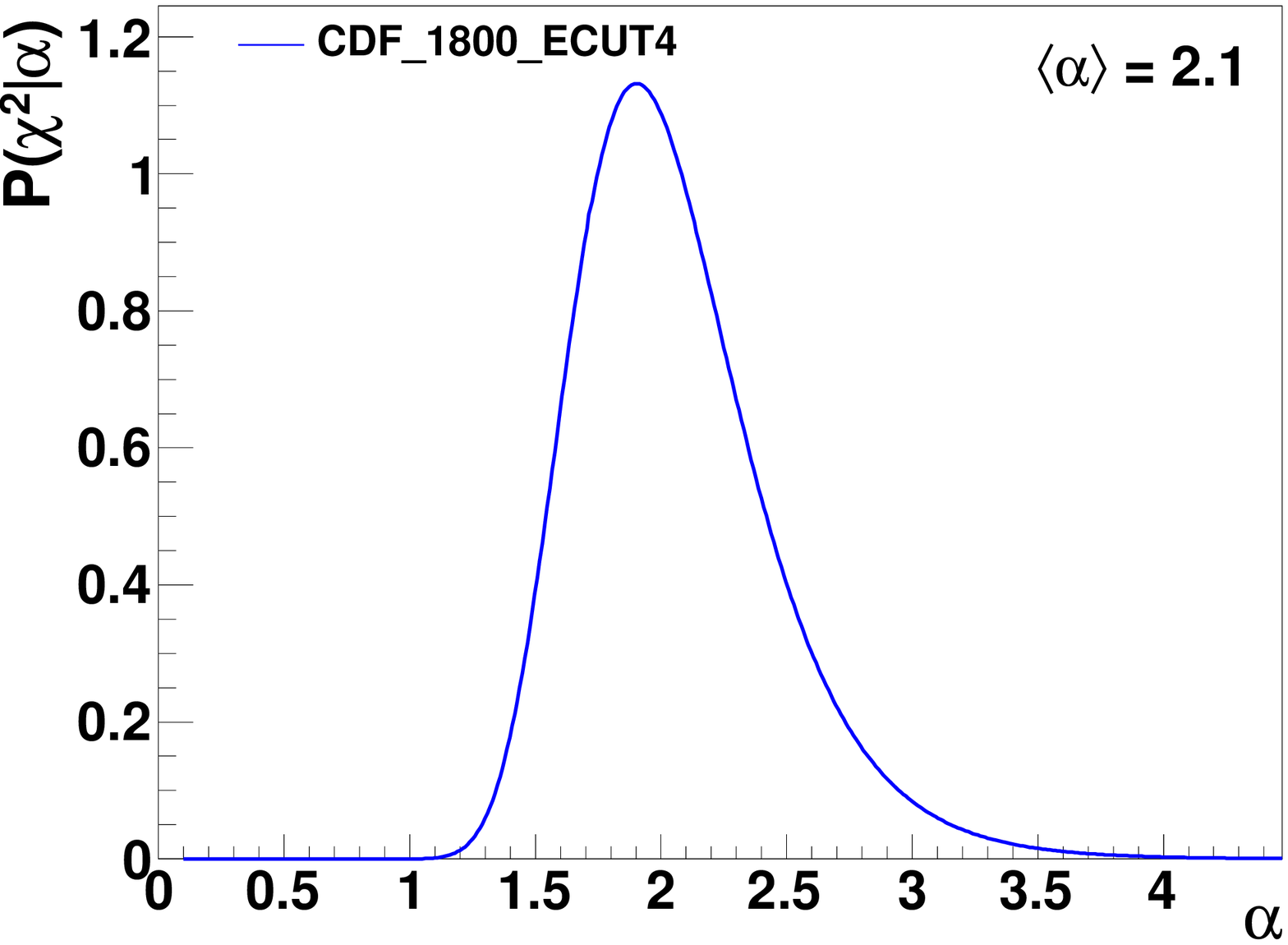}
\epsfig{width=0.32\textwidth,figure=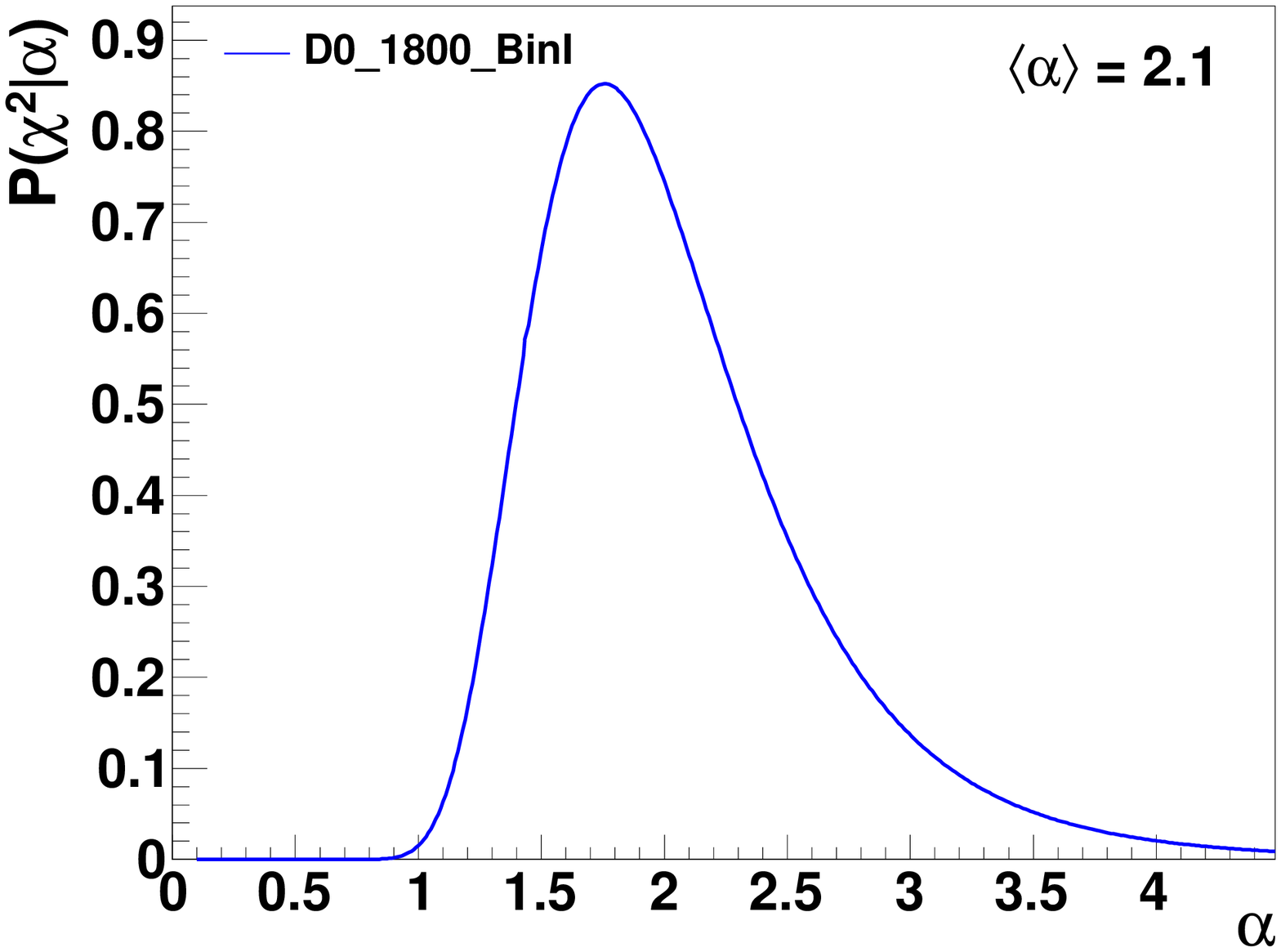}
\caption{\small Distribution of the $\alpha$ rescaling variable, Eq.~(\ref{eq:rescaling}),
for those datasets that show poor agreement with NLO pQCD.
From top-to-bottom and left-to-right, we show two UA1-546~GeV
datasets ($\chi^2$~=~6.1 and 7.4 respectively), one UA1
($\chi^2$~=~5.9) and UA2 ($\chi^2$~=~3.0) datasets at 630~GeV, and
finally two Tevatron Run-I CDF and D0 sets ($\chi^2$~=~3.6 and 3.1 respectively).
\label{fig:palpha2}}
\end{figure}

%Let us discuss in more detail the results of Table~\ref{tab:chi2datasets}. 
Looking in more detail, it is worth noticing that for those few systems which are not well reproduced by NLO pQCD, 
there are always other measurements\footnote{The only exception to this are two UA1 measurements at
546~GeV (Fig.~\ref{fig:datatheo_200_546}) for which the large $\chi^2$ is just driven by a single outlier
data-point at the highest $\ETg$ measured.} covering similar ($\sqrts$,$\ETg$,$y_\gamma$) domains 
with a good $\chi^2$. This fact indicates that in such cases the problem
is not likely related to the theoretical prediction but of experimental origin.
%(such as e.g. the photon energy calibration). 
In other words, the issue is not
data--theory compatibility but rather and inconsistency problem
between measurements covering the same kinematics.
Note that in the cases where the $\chi^2$ is poor, the reweighted
$\chi^2_{\rm rw}$ is only slightly better, which confirms that the
data are not consistent with the theory even after refitting.
To quantify better this effect, we can make use of the probability distribution
for the rescaling variable $\alpha$ discussed in Sect.~\ref{sec:rw}.
The last column of Table~\ref{tab:chi2datasets} lists the mean value of the $\mathcal{P}(\alpha)$ 
distribution\footnote{It is easy to check that if the underlying distribution
is a $\chi^2$ distribution, the mean value of $\mathcal{P}(\alpha)$ is given by $\la \alpha \ra =1+1/2N_{\rm dat}$ 
with $N_{\rm dat}$ the number of data points of that particular system.}, 
Eq.~(\ref{eq:rescaling}), for all systems.
In Fig.~\ref{fig:palpha2} we show the $\mathcal{P}(\alpha)$ distribution for those systems in Table~\ref{tab:chi2datasets}
with initial $\chi^2\ge 3$. In all cases $\mathcal{P}(\alpha)$ peaks at a value of $\alpha$ above $\sim$2. 
This fact points to %confirms that the origin of the discrepancy of these datasets with NLO pQCD is
an underestimation of the experimental uncertainties in the corresponding measurements.\\

%%%%%%%%%%%%%%%
%%%%%%%%%%%%%%%%

\begin{figure}[ht!]
\centering
\epsfig{width=0.24\textwidth,figure=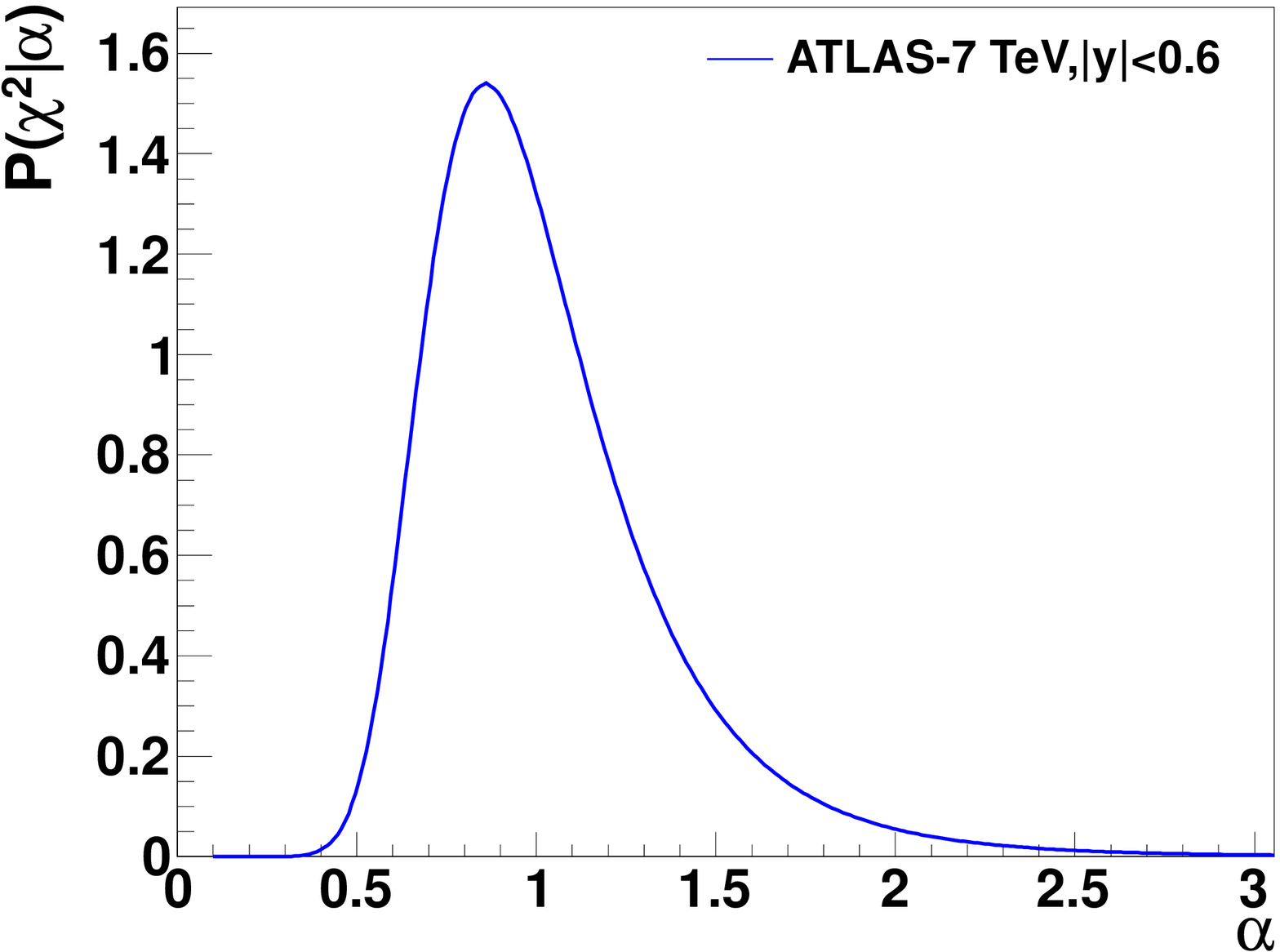}
\epsfig{width=0.24\textwidth,figure=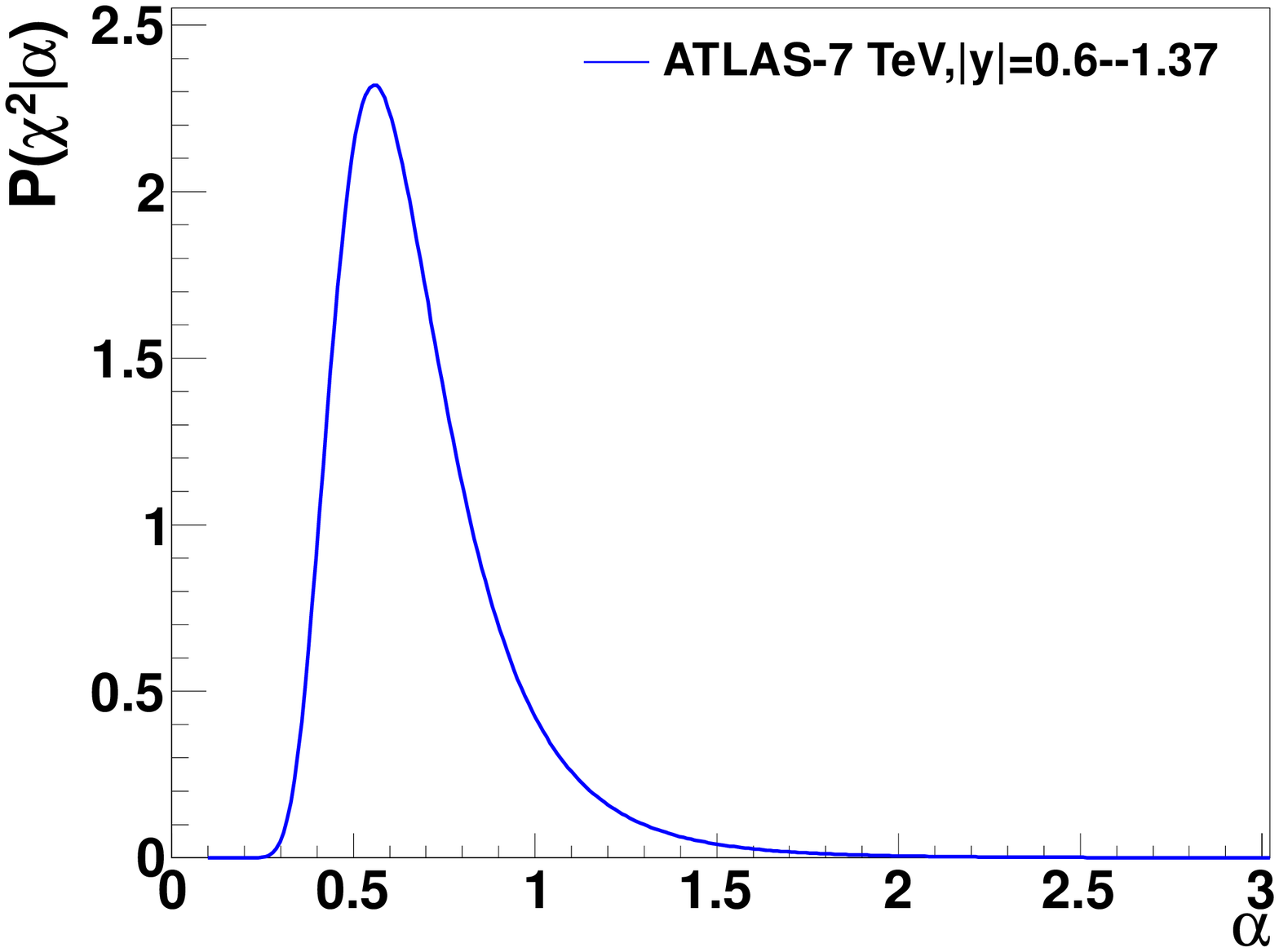}
\epsfig{width=0.24\textwidth,figure=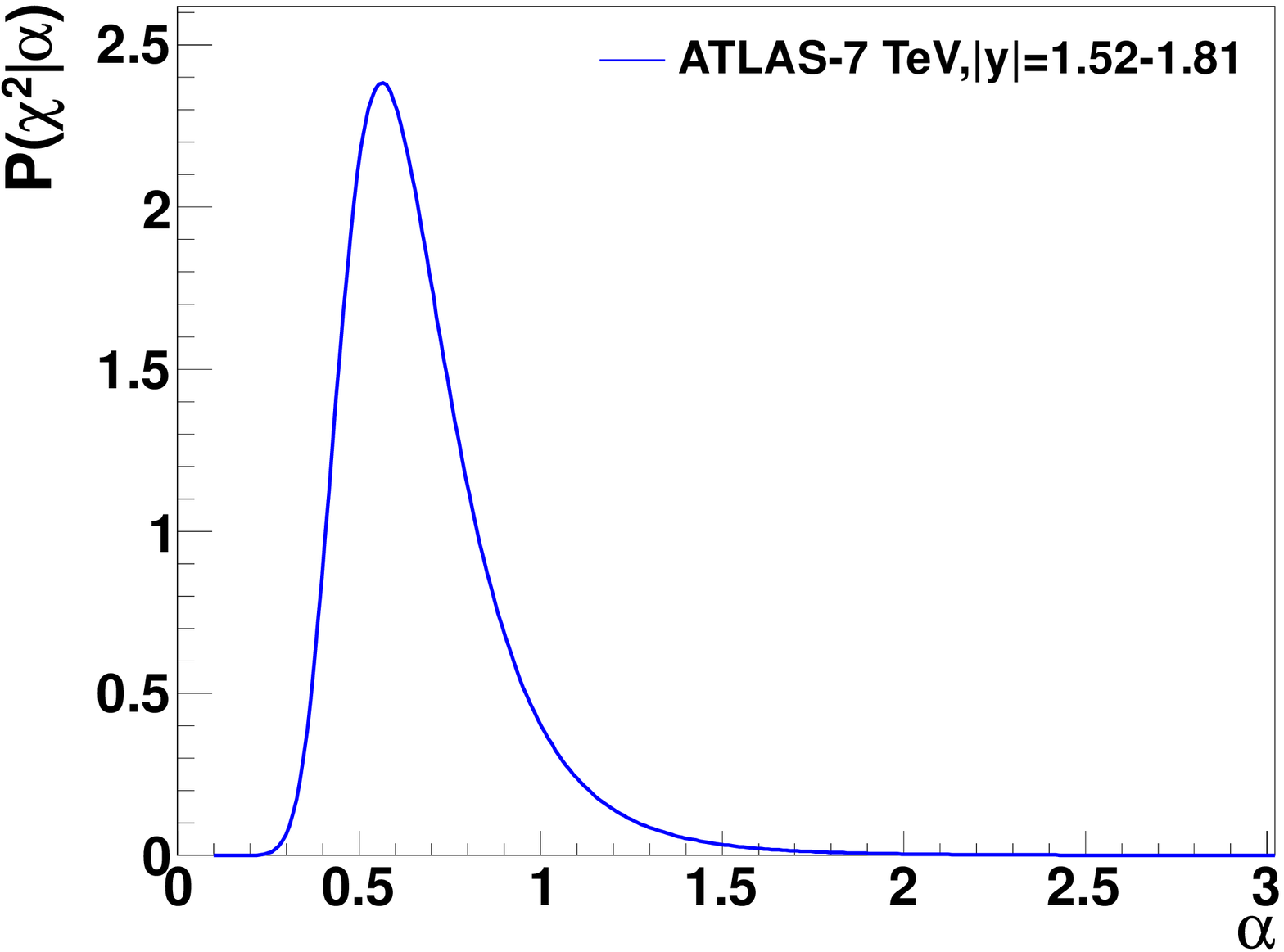}
\epsfig{width=0.24\textwidth,figure=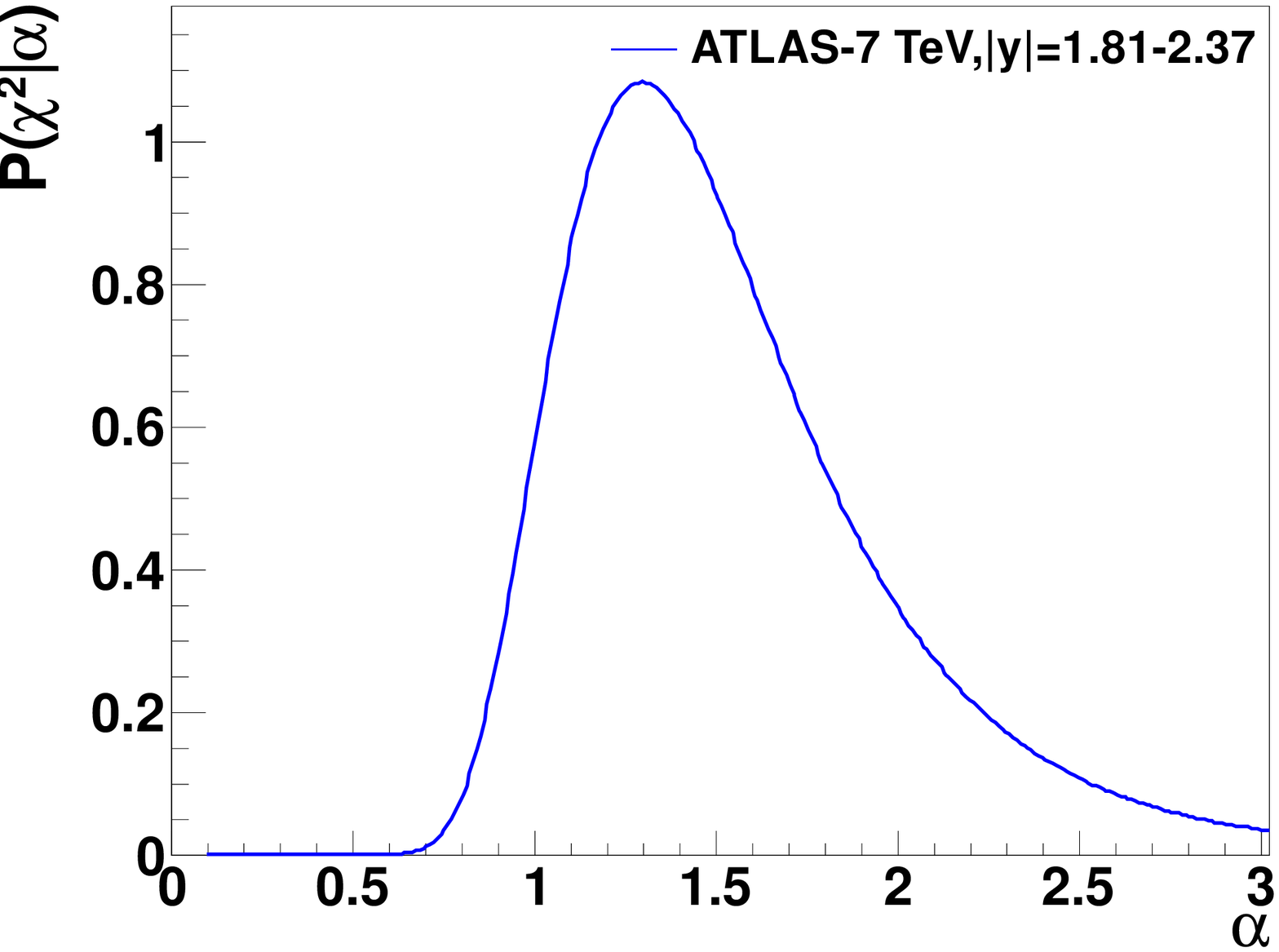}\\
\epsfig{width=0.24\textwidth,figure=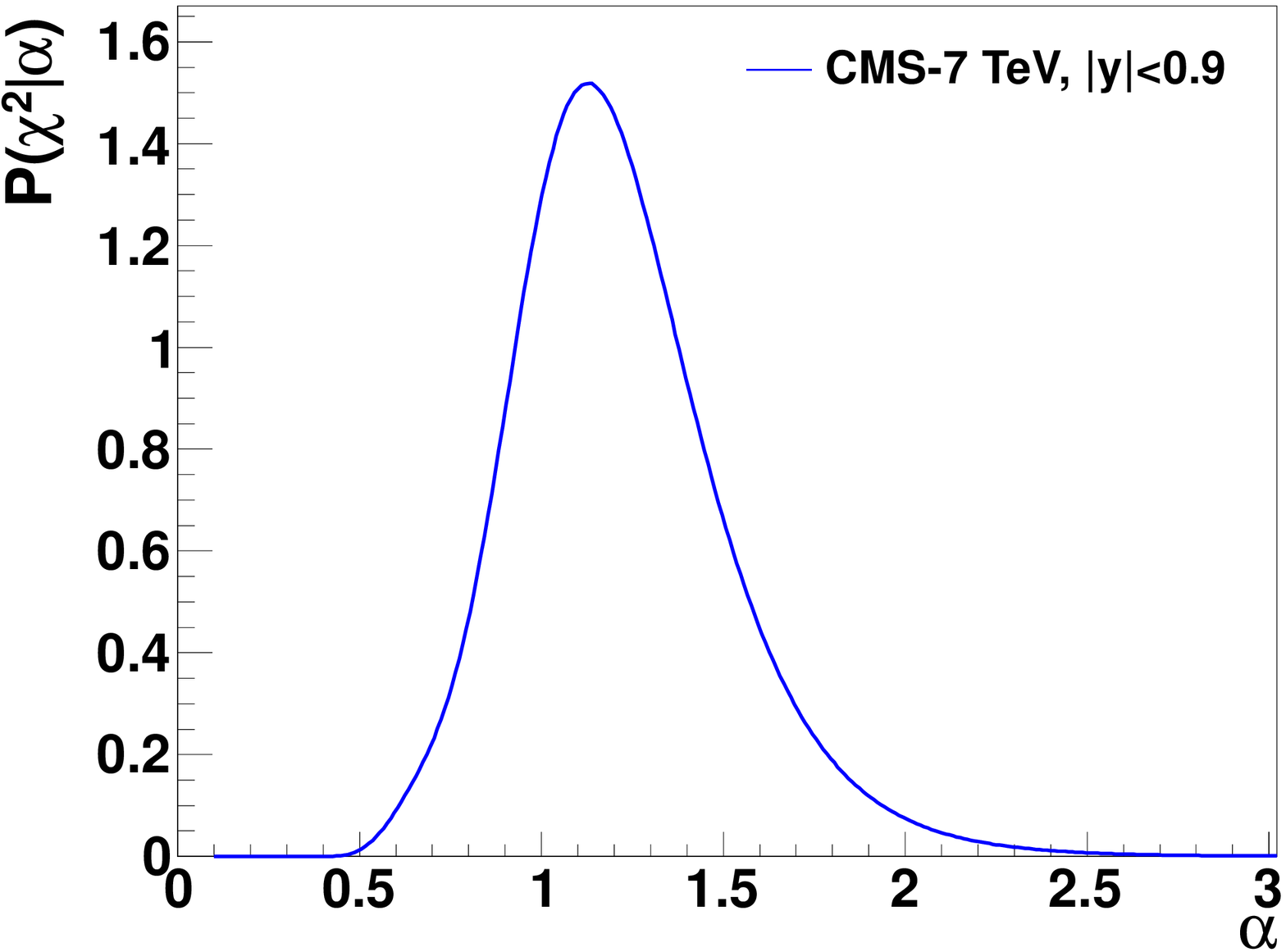}
\epsfig{width=0.24\textwidth,figure=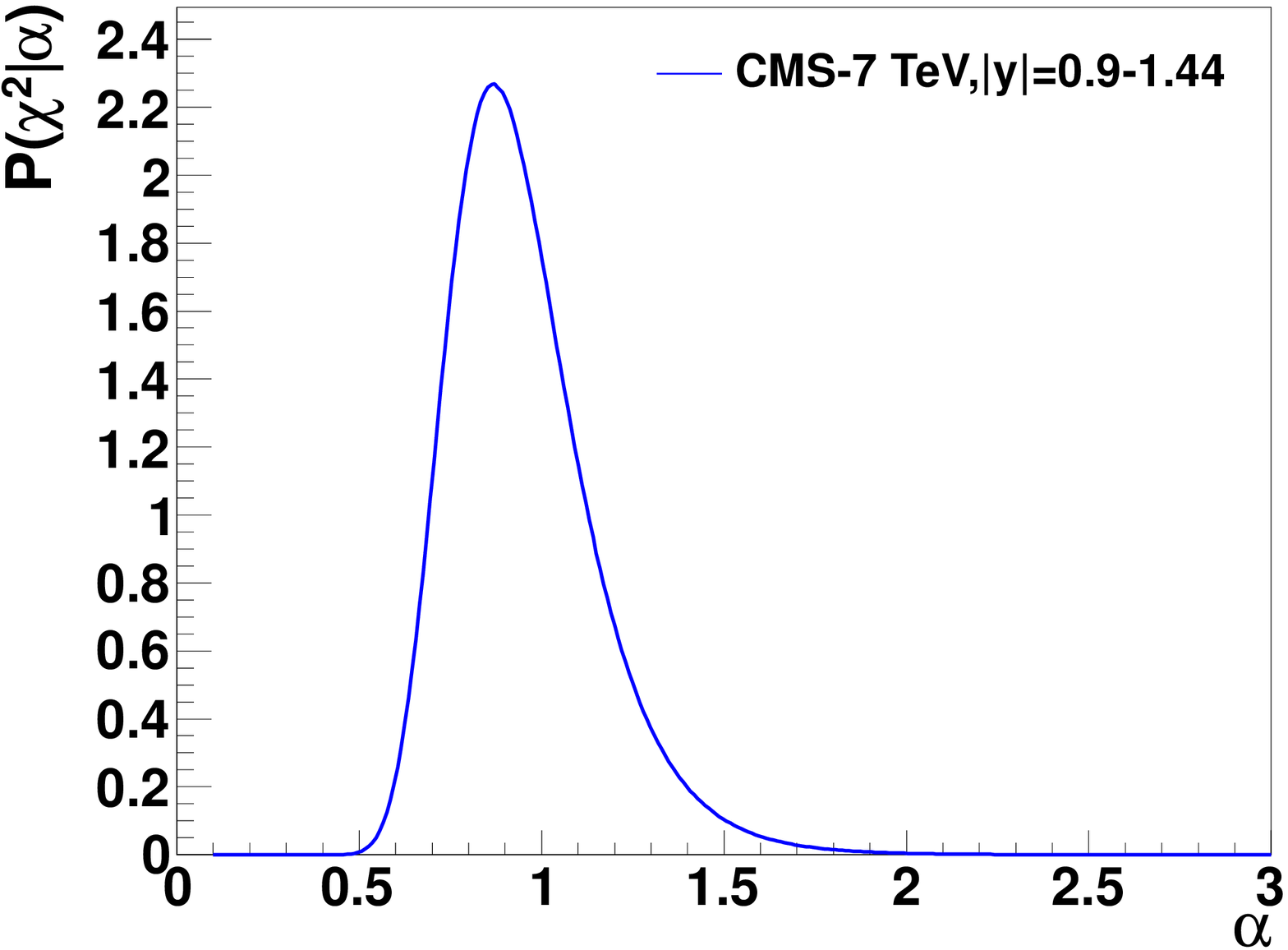}
\epsfig{width=0.24\textwidth,figure=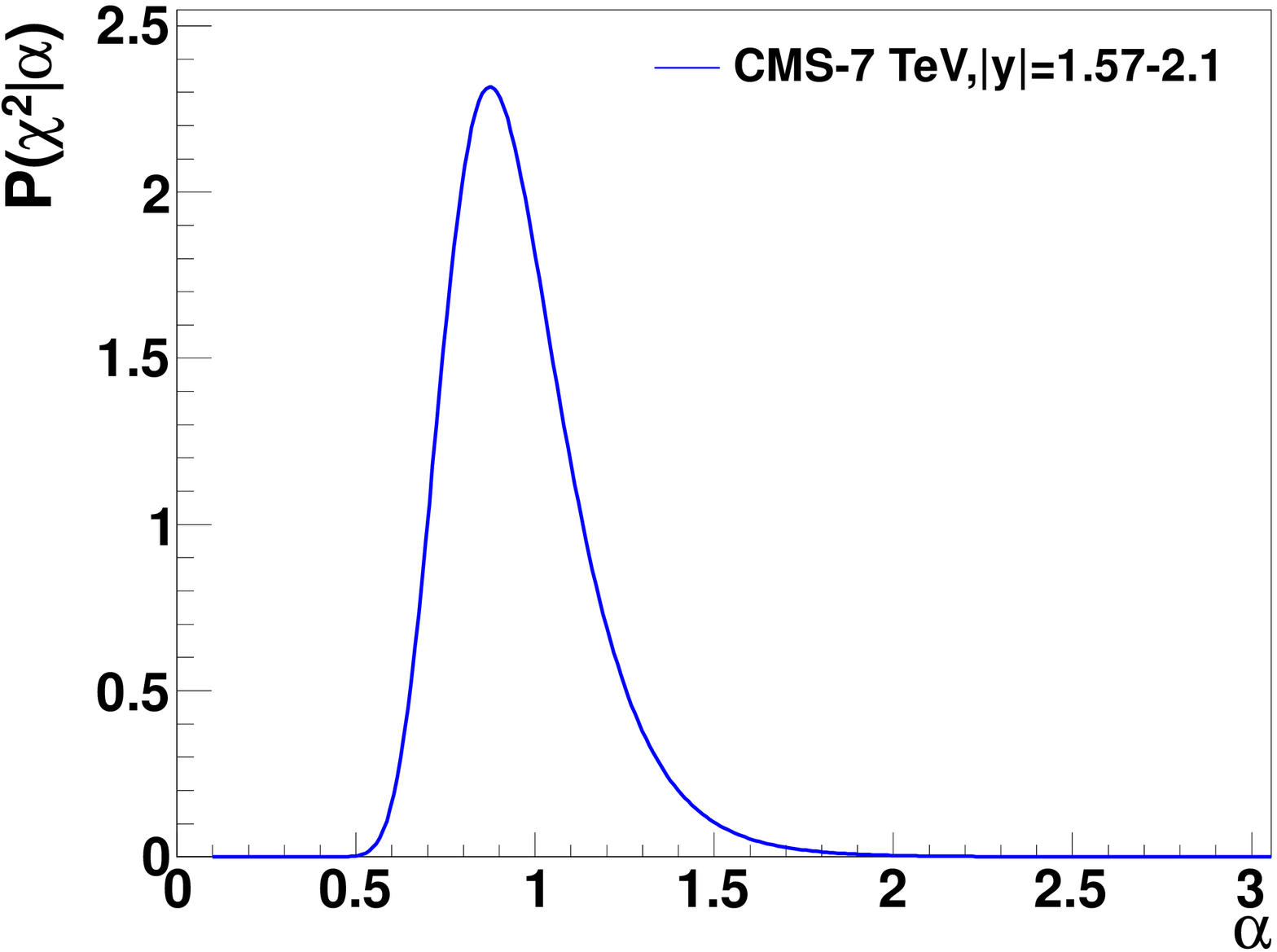}
\epsfig{width=0.24\textwidth,figure=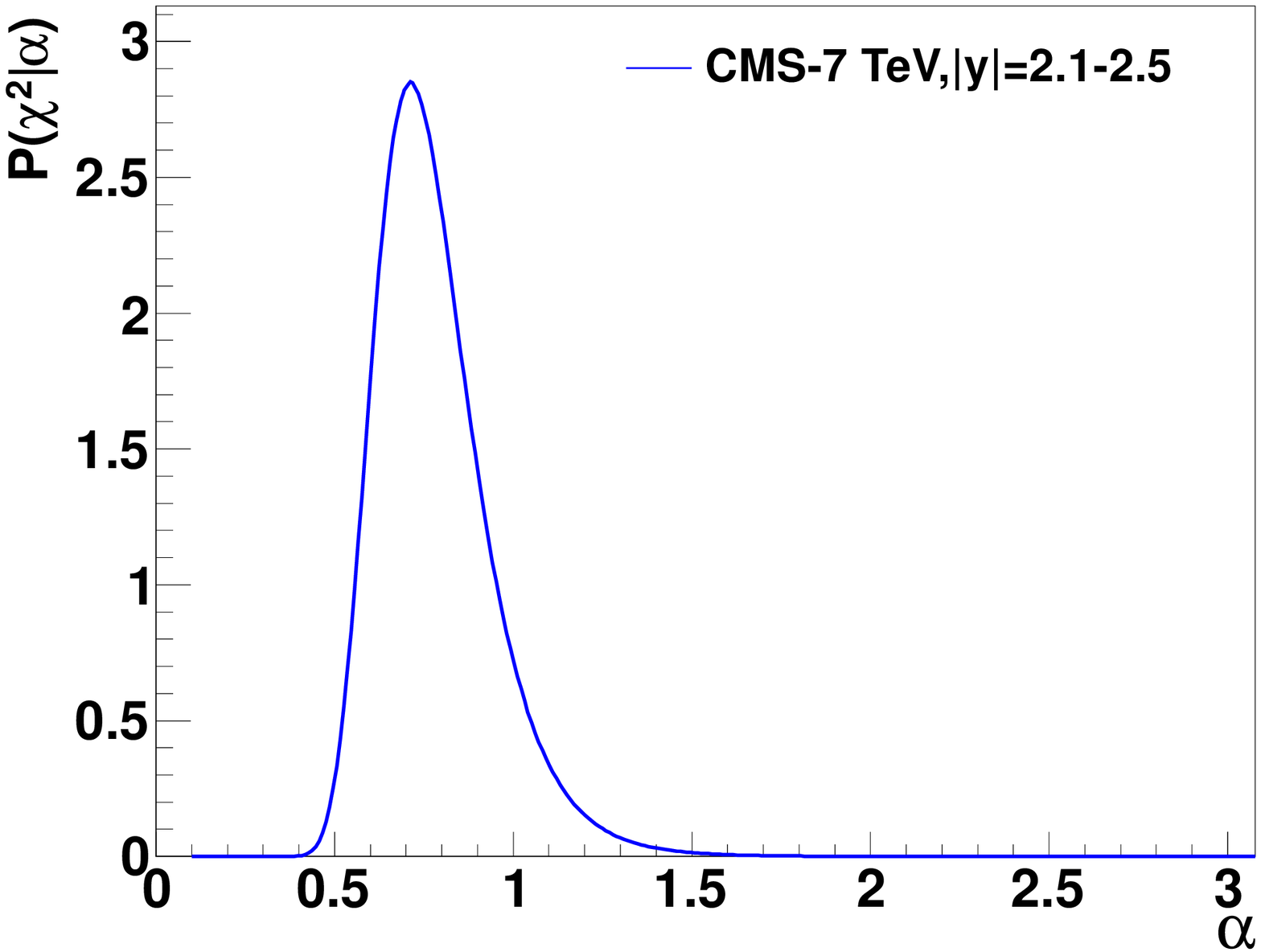}
\caption{\small Distribution of the $\alpha$ rescaling variable, Eq.~(\ref{eq:rescaling}),
 for each one of the rapidity bins of the 36~pb$^{-1}$ datasets
from ATLAS (top plots) and CMS (bottom plots). \label{fig:palpha}}
\end{figure}

As a comparison, in Fig.~\ref{fig:palpha} we show the $\mathcal{P}(\alpha)$ 
distributions for the rapidity bins of the 36~pb$^{-1}$ results from ATLAS and CMS. 
In all cases $\mathcal{P}(\alpha)$  peaks close to one, confirming the consistency 
of these datasets with NLO pQCD and the proper estimation of their associated experimental errors.

%%%%%%%%%%%%%%%
%%%%%%%%%%%%%%%%

%%%%%%%%%%%%%%%%%%%%%%%%%%%%%%%%%%%%%%%%%%%%%%%%%%%%%%%%%%%%%%%%%%%%%%%%%%%%%%%%%%%%%%%%

\subsection{Sensitivity of isolated-$\gamma$ to the proton PDFs}

\begin{figure}[htbp!]
\centering
\epsfig{width=0.49\textwidth,figure=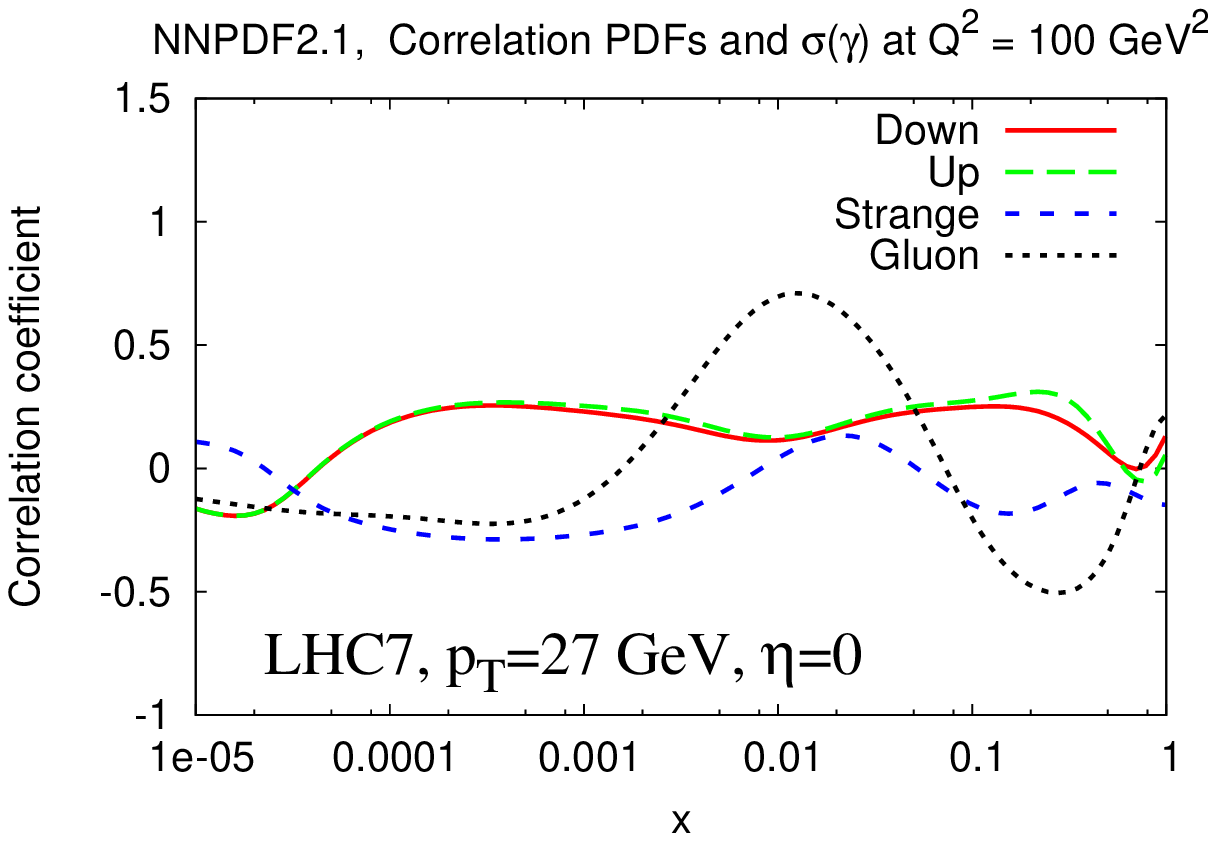}
\epsfig{width=0.49\textwidth,figure=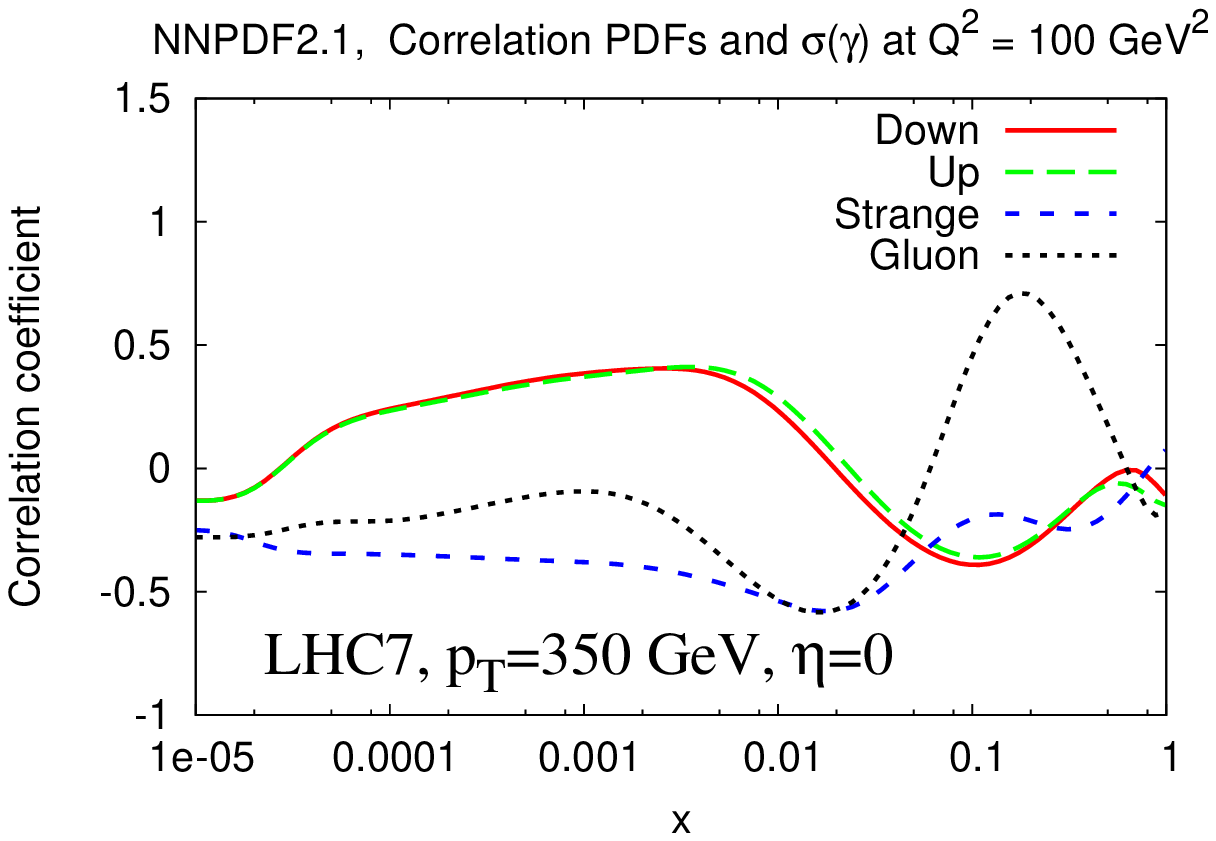}
\epsfig{width=0.49\textwidth,figure=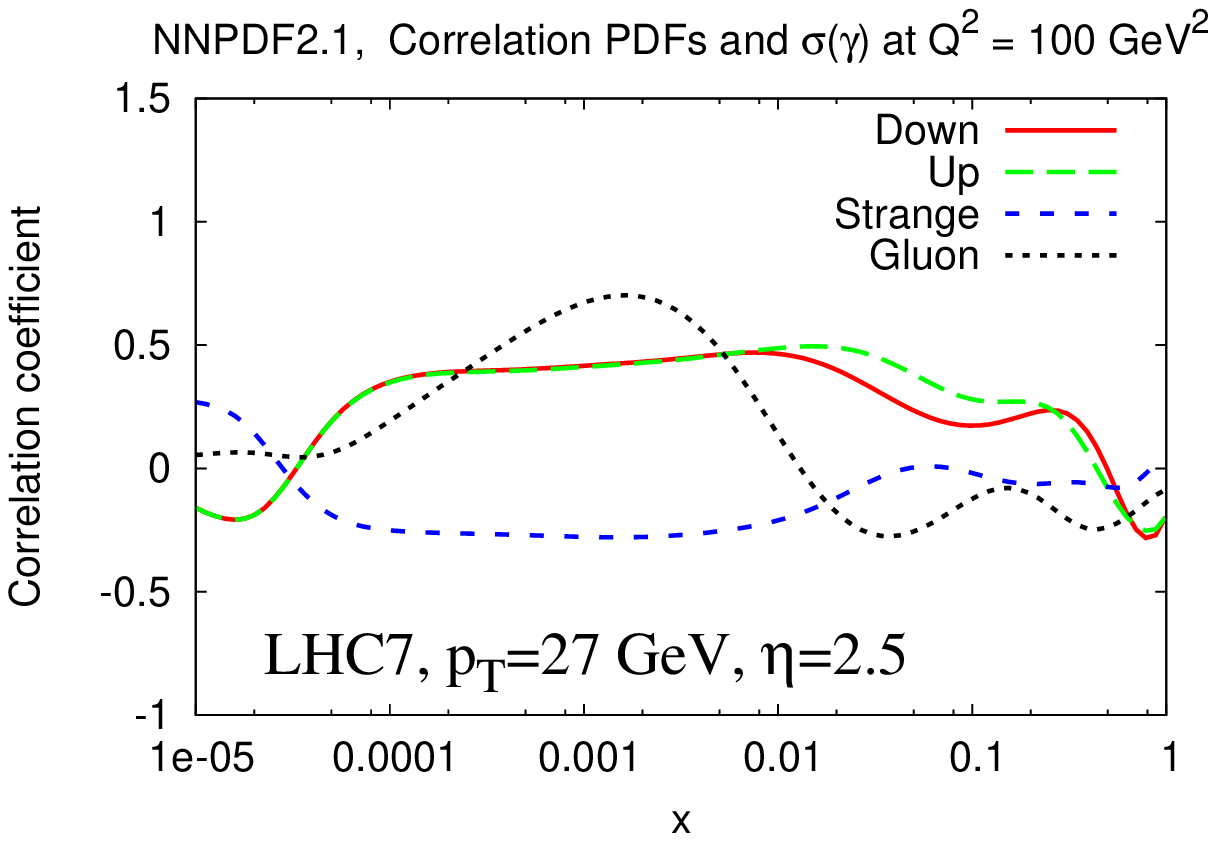}
\epsfig{width=0.49\textwidth,figure=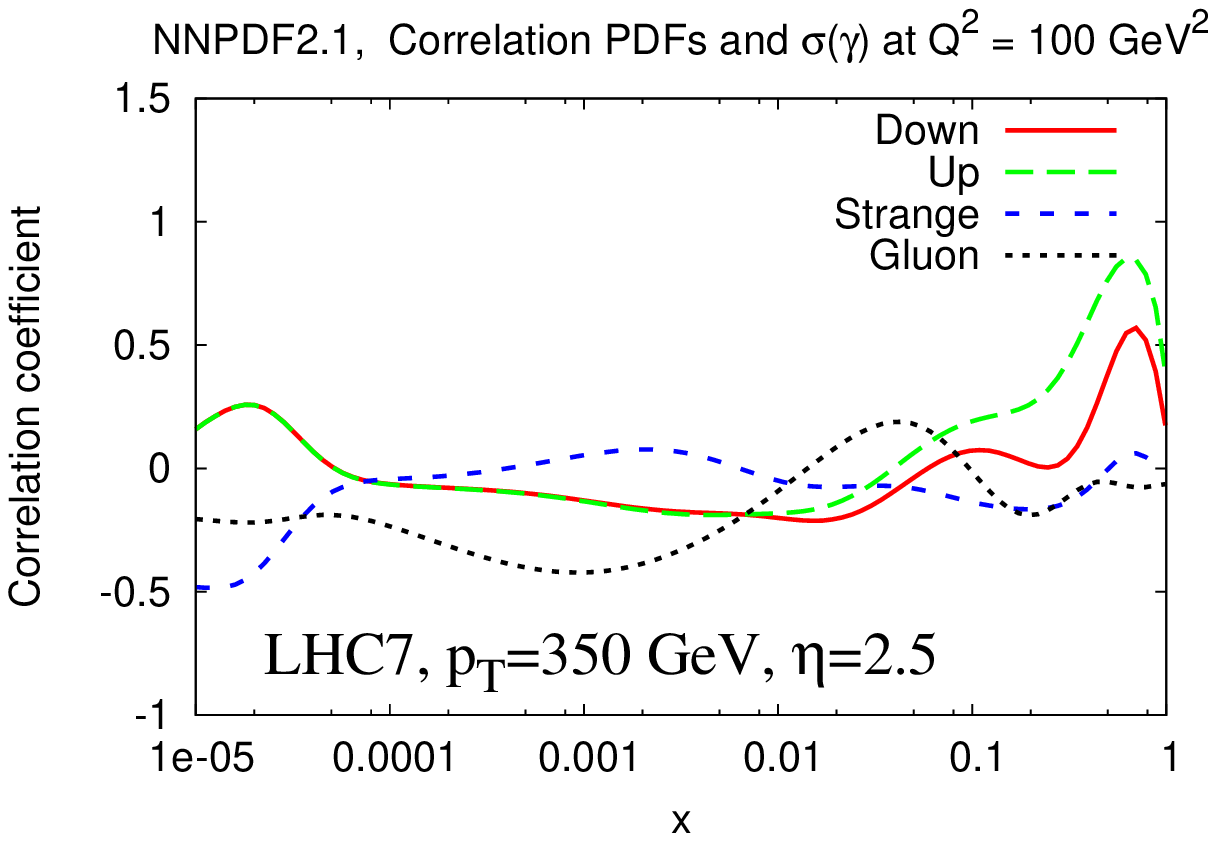}
\caption{\small Correlation between the isolated-$\gamma$ cross section
and various flavours of the NNPDF2.1 parton densities for different ($\ETg$,$y_\gamma$) kinematical ranges at the LHC.
Top (bottom) plots show the correlation at central (forward) rapidities.
Left (right) plots show the correlation for typical low (high) $\ETg$ values.
\label{fig:correlations}}
\end{figure}

%In order to quantify in more detail 
The dependence of isolated-$\gamma$ production 
at different kinematics ranges on the individual flavour of the underlying parton densities
%it is useful to determine the sensitivity of isolated-photon production to the gluon PDF, 
%one can compute 
can be quantified by computing the correlation coefficient between each one of the light-quark and gluon 
distributions and the NLO cross sections as discussed in~\cite{Ball:2011mu}.
%for different kinematics of collider photon production.
Correlation profiles are shown in Fig.~\ref{fig:correlations} for four representative values of 
the photon energy and rapidity typical of the LHC measurements.
Isolated photons at central rapidities (top panels) have a dominant sensitivity to $g(x,Q^2)$ 
in a range between $x$~=~0.01 and  $x$~=~0.2 for increasing photon transverse energy.
At forward rapidities (bottom panels) isolated-$\gamma$ are sensitive to the gluon
densities at $x$ values around 10$^{-3}$ for low $\ETg$, but the PDF sensitivity shifts 
to the valence $u$-quarks as the photon energy increases. Being able to carry out measurements
at even more forward rapidities, e.g. down to $y_\gamma\approx$~5 accessible in LHCb~\cite{DeLorenzi:2010zt}
and/or at low-$\ETg$ accessible via photon-conversions in ALICE~\cite{Marin:2008zz},
would be very interesting in order to probe $x$ values below $10^{-4}$
where $g(x,Q^2)$ is only weakly constrained~\cite{Rojo:2009ut}  
by HERA $F_2$, $F_L$, $F_{2,c}$ structure functions and by the momentum sum-rule.%\\
%I believe that the data have an impact forward pq inferior
%first error bars are larger, second only depend on gluon
%data at low PT and the third pq gluó in 0001 is much better
%determined by the HERA data that large x.
%\item Yes, but combined HERA-I is very precise, and you have data in F2c month
%and the momentum sum rule. It is only for $ x $ <1e-4 errors gluon becoming more substantial.

%%%%%%%%%%%%%%%%%%%%%%%%%%%%%%%%
%%%%%%%%%%%%%%%%%%%%%%%%%%%

\begin{figure}[ht!]
\centering
\epsfig{width=0.49\textwidth,figure=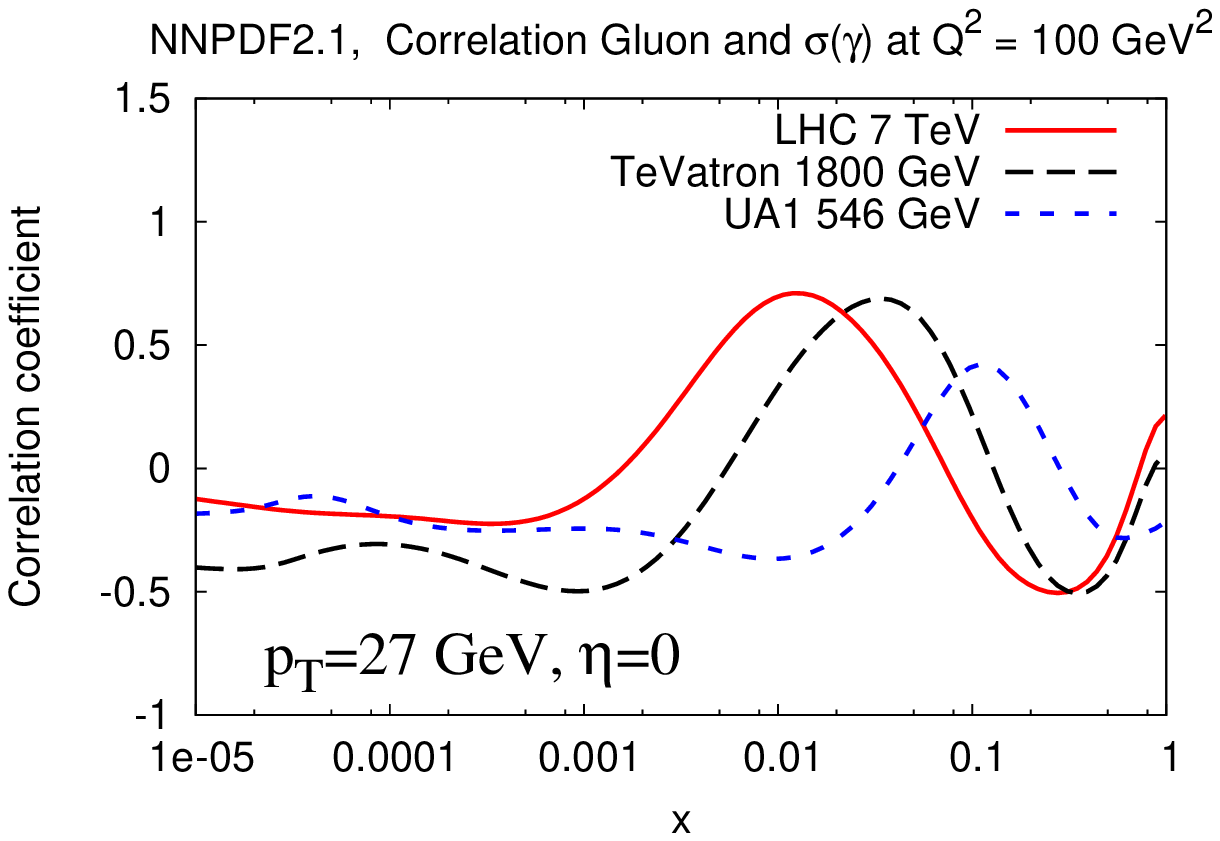}
\epsfig{width=0.49\textwidth,figure=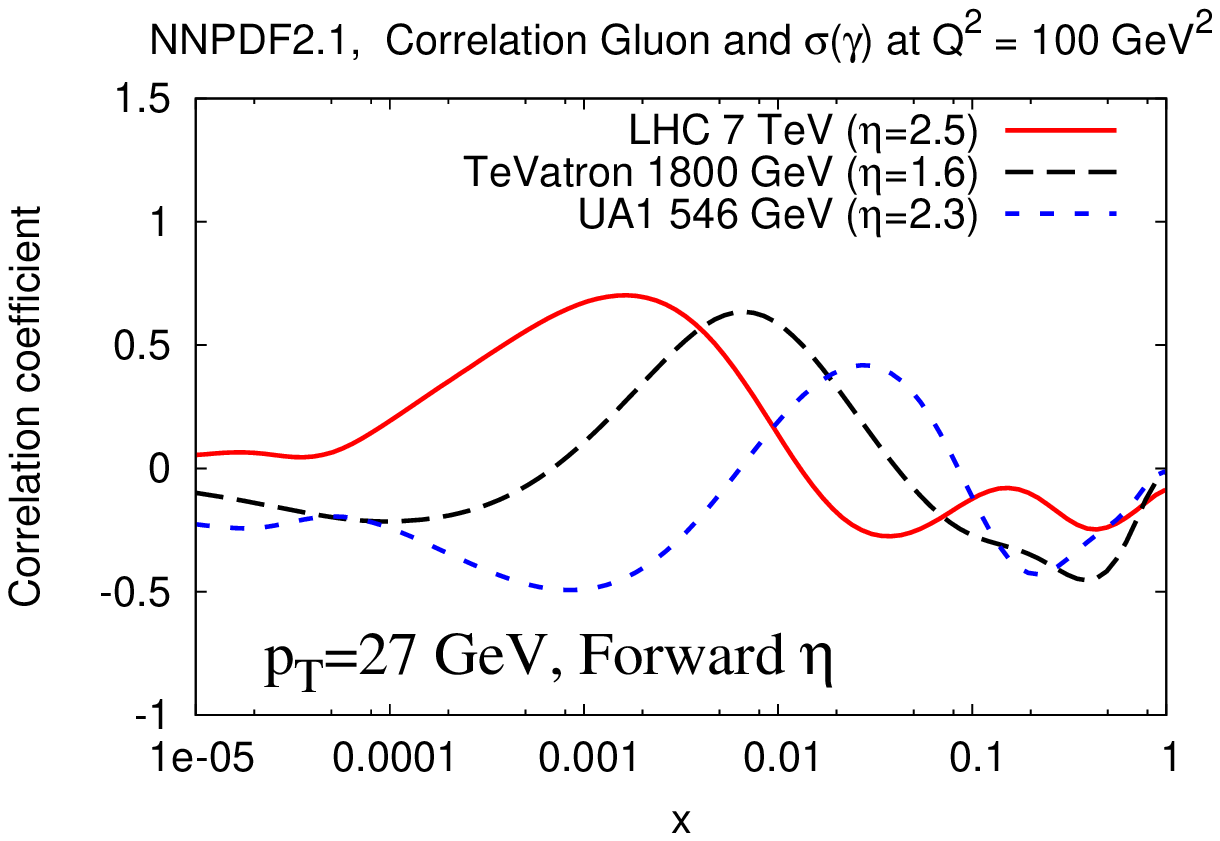}
\caption{\small Correlations between the  gluon PDF and the isolated-$\gamma$ cross section
at a fixed $\ETg$~=~27~GeV in \pp\ and \ppbar\ collisions at central (left) and forward (right)
rapidities for different collider energies.
\label{fig:correlations2}}
\end{figure}

In Fig.~\ref{fig:correlations2} we plot the sensitivity  of the isolated-$\gamma$ data 
to the proton gluon at a fixed value of the photon transverse energy ($\ETg$~=~27~GeV) 
for various collider energies at central (left) and forward (right) rapidities. 
At midrapidity, the peak of the correlations shifts towards larger $x$ values for decreasing $\sqrts$,
up to about $x\approx$~0.1 at $\cO{\rm 500\;GeV}$. At forward rapidities the LHC data
are more sensitive to the gluon density than measurements at \spps\ and Tevatron because 
the former involve \pp\ collisions (dominated by $qg$-Compton processes) rather than \ppbar\ collisions 
(with \qqbar-annihilation playing a larger role, Fig.~\ref{fig:subproc}), and because the ($\ETg$,$y_\gamma$)
phase-space covered by ATLAS and CMS is much larger than that at lower c.m. energies.\\

%%%%%%%%%%%%%%%%%%%%%%%%%%%%%%%%
%%%%%%%%%%%%%%%%%%%%%%%%%%%

%%%%%%%%%%%%%%%%%%%%%%%%%%%%%%%%%%%%%%%%%%%%%%%%%%%%%%%%%%%%%%%%%%%%%%%%%%%%%%%%%%%%%%%%

\subsection{Impact of isolated-photon data on the gluon}

%As discussed in the previous Section, isolated-photon data are particularly sensitive to the
%gluon distribution at $x$ values between 0.01 and 0.2.
In order to quantify the constraints that the various photon datasets impose on the gluon PDF,
the results of the $\chi^2$-analyses listed in Table~\ref{tab:chi2datasets} 
are combined into seven groups according to each collider energy :
%\footnote{We don't consider the CMS 2.76~TeV data with small statistics .}: 
200, 546, 630~GeV and 1.8, 1.96, 2.76 and 7~TeV. In the  
grouping, we discard the six datasets with $\chi^2 \ge 3$ which have underestimated experimental
uncertainties (Fig.~\ref{fig:palpha2}) and only result in a loss of accuracy of the reweighting
method without any impact on the PDFs whatsoever.\\

The effective number of replicas $N_{\rm eff}$ after reweighting, Eq.~(\ref{eq:effective}), 
for each collider energy is summarised in Table~\ref{tab:effective}. As expected,
the more constraining datasets, i.e. those with smallest $N_{\rm eff}$, are those from the LHC at 7~TeV. 
The measurement at 2.76~TeV has only 5 data-points fully consistent with NLO with basically no impact on the PDFs. 
Of the rest, only the Tevatron Run-II and the 630~GeV datasets seem
to have also some, albeit small, reduction of $N_{\rm eff}$. % impact on  $g(x,Q^2)$.\\ %, albeit smaller than in the LHC case.

%%%%%%%%%%%%%%%%%%%%%%%%%%%%%%%
\begin{table}[htbp!]
\centering
%\begin{tabular}{|c|c|c|c|c|c|c|}
\begin{tabular}{cccccccc}
\hline
 $\sqrts$ (TeV) & 0.2  & 0.546 & 0.630 & 1.8  & 1.96 & 2.76  & 7 \\ %\hline
$N_{\rm eff}$   & 99.6 & 99    & 95    & 99.8 & 96   & 96 & 87 \\ \hline
\end{tabular}
\caption{\small Effective number of replicas $N_{\rm eff}$, 
Eq.~(\ref{eq:effective}), left for each one of the collider energies
considered, after inclusion of the isolated-$\gamma$ data. 
The starting number of NNPDF replicas for this analysis is $N_{\rm rep}=100$.
\label{tab:effective}}
\end{table}
%%%%%%%%%%%%%%%%%%%%%%%%%%%%%%%

\begin{figure}[htbp!]
\centering
\epsfig{width=0.32\textwidth,figure=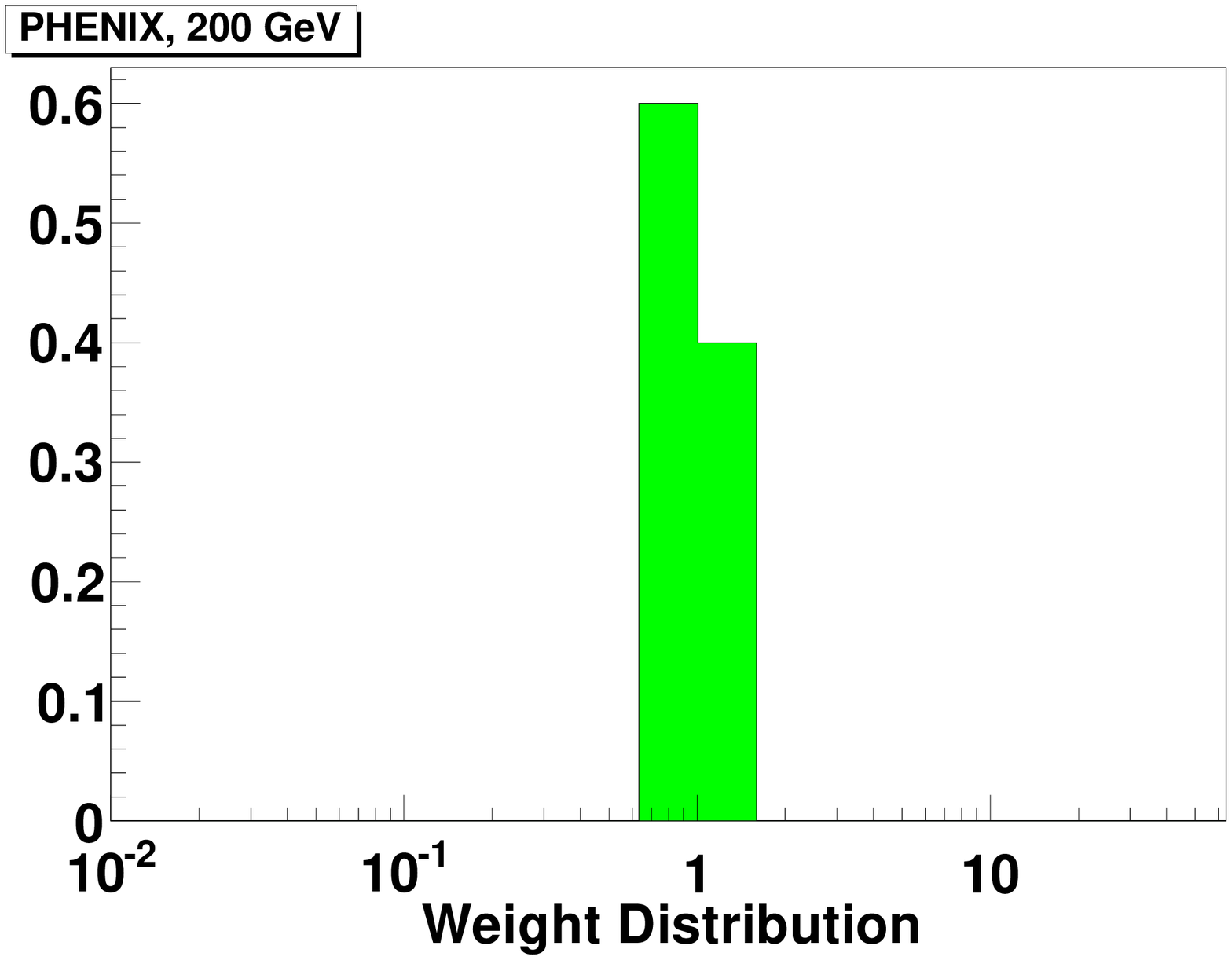}
\epsfig{width=0.32\textwidth,figure=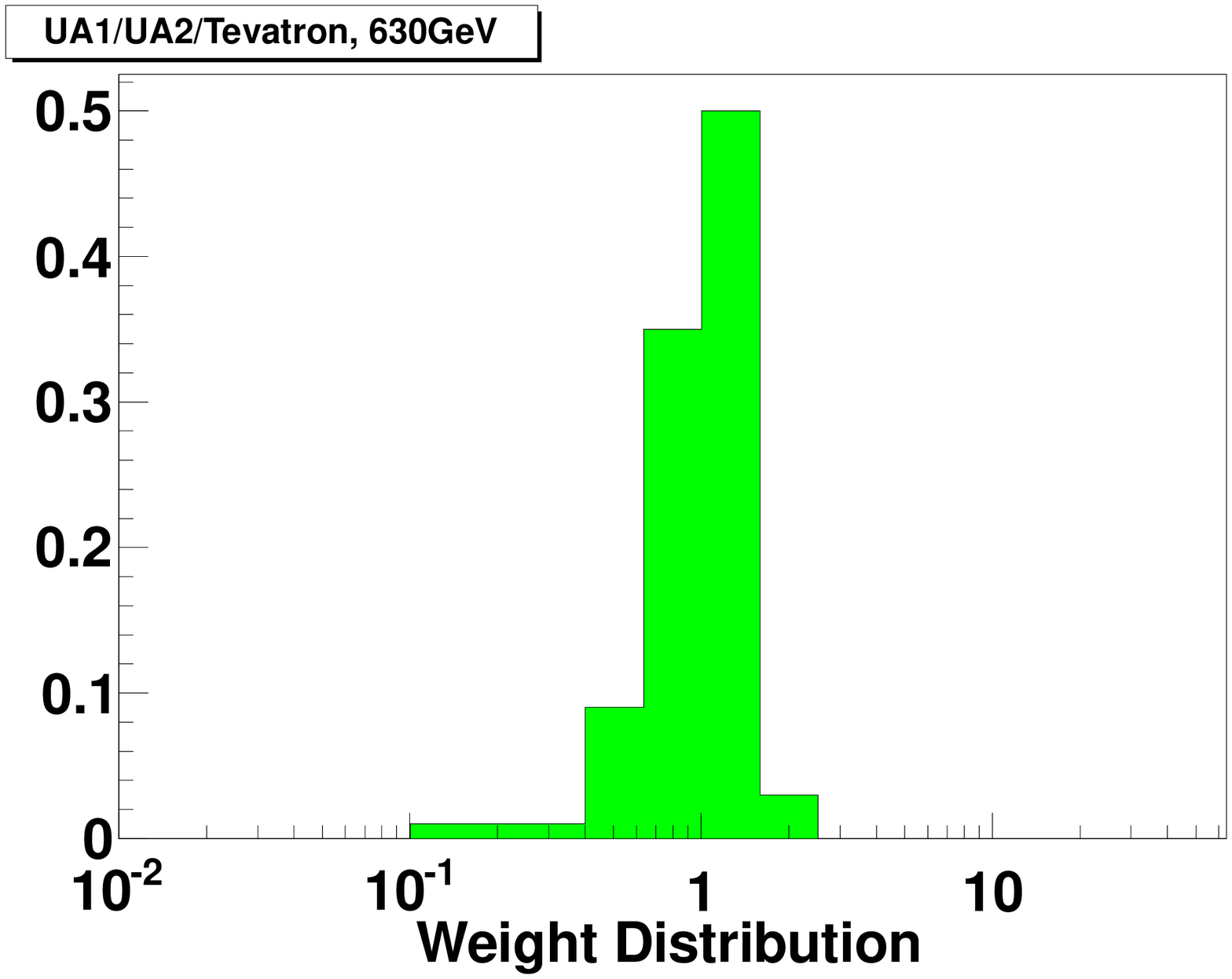}
\epsfig{width=0.32\textwidth,figure=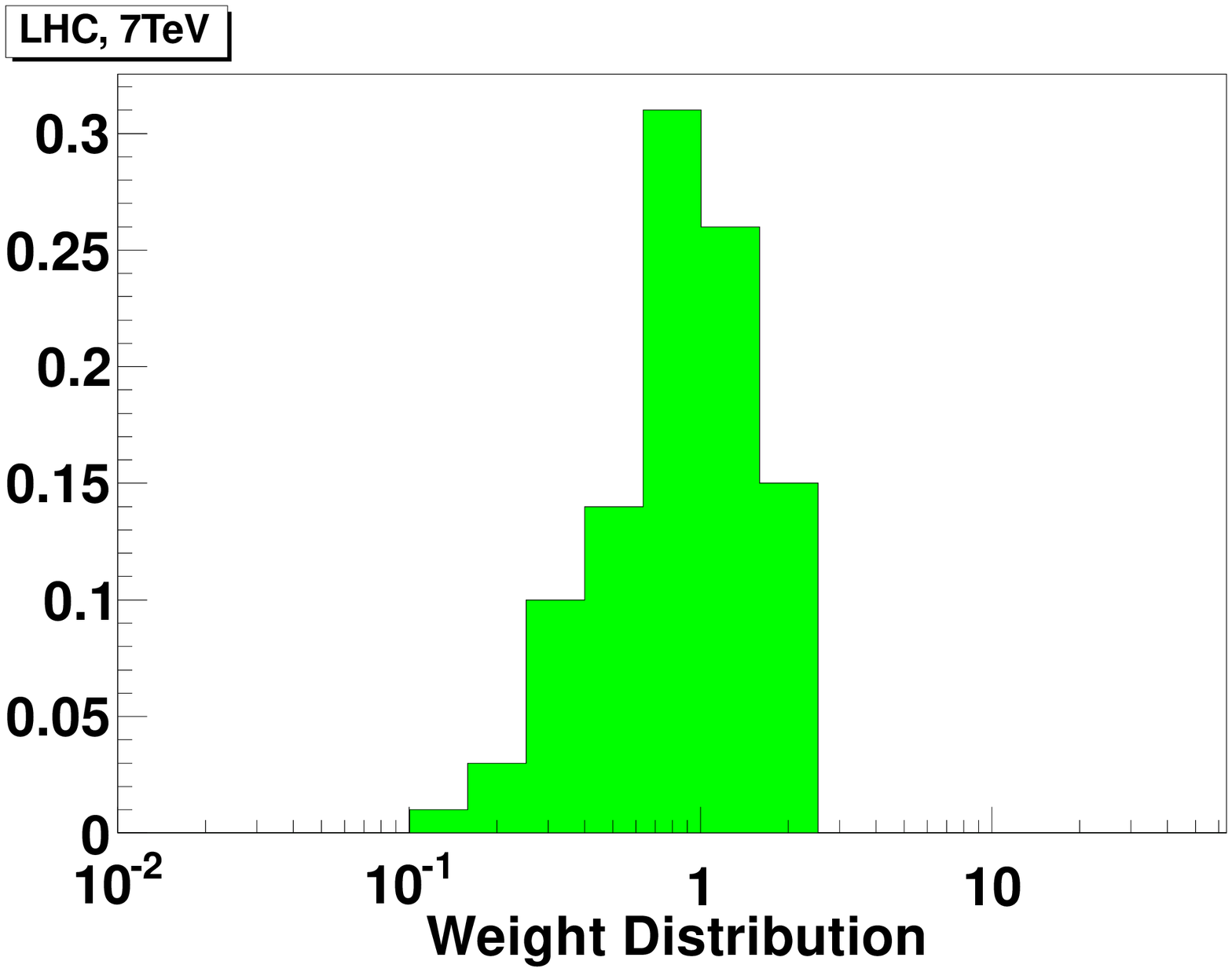}
\caption{\small Distribution of NNPDF replicas weights, Eq.~(\ref{eq:weights}), for three
representative collider energies: PHENIX data at $\sqrts$~=~200~GeV
(left), $\sqrts=630$~GeV data (center), and LHC results at $\sqrts$~=~7~TeV (right).
\label{fig:weights}}
\end{figure}

In Fig.~\ref{fig:weights} we show the 100 replica weights, Eq.~(\ref{eq:weights}), 
for three representative collider energies: PHENIX at $\sqrts$~=~200~GeV,
UA1/UA2/Tevatron at $\sqrts$~=~630~GeV, and the LHC data at $\sqrts$~=~7~TeV. 
For the least constraining dataset (the PHENIX data with $N_{\rm eff}$ closer to $N_{\rm rep}$)
all replicas have essentially the same weight. 
For the other two datasets the distributions become broader, specially for LHC at 7~TeV, 
showing that the impact of these measurements on the PDFs is larger -- some replicas 
are preferred (larger weight) than others (smaller weight) when confronted to the photon results.\\

%%%%%%%%%%%%%%%
%%%%%%%%%%%%%%%%

%%%%%%%%%%%%%%%
\begin{figure}[htbp!]
\centering
\epsfig{width=0.49\textwidth,figure=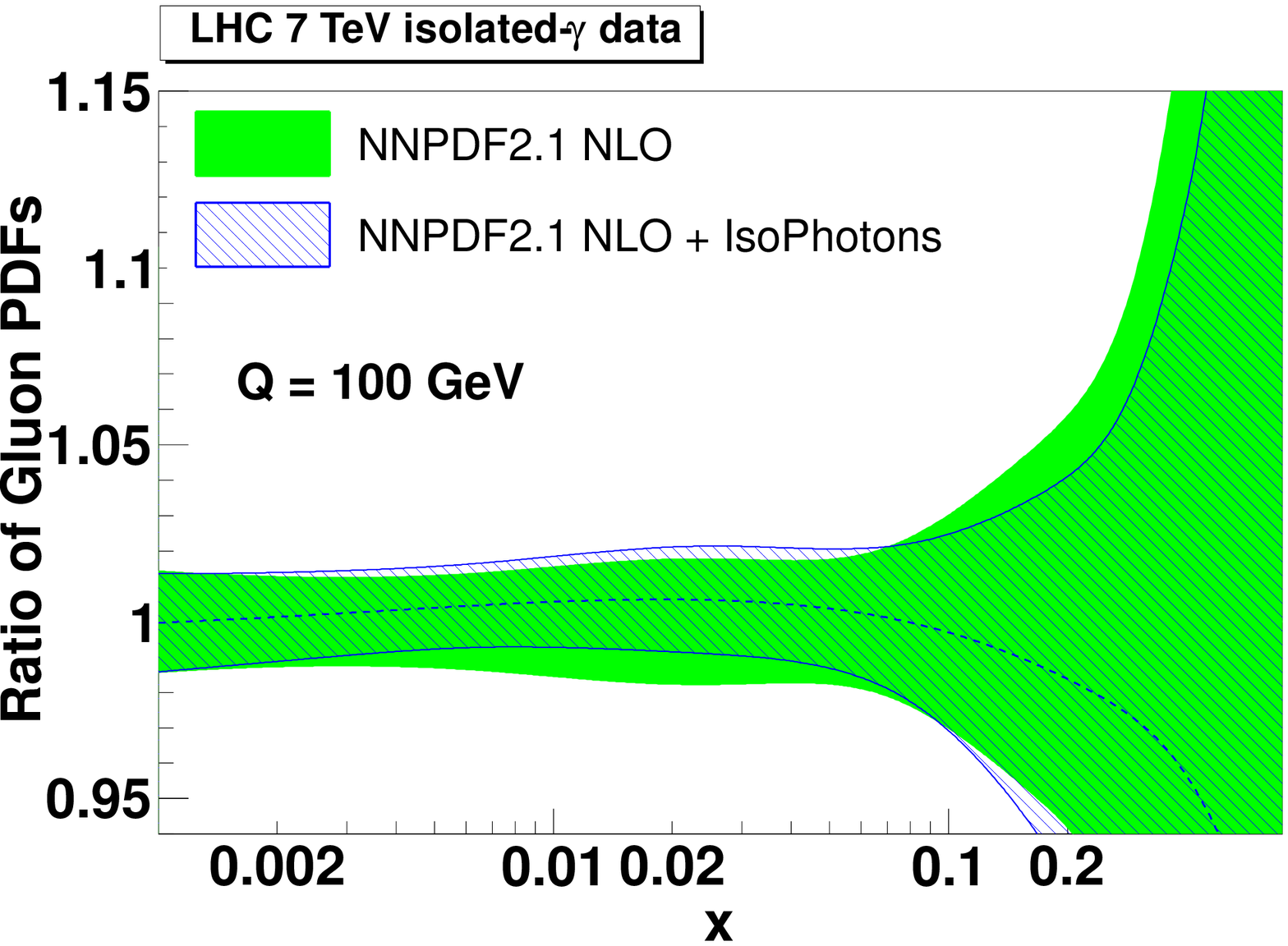}
\epsfig{width=0.49\textwidth,figure=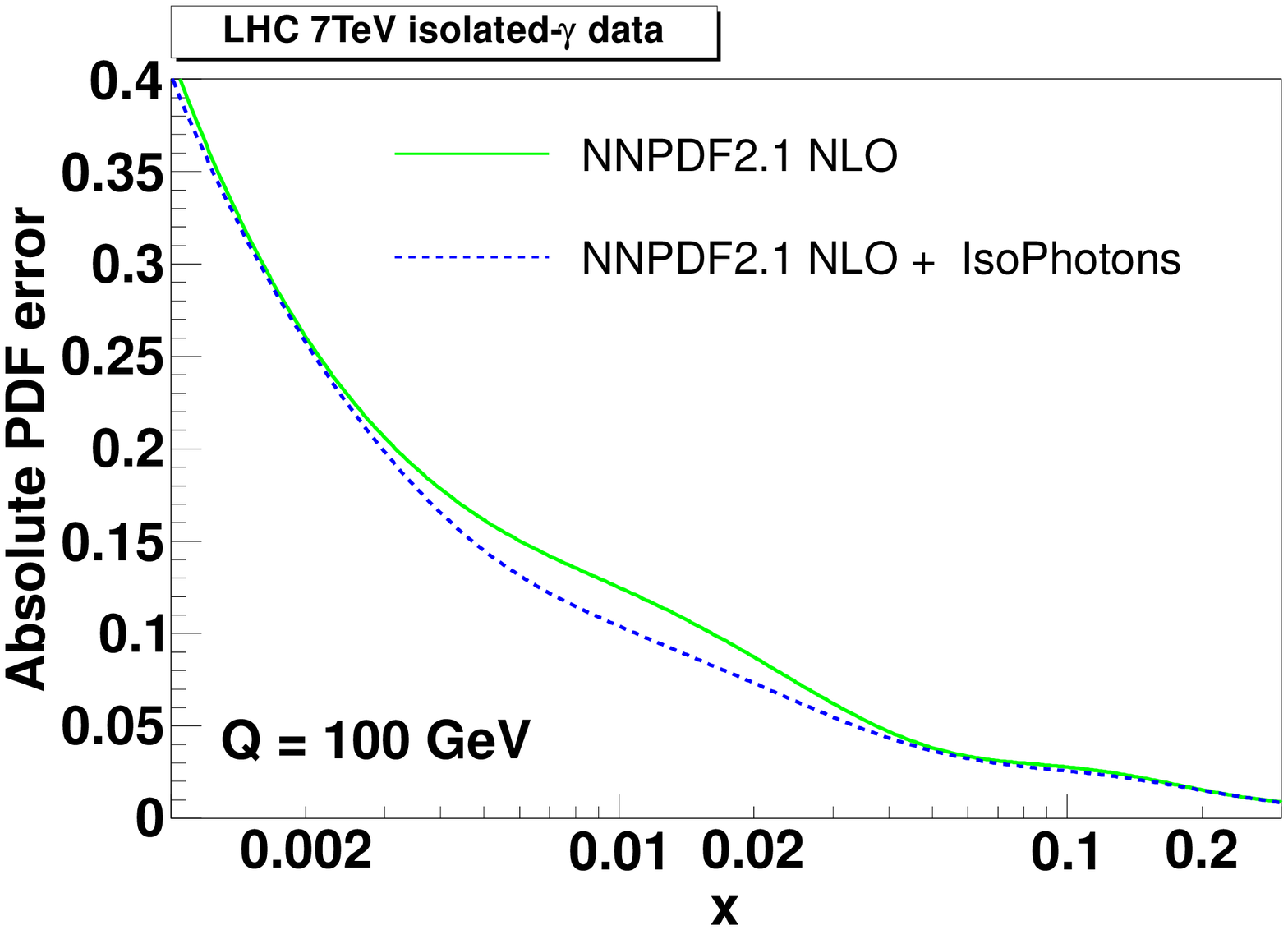}\\
\epsfig{width=0.49\textwidth,figure=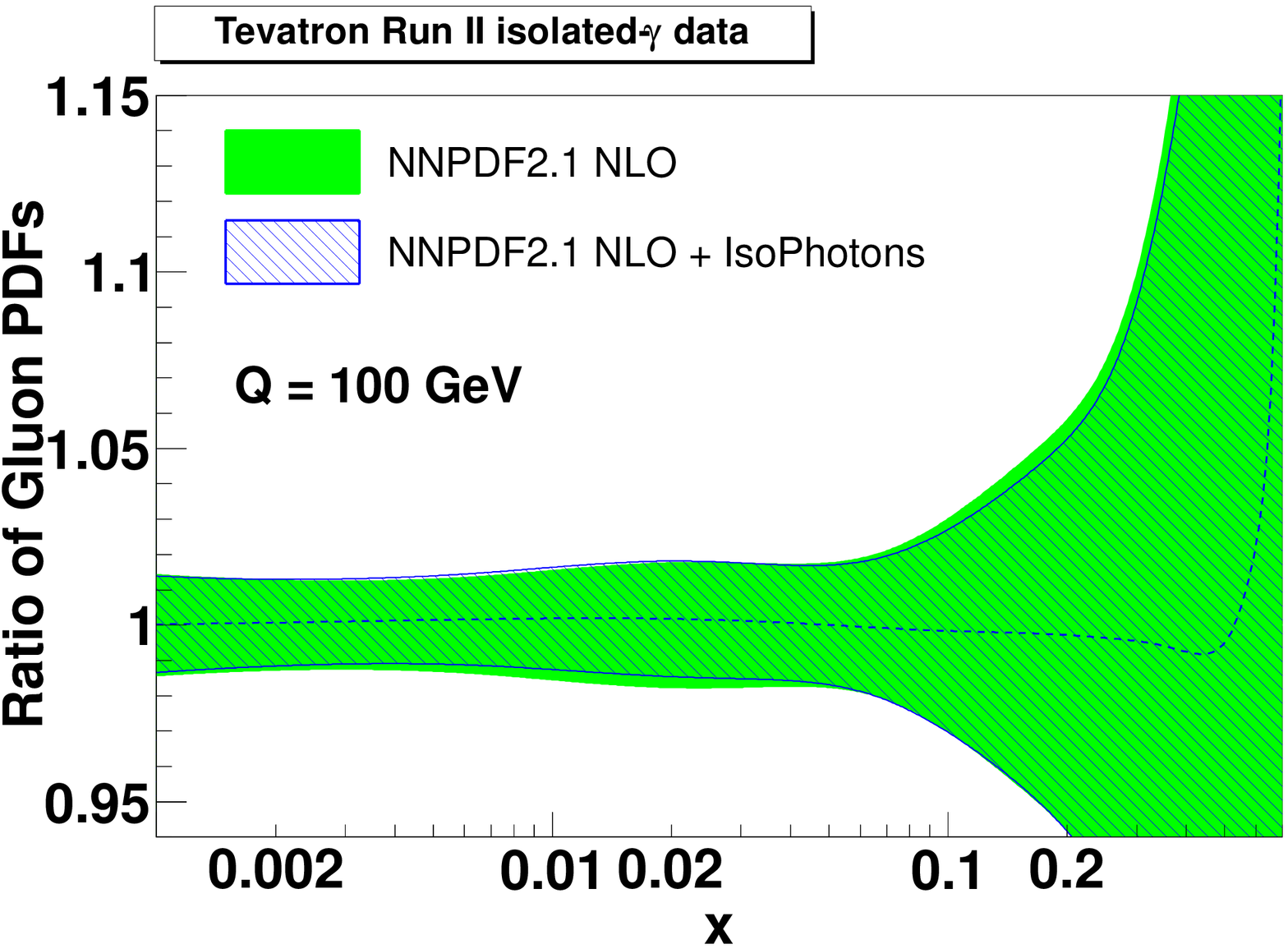}
\epsfig{width=0.49\textwidth,figure=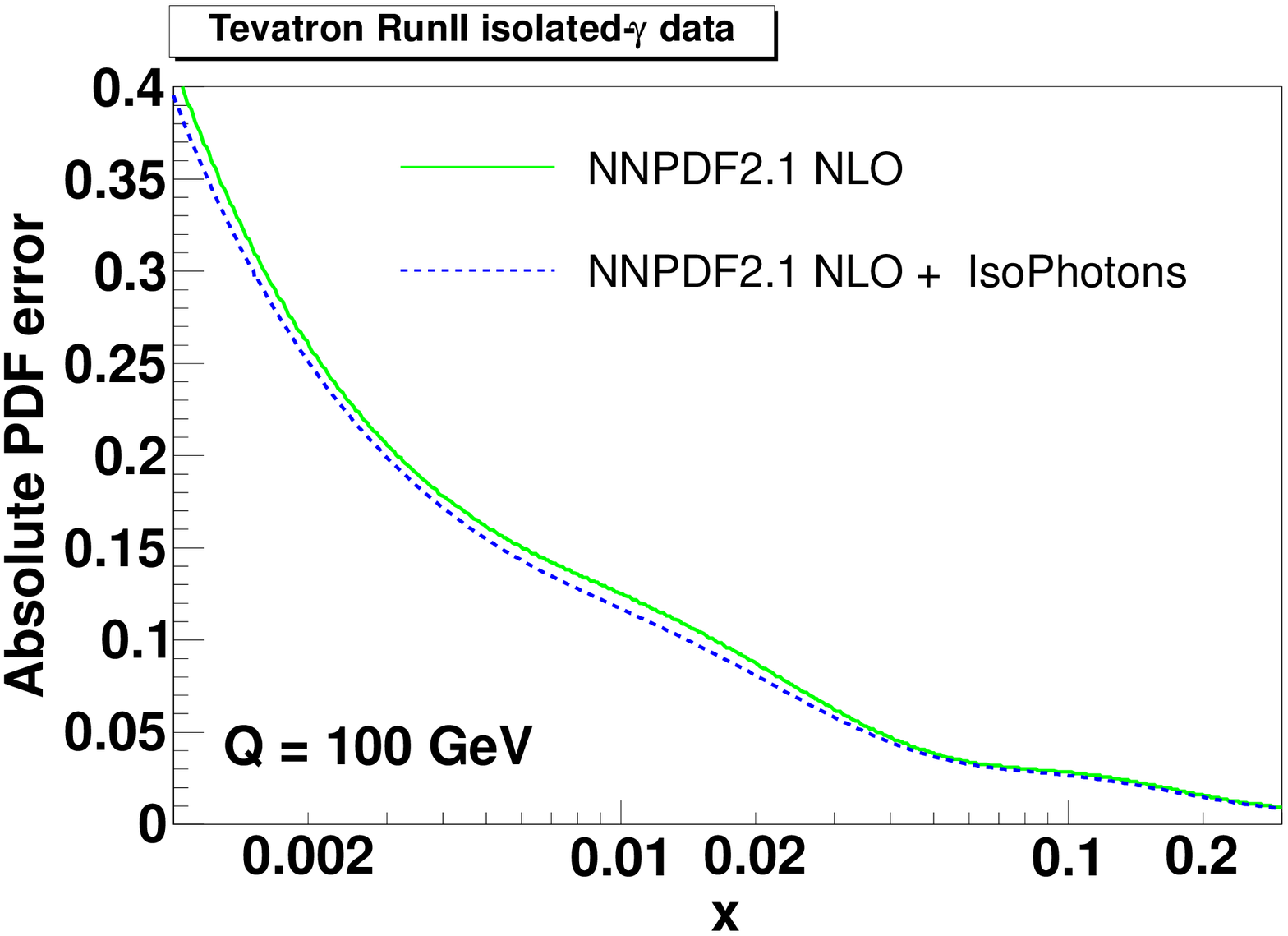}\\
\epsfig{width=0.49\textwidth,figure=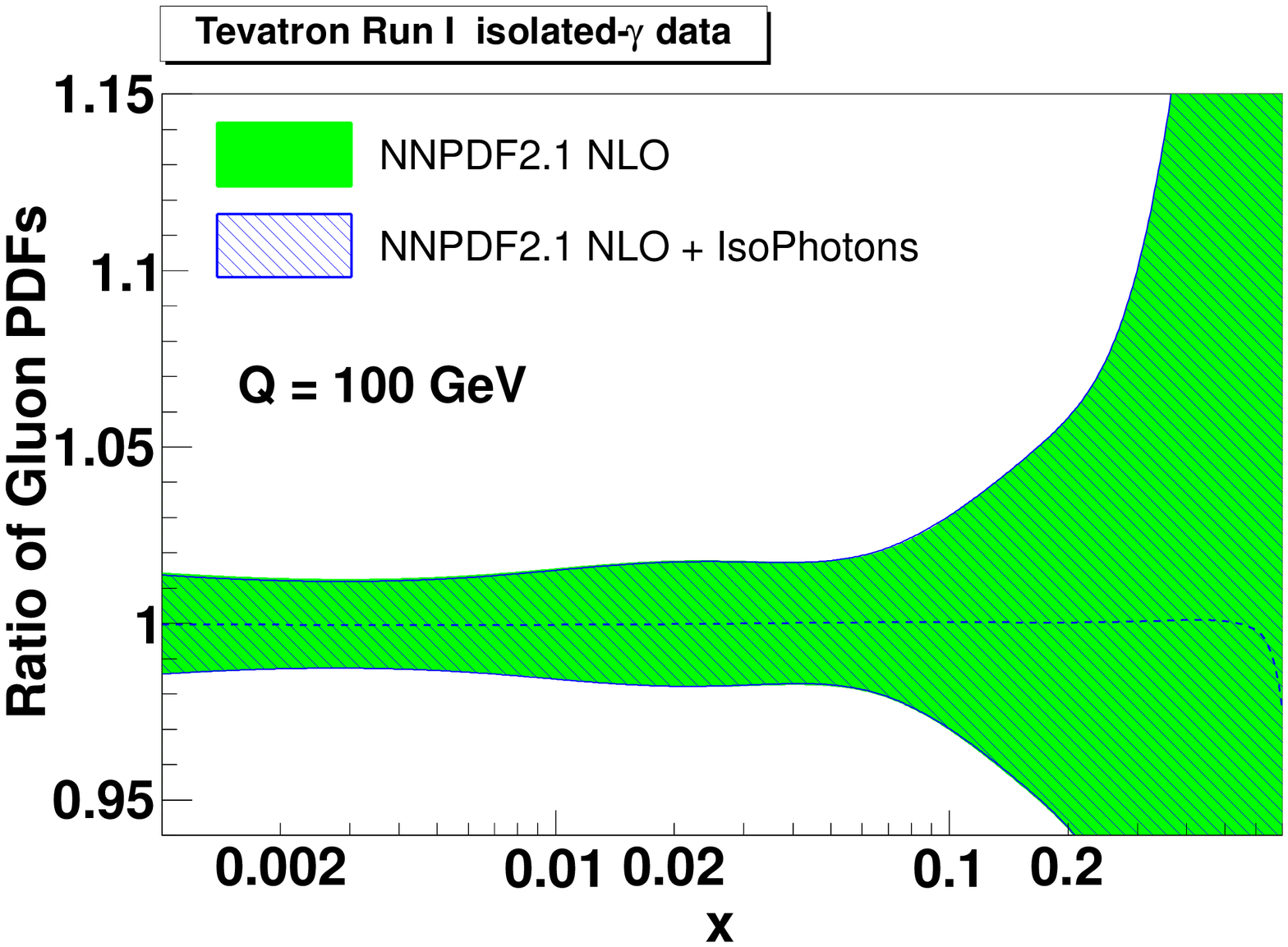}
\epsfig{width=0.49\textwidth,figure=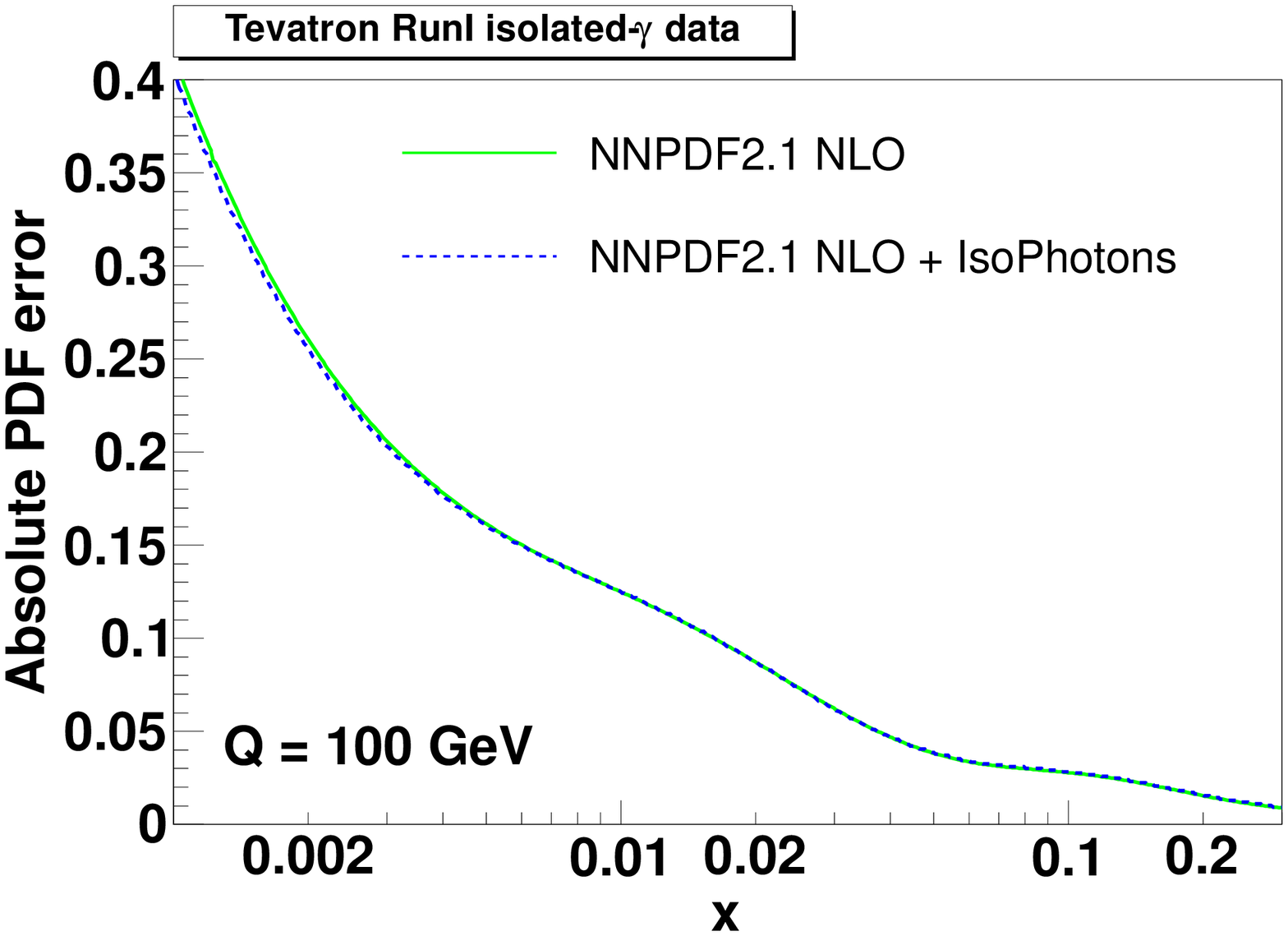}
\caption{\small Comparison between the NNPDF2.1 NLO
gluon before (green solid band) and after (dashed blue area) 
inclusion of the isolated-$\gamma$ data from 
(top to bottom): LHC-7~TeV, Tevatron Run-II at 1.96~TeV, and Run-II at 1.8~TeV. 
The left plots show the ratio between the original and the new $g(x,Q^2)$ 
while the right panels indicates the reduction of absolute $g(x,Q^2)$ 
uncertainties thanks to the photon data. PDFs are valued at $Q=100$~GeV, a typical LHC scale.
\label{fig:gluon}}
\end{figure}
%%%%%%%%%%%%%%%%

%%%%%%%%%%%%%%%
\begin{figure}[htbp!]
\centering
\epsfig{width=0.49\textwidth,figure=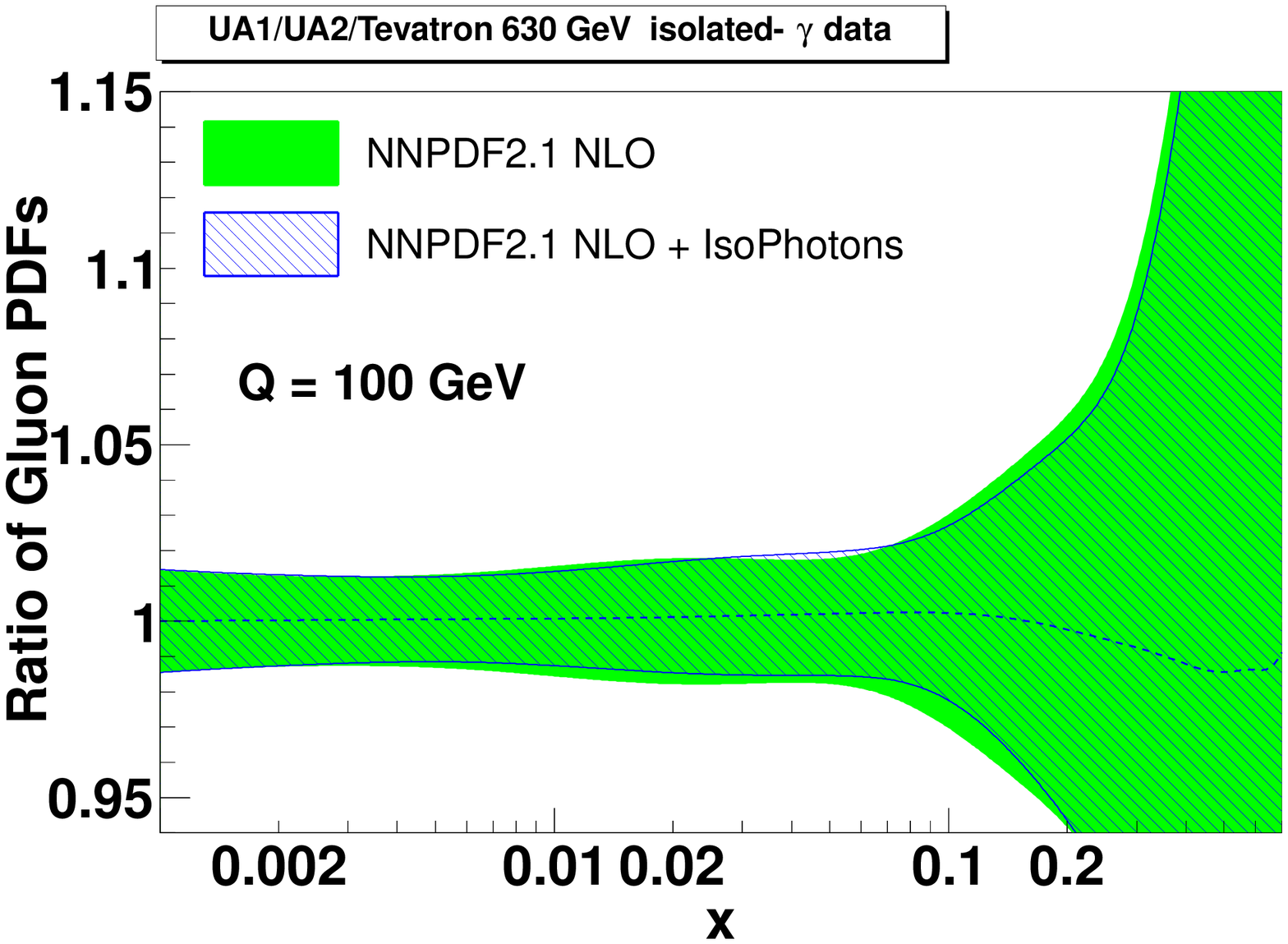}
\epsfig{width=0.49\textwidth,figure=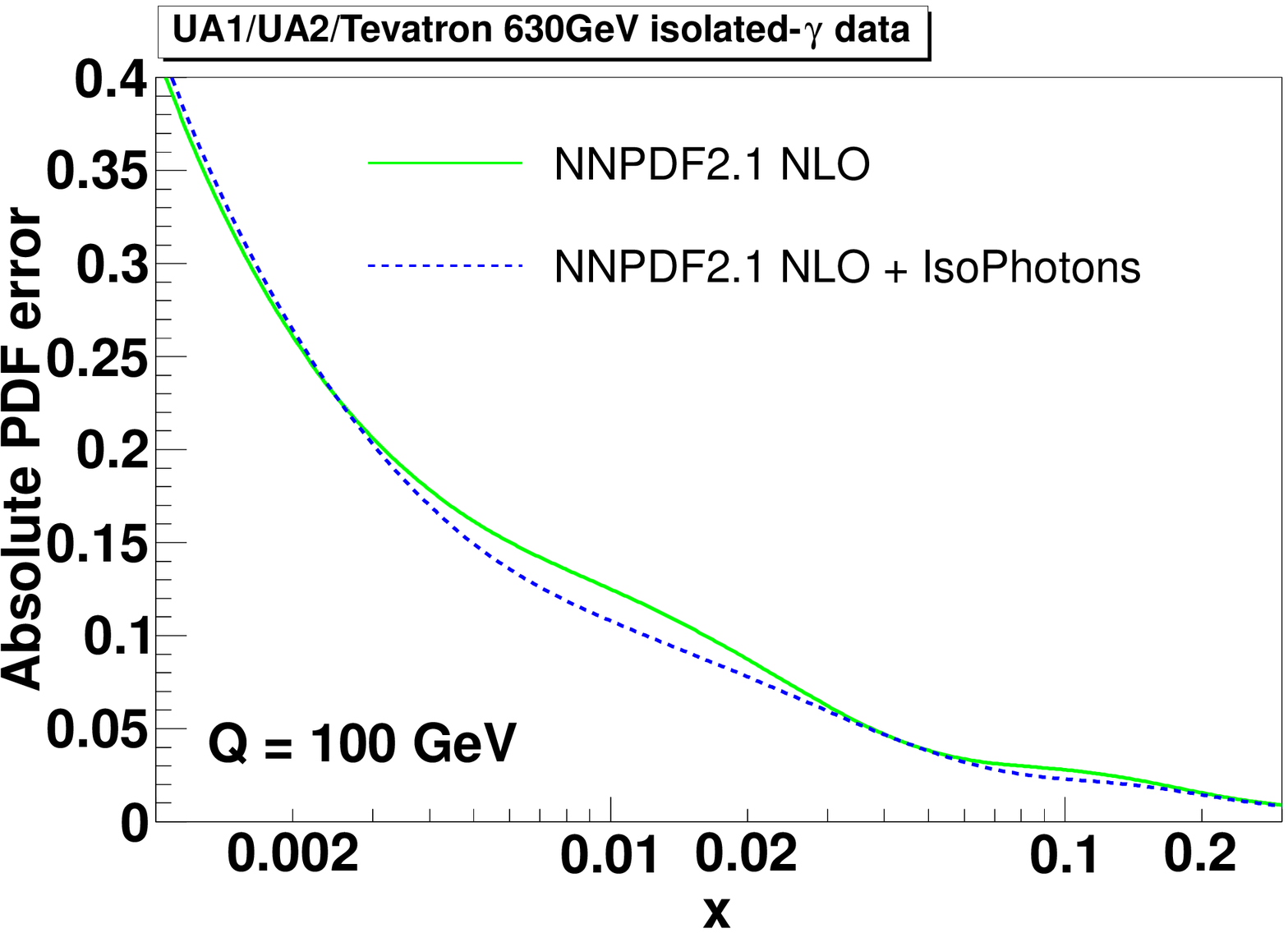}\\
\epsfig{width=0.49\textwidth,figure=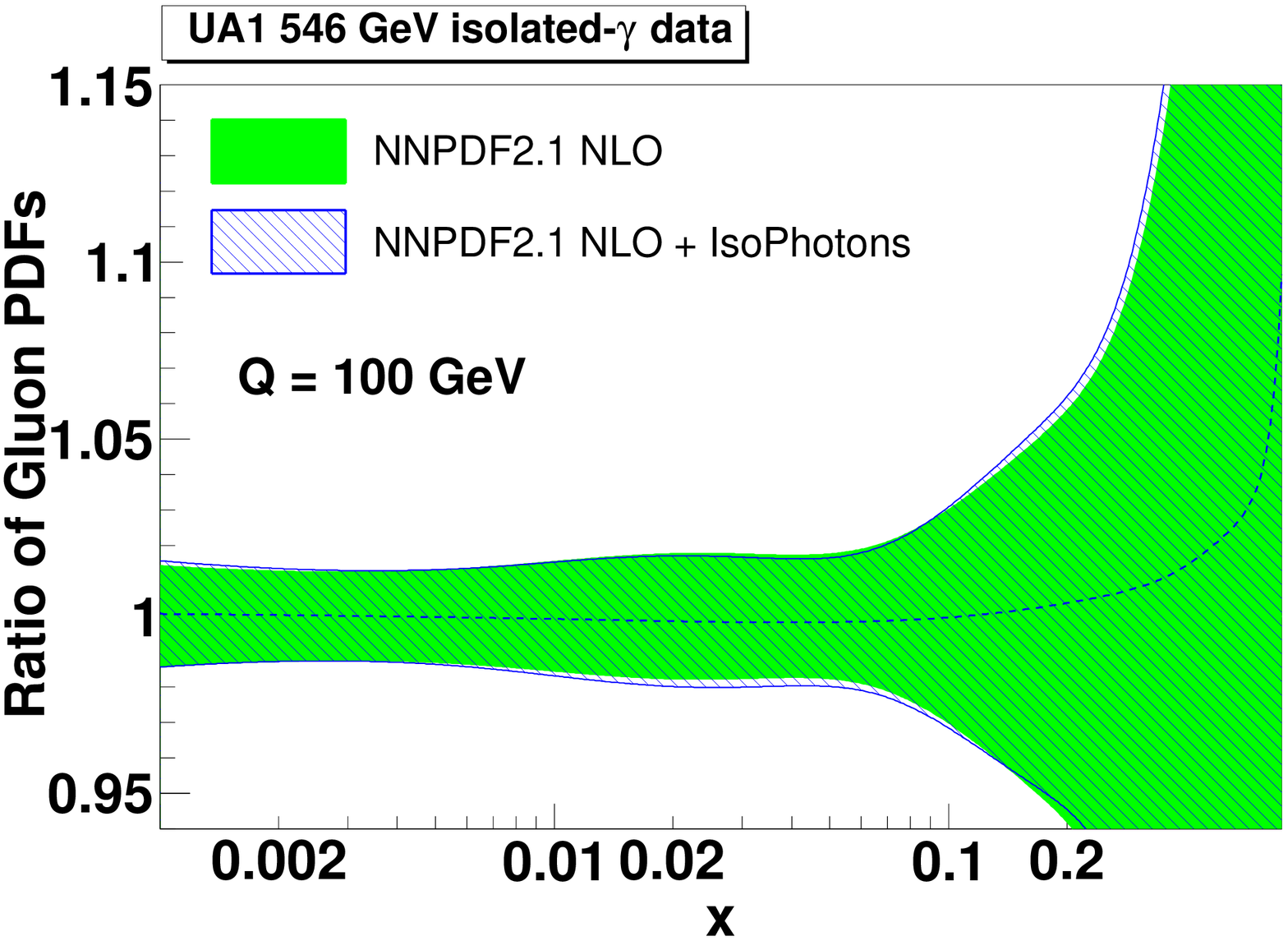}
\epsfig{width=0.49\textwidth,figure=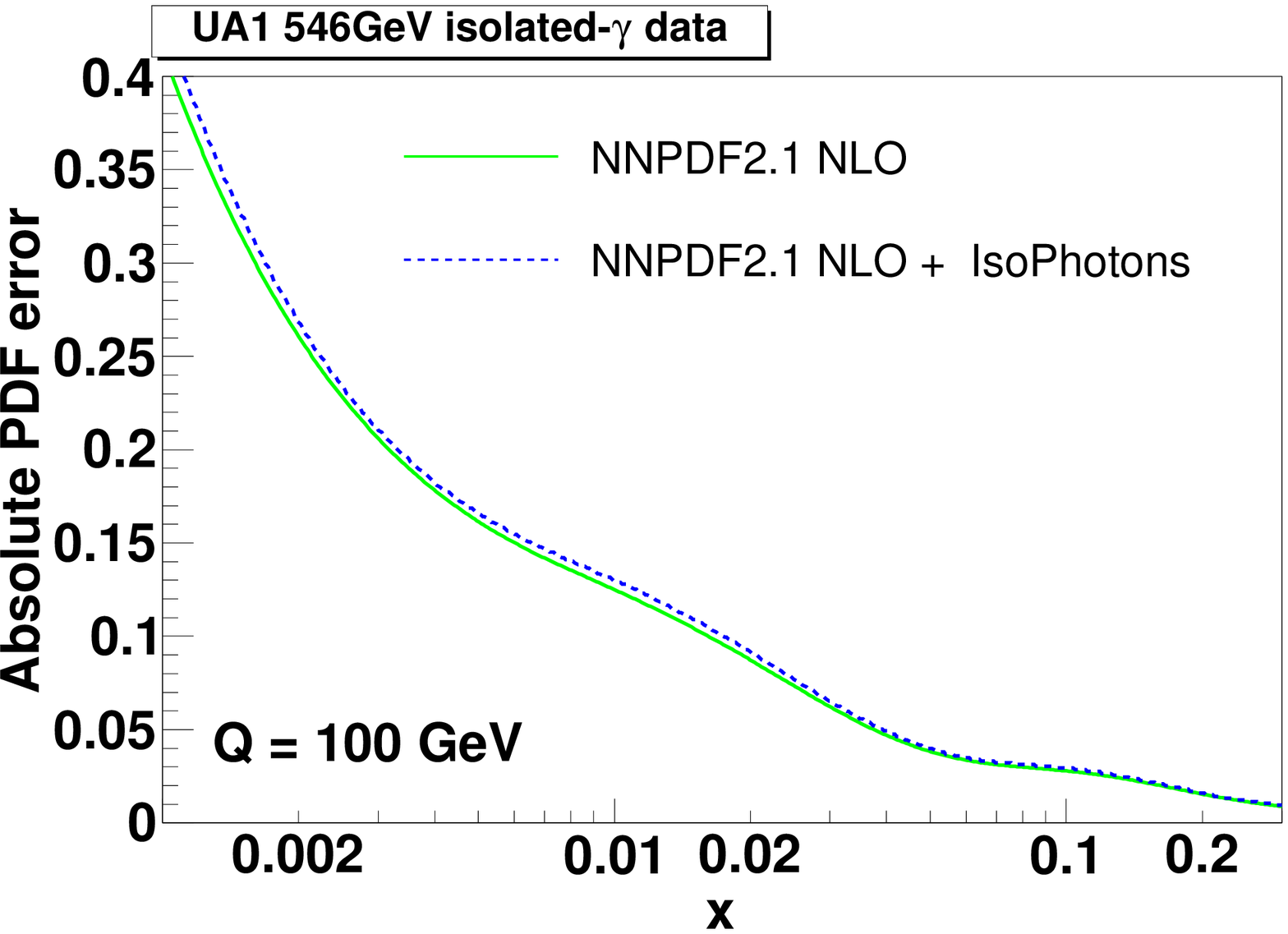}\\
\epsfig{width=0.49\textwidth,figure=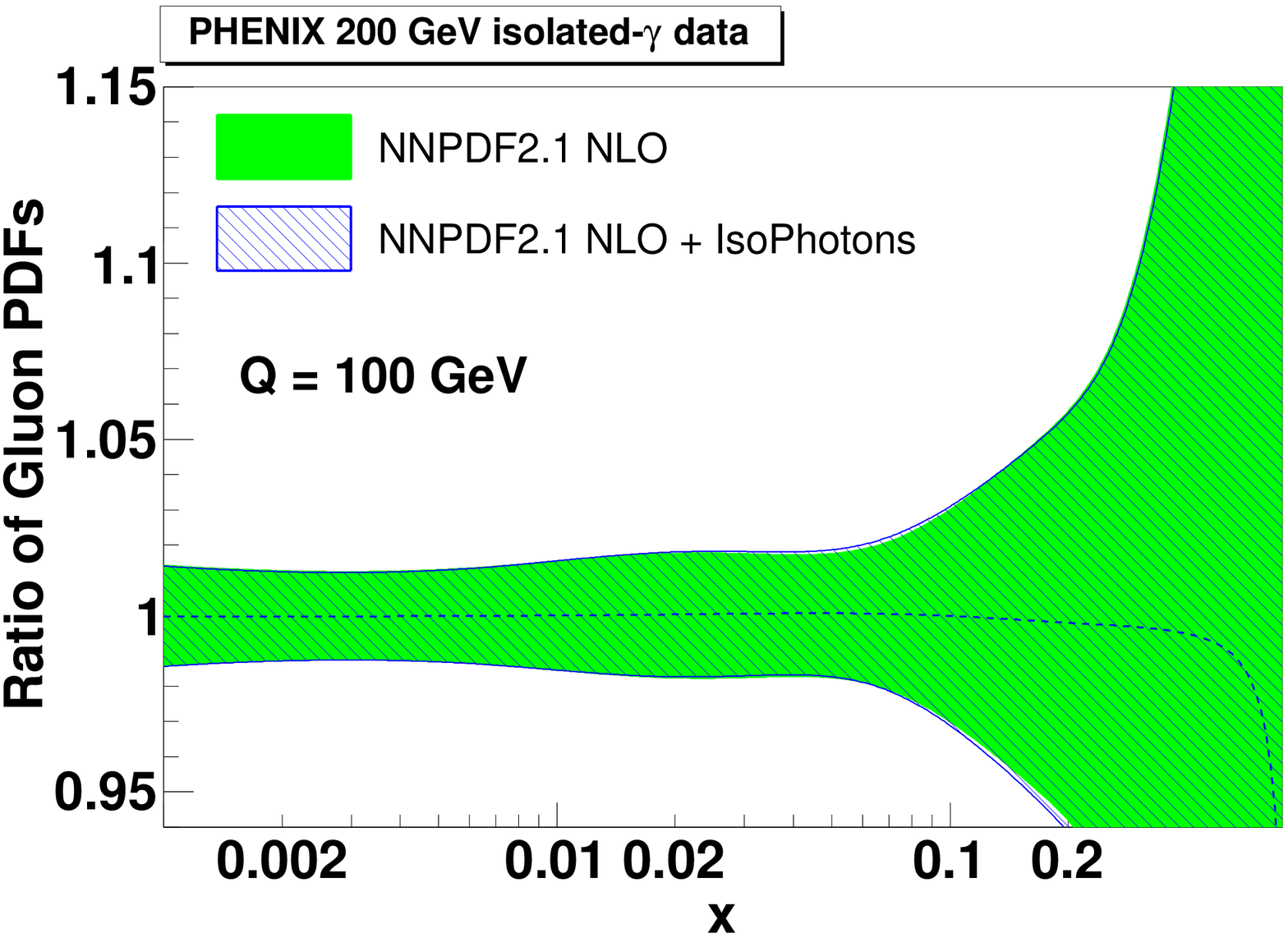}
\epsfig{width=0.49\textwidth,figure=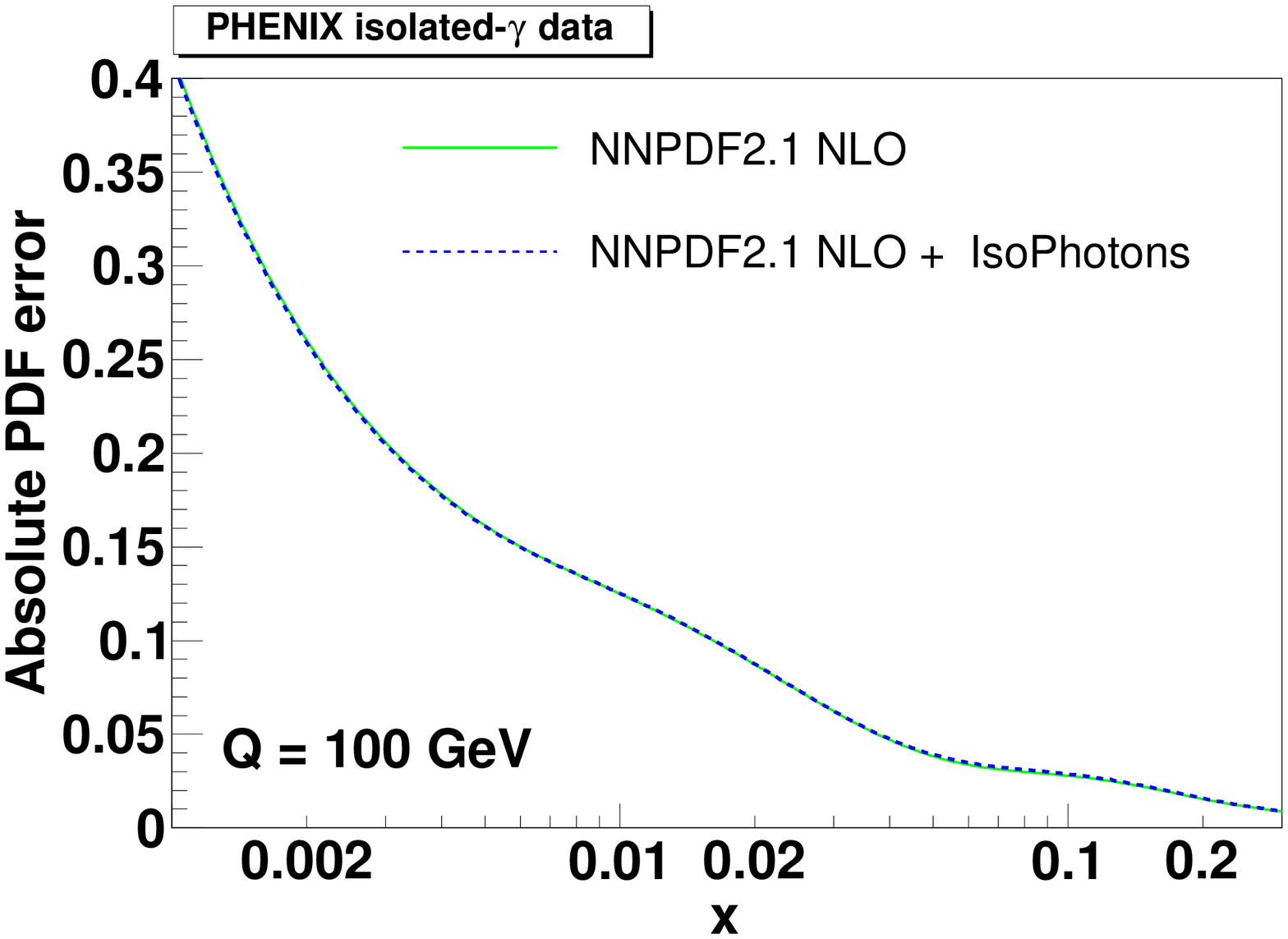}
\caption{\small Same as Fig.~\ref{fig:gluon} 
for the remaining collider data (top to bottom): UA1/UA2/Tevatron at 630~GeV, UA1-546~GeV, and PHENIX-200~GeV.
\label{fig:gluon2}}
\end{figure}
%%%%%%%%%%%%%%%%

\begin{figure}[htbp!]
\centering
\epsfig{width=0.325\textwidth,figure=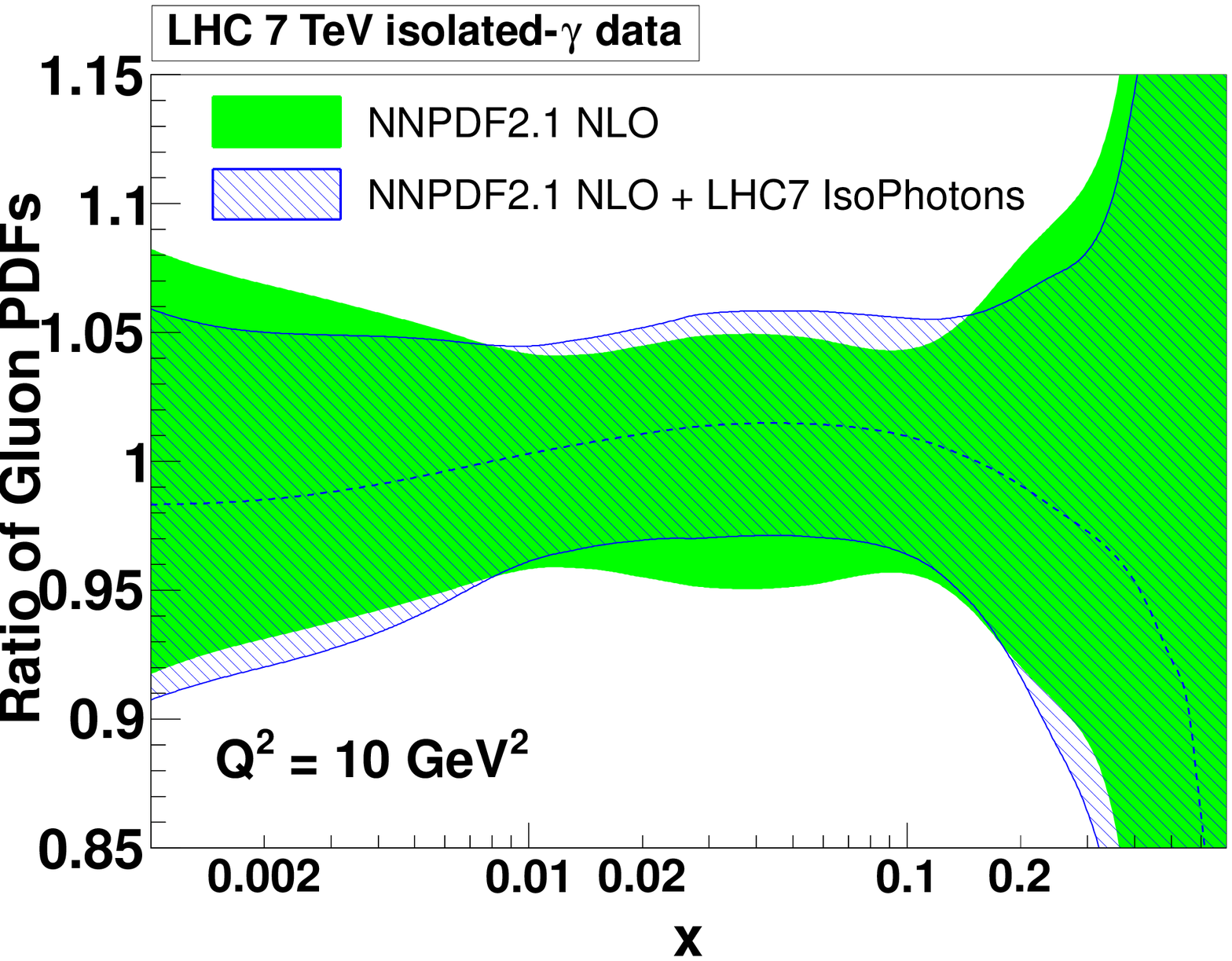}
\epsfig{width=0.325\textwidth,figure=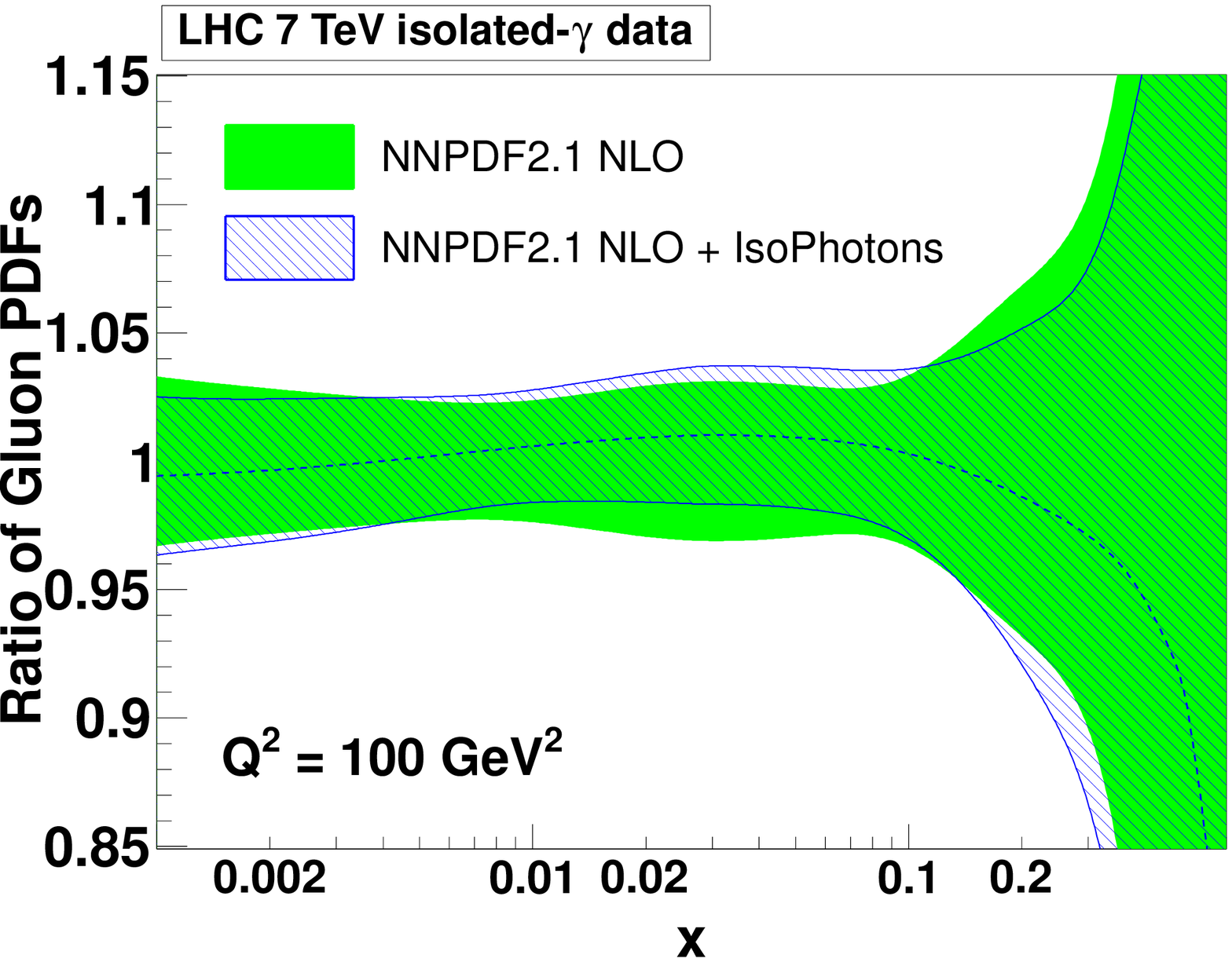}
\epsfig{width=0.325\textwidth,figure=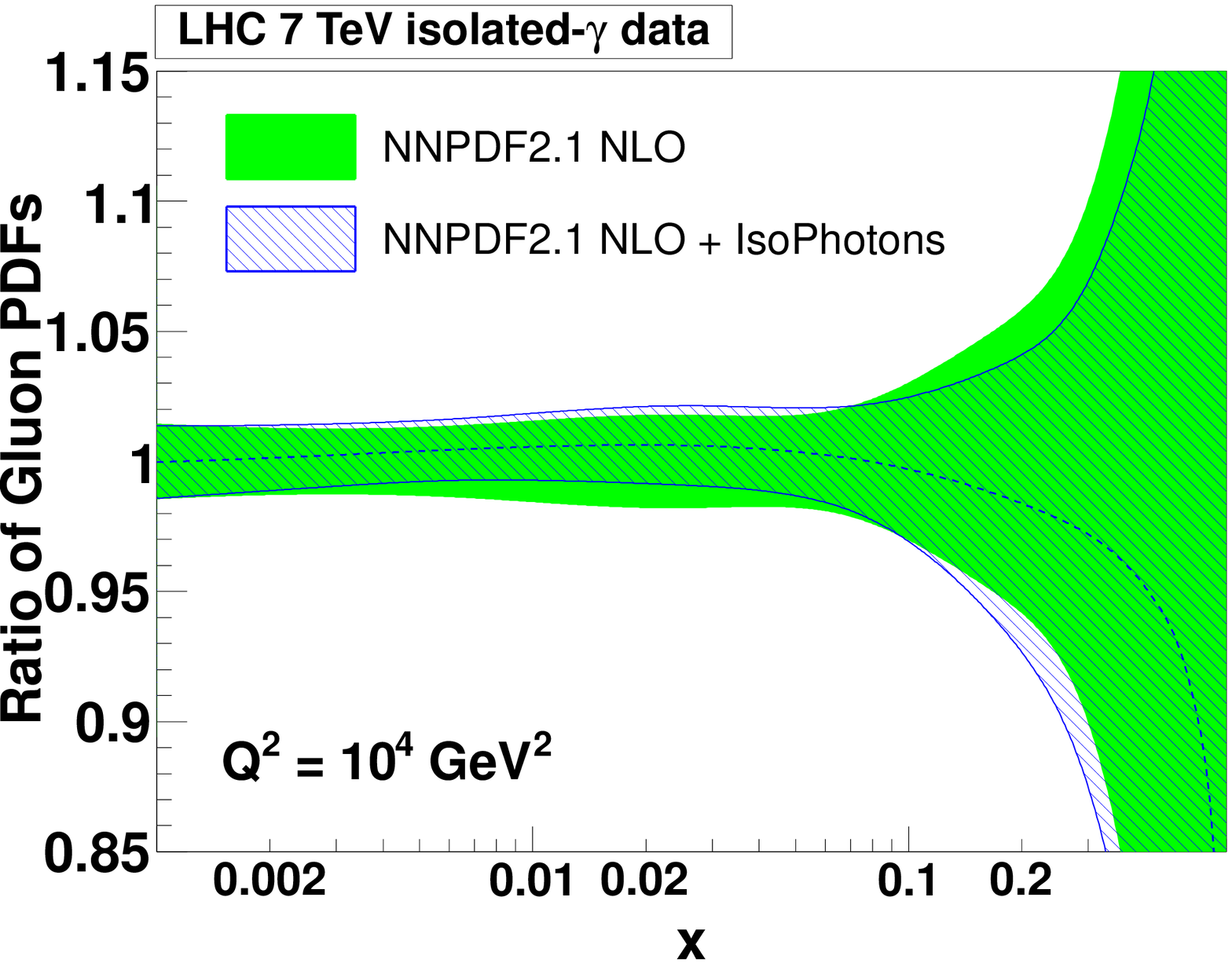}
\caption{\small Comparison between the NNPDF2.1 NLO
gluon before (green solid band) and after (dashed blue area) 
inclusion of the LHC-7~TeV isolated-$\gamma$ data,
with PDFs evaluated (left to right) at $Q$~=~3.16, 10, and 100~GeV.
\label{fig:gluon-scales}}
\end{figure}

The direct quantification of the impact on the $g(x,Q^2)$ distribution is shown 
in Figs.~\ref{fig:gluon}--\ref{fig:gluon-scales} where the NNPDF2.1 NLO gluon
is shown before and after including the isolated-photon data with reweighting.
In Figs.~\ref{fig:gluon}--\ref{fig:gluon2} the PDFs are evaluated at a typical LHC scale of $Q$~=~100~GeV.
We show both the relative improvement (left panels, ratio over the original 
gluon PDF) as well as the reduction in the absolute PDF errors (right panels).
Only the LHC data lead to a significant PDF uncertainty reduction, of up to 20 percent, localised at medium %and moderately large $x$, 
$x\approx$~0.002 to 0.05 (Fig.~\ref{fig:gluon-scales}). 
The Tevatron Run-II and the 630-GeV measurements bring rather small improvements
around $x\approx$~0.01--0.02, while the other datasets have negligible impact on the gluon central value and
associated uncertainties. 
This is consistent with the expectation that the effects should be maximal in phase-space regions 
where the photon production cross section depends more strongly on the gluon distribution (Fig.~\ref{fig:correlations2}). 
In the cases where the PDF uncertainties are reduced, the central value of $g(x,Q^2)$ is essentially 
unaffected (Figs.~\ref{fig:gluon}--\ref{fig:gluon2} left) indicating that the large-$x$ gluon determined from 
the Tevatron jet data is consistent with the large-$x$ gluon constrained by the LHC photon results. 
This is an important cross-check of the validity of pQCD factorisation and of the PDF 
universality using cross sections measured at hadronic colliders.
The effect of the LHC data on the gluon is similar at lower scales, where the impact of photon
data is shifted to somewhat larger values of $x$ as dictated by DGLAP evolution, 
as exemplified in Fig.~\ref{fig:gluon-scales}.\\

%%%%%%%%%%%%%%%
%%%%%%%%%%%%%%%%

\begin{figure}[htbp!]
\centering
\epsfig{width=0.60\textwidth,figure=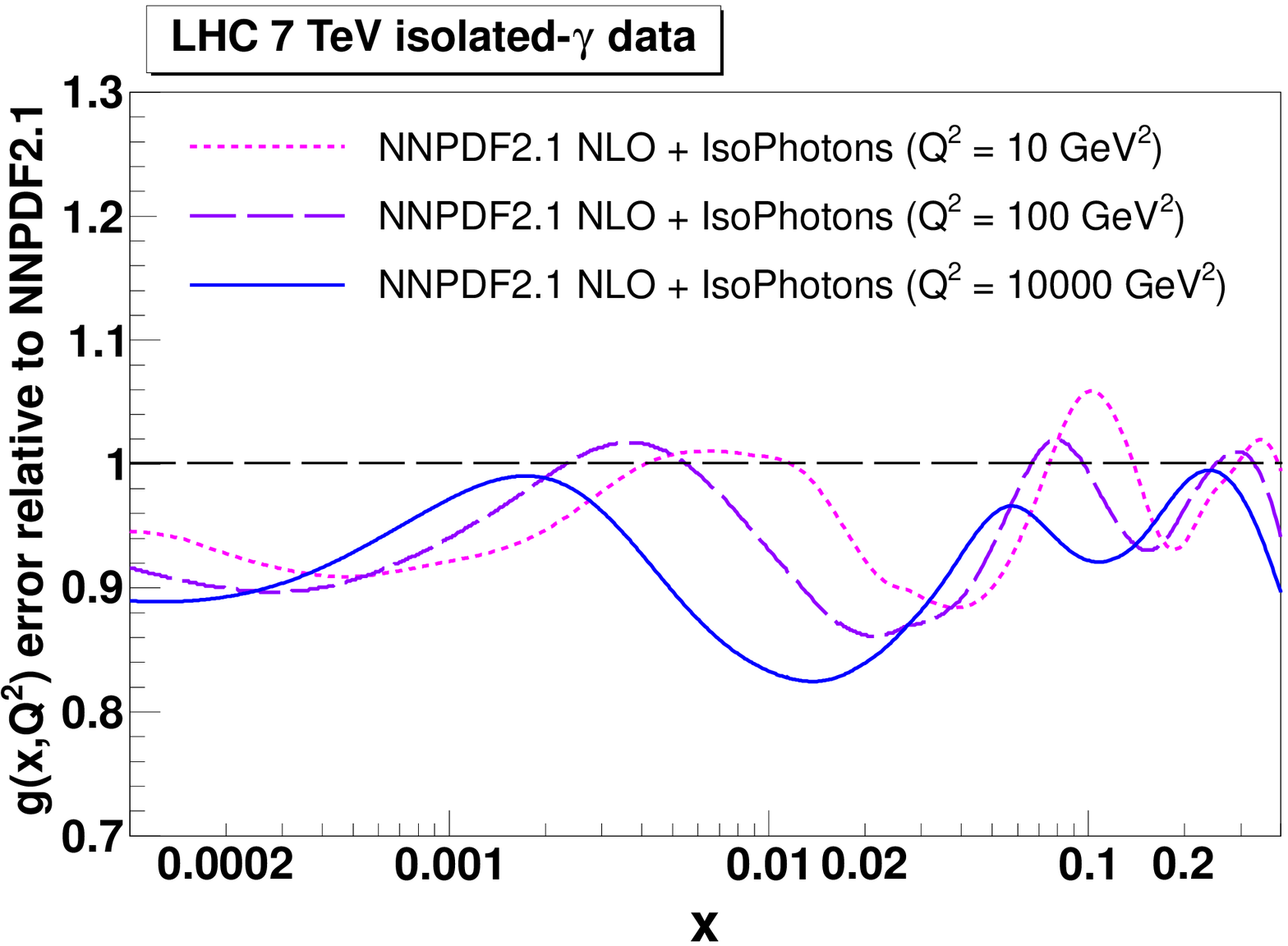}
\caption{\small Relative reduction of the NNPDF2.1 NLO gluon distribution uncertainty at scales $Q$~=~3.16, 10, and
  100~GeV, after inclusion of the LHC isolated-photon data via reweighting.
\label{fig:gluon-err-red}}
\end{figure}

The relative reduction of the uncertainties in the gluon distribution is better illustrated
in Fig.~\ref{fig:gluon-err-red}. For three scales $Q$~=~3.16, 10, and 100~GeV, the
maximum $g(x,Q^2)$ uncertainty reduction is 10--20\% in the region $x\approx$~0.01--0.05. 
For the lower scales, in some small regions of $x$, the PDF errors can be slightly increased when the photon
data are included. This is just a fluctuation effect, due to the limited statistics of the reweighting
procedure, that is quickly washed out by DGLAP evolution at higher scales.\\

%%%%%%%%%%%%%%%
%%%%%%%%%%%%%%%%

As a last check, we have confirmed that the constraints on the 
quark PDFs from the current isolated-$\gamma$ data are essentially negligible. 
This is not unexpected since light-quarks distributions are known more accurately 
than the gluon PDF in the kinematical region relevant for photon production (Fig.~\ref{fig:correlations}). 
However, the impact of the isolated-photon measurements would certainly be more
important in the so--called ``collider only'' fits~\cite{Ball:2011uy}, 
where the use of data from colliders alone leads to larger PDF
uncertainties on the quark sector (constrained mostly by fixed-target data in
pre-LHC PDF sets), since in this case the $q,\bar q$ densities could be also 
directly constrained.

%%%%%%%%%%%%%%%%%%%%%%%%%%%%%%%%%%%%%%%%%%%%%%%%%%%%%%%%%%%%%%%%%%%%%%%%%%%%%%%%%%%%%%%%
%\clearpage

\section{Predictions of the gluon-fusion Higgs boson production cross section}
\label{sec:higgs}

The dominant production channel for the SM Higgs boson at the LHC is gluon fusion~\cite{Dittmaier:2011ti,Dittmaier:2012vm} 
and thus theoretical predictions depend strongly on the gluon PDF choice as well as 
on the associated value of $\as$. Recently there has been an intense discussion 
in the literature~\cite{Baglio:2010ae,Baglio:2011wn,Alekhin:2011ey}
about the theoretical uncertainties to assign to the $\sigma(g\,g \to H)$ cross section
at hadron colliders. This has been motivated by the fact that predictions
from non-global PDF sets lead to very different results
than global PDF fits which are in reasonable agreement among each other~\cite{Watt:2011kp}. 
This has of course implications for the current Tevatron and 
LHC Higgs exclusion limits~\cite{CDFandD0:2011aa,Collaboration:2012si,Collaboration:2012tx}. 
Dedicated studies~\cite{Thorne:2011kq,NNPDF:2011aa,Ball:2011us} 
have shown that, within a global fit, the gluon PDF and the strong coupling
are stabilised by the inclusive jet data. It is thus of outmost importance
to find additional observables sensitive to $g(x,Q^2)$ 
to further improve the predictions for gluon-gluon Higgs cross sections, 
and  in this context isolated-photons appear as a promising candidate.\\

As shown in the previous section, the LHC isolated-photon data leads to a 
reduction of the $g(x,Q^2)$ uncertainties at around $x\approx$~0.02. 
Since this kinematical region is relevant for many important gluon--driven processes at the 
LHC, we study how the PDF uncertainties involved in the determination of the cross sections for various
processes, from Higgs boson to top-pair production~\cite{Demartin:2010er}, can be potentially reduced thanks
to the inclusion of isolated-$\gamma$ data. We have decided to consider only 7-TeV photon results 
since the constraints from measurements at the other c.m. energies are much milder and, in addition, 
LHC is the only collider for which no dataset needs to be discarded.
We show also that the central cross section predictions for Higgs production, 
mostly driven by the inclusive jet data in global PDF fits, remain stable when
isolated-photon data are included.\\

%%%%%%%%%%%%%%%%%%%%%%%%%%%%%%%
\begin{table}[htbp!]
\centering
\small
\begin{tabular}{c c c c c}\hline
Process / Cross section & $gg \to H$(120) & $gg \to H$(160) & $gg \to H$(200) & $gg\to H$(500) \\ \hline
NNPDF2.1             & 11640 $\pm$ 181 fb & 6052 $\pm$ 103 fb & 3494 $\pm$ 66 fb & 219.3 $\pm$ 8.3 fb  \\
NNPDF2.1 + LHC IsoPhotons &  11701 $\pm$ 140 fb  & 6073 $\pm$ 86 fb & 3504 $\pm$ 56 fb & 218.4 $\pm$ 7.6 fb \\
\hline \\ [1.ex] 
\hline
%Process & \hspace{1cm} $t\bar{t}$ \hspace{1cm} & \hspace{0.8cm} $t\bar{t}H$ \hspace{0.8cm} & \hspace{0.8cm} $WH$ \hspace{0.8cm} & \hspace{0.8cm} $ZH$ \hspace{0.8cm} \\ \hline
Process / Cross section & $t\bar{t}$ & $t\bar{t}\,H$(120) & $W\,H$(120) & $Z\,H$(120) \\ \hline
NNPDF2.1 & 162  $\pm$ 51 pb &  114 $\pm$ 5 fb & 447 $\pm$ 9 fb & 364 $\pm$ 6 fb  \\
NNPDF2.1 + LHC IsoPhotons & 162  $\pm$ 47 pb  &  113 $\pm$ 4 fb &  448 $\pm$ 9 fb  & 365 $\pm$ 6 fb\\\hline
\end{tabular}
\caption{\small NLO cross sections in \pp\ collisions at 7~TeV obtained with {\sc mcfm} using the NNPDF2.1
  set, before and after including the LHC isolated-photon data. 
Top: Higgs production in gluon-gluon fusion for different values of $M_{\rm H}$. 
Bottom: QCD top-pair production, and Higgs production ($M_{\rm H}$~=~120~GeV) associated with top-pairs and electroweak bosons.
\label{tab:xsec}}
\end{table}
%%%%%%%%%%%%%%%%%%%%%%%%%%%%%%%

\begin{figure}[htbp!]
\centering
\epsfig{width=0.63\textwidth,figure=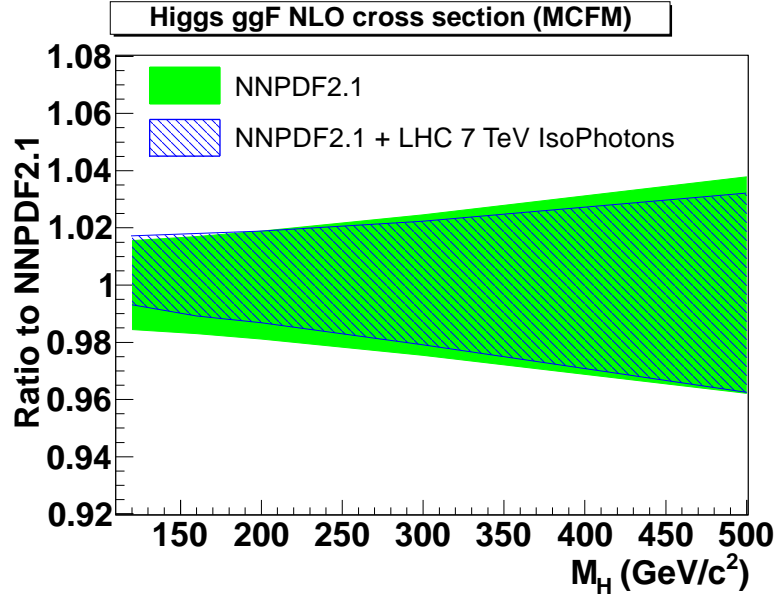}
\caption{\small \label{fig:dat-th} Ratio of Higgs
production gluon-fusion cross sections with NNPDF2.1 NLO PDFs before
and after including the LHC isolated-photon data.}
\end{figure}

The production cross sections for the following processes have been computed at
NLO\footnote{The NNLO/NLO K--factor, when available, is know to depend mildly on the choice of PDF. Therefore,
for the study of interest here (impact of the PDF error reduction on LHC cross sections)
NLO calculations are sufficient.} with the {\sc mcfm} code~\cite{MCFMurl,Campbell:2004ch,Campbell:2002tg}
for \pp\ collisions at $\sqrts$=~=7~TeV:
(i) Higgs boson in gluon fusion ($gg \to H$) for masses $M_{\rm H}=120$, 160, 200, 300 and
500~GeV, Higgs boson in association with (ii) top quark pairs ($t\bar{t}H$), and 
(iii) $W$ and $Z$ ($WH$, $ZH$); and (iv) top quark pair production ($t\bar{t}$).
All these processes are, either as signal or as background, relevant for Higgs 
searches at the LHC. The results for the cross sections obtained with
the NNPDF2.1 NLO set, before and after including the LHC isolated
photon data, are summarised in Table~\ref{tab:xsec}.
The results for Higgs boson production in gluon-gluon fusion
are also shown in Fig.~\ref{fig:dat-th}. While the most updated searches
exclude the SM Higgs boson with masses outside the
115--127~GeV range~\cite{Collaboration:2012si,Collaboration:2012tx}, 
we nevertheless show a wider mass range to illustrate the impact of the photon data.\\

%%%%%%%%%%%%%%%%%%%%%%%%%%%
%%%%%%%%%%%%%%%%%%%%%%%%%%%%

As one can see from Table~\ref{tab:xsec} and Fig.~\ref{fig:dat-th}, 
the inclusion of isolated-photon results in the NNPDF analysis leads to 
reduced PDF uncertainties in the %$g\,g -$Higgs 
gluon-gluon Higgs production cross section,
which is maximal for low Higgs masses (in the region not yet excluded
by the ATLAS and CMS limits) and can be as large as 20\%.
For other processes the improvements are much more modest. Indeed, processes with a pair of top-quarks
are produced with much larger virtualities than a (low-mass) Higgs and have reduced gluon uncertainties.
The same holds for processes with associated electroweak boson production which are dominated by quark PDFs.
In all cases, the central prediction is in good agreement with the reference NNPDF2.1
results, and thus consistent with the information obtained
from the inclusive jet data in a global fit.
All in all, these results demonstrate that isolated-photon data can be useful to reduce
the theoretical (PDF) uncertainties in the cross section for the many gluon-induced processes at the LHC.\\
%leading gluon-gluon Higgs production channel at the LHC. 
%If in addition isolated-$\gamma$ data are used to consistently determine the strong coupling, 
%the combined PDF+$\as$ uncertainty would be even smaller.

%%%%%%%%%%%%%%%%%%%%%%%%%%%%%%%%%%%%%%%%%%%%%%%%%%%%%%%%%%%%%%%%%%%%%%%%%%%%%%%%%%%%%%%%
\clearpage

\section{Summary and outlook}
\label{sec:summary}

Using up-to-date NLO pQCD theoretical calculations for isolated-photon production, 
as implemented in the \jetphox\ program combined with the NNPDF2.1 parton densities
and its associated PDF reweighting technique, we have quantified the impact on the gluon density 
of all the existing isolated-$\gamma$ data measured in \pp\ and \ppbar\ collisions at collider energies. 
The main results of this work can be summarised as follows:
\begin{itemize}
\item NLO pQCD provides generally a good description of the isolated-photon measurements from 200~GeV up to 7~TeV 
in a wide kinematic range of photon transverse energies and rapidities. A few outliers exist which can be identified 
as arising from problems in experimental datasets with underestimated systematic uncertainties.

\item The ATLAS and CMS measurements at the LHC, amounting to more than one hundred data-points, 
are consistent with each other and with pQCD and precise enough to moderately constrain the gluon 
PDF in the region $x\approx 0.002 - 0.05$ in a large range of scales $Q^2$~=~10--10$^4$ GeV$^2$. 
The central value of the gluon distribution is unmodified but its uncertainty is reduced by up to 20\% around
$x\approx$~0.02.

\item New PDFs including the LHC photon data lead to improved predictions for low mass Higgs production in the
  gluon-fusion channel, with central values of the theoretical cross section unmodified but with associated
  PDF uncertainties decreased by as much as 20\%. 
\end{itemize}

These results confirm that there is no reason not to include isolated-photon data measured at LHC energies 
into global-fit PDF analyses, on a similar footing as the current inclusive jet production data. 
The main difficulty to systematically carry out such a program is the slowness of the NLO pQCD codes to
compute the corresponding theoretical predictions, and thus an eventual fast interface 
{\it \`a la} {\sc ApplGrid} or \textsc{FastNLO} would be very advantageous.\\

Various possible future developments of this study are outlined next.
First, from the experimental side, measurements of the inclusive single isolated-photon spectra at high-$\ETg$
exploiting the much larger luminosity collected in 2011 by the ATLAS and CMS collaborations at 7 TeV, as well as data at 8~TeV in 2012 and eventually at 14~TeV 
%by all LHC experiments (we note that ALICE has potentially interesting capabilities at low-$\ETg$~\cite{Ichou:2010zz}
%and that LHCb can cover a much forward rapidity range)
would be very beneficial. The addition of a few hundred new data-points over a large
kinematic range, possibly including the enhanced $\ETg$--$y_\gamma$ range covered by ALICE and LHCb,
will certainly lead to even stronger constraints on the gluon density. 
The measurement of ratios of photon cross sections at 8 (or 14) to 7~TeV, where part of the experimental
uncertainties cancel out, and/or of the double-differential $\gamma$-jet spectra (with constrained kinematics by
the concurrent measurement of the photon and jet) can also provide new important constraints on $g(x,Q^2)$.
Second, at the lowest end of collider energies, the possibility of measuring isolated-$\gamma$ production in
the full range of the RHIC c.m. energies ($\sqrts\approx$~~20--500~GeV) would be also very useful to clarify
for once the long-standing disagreement between fixed-target data (as well as the oldest ISR and a few of the
\spps\ results) and NLO calculations. In all cases, the use of a smooth-cone prescription for the photon
isolation~\cite{Frixione:1998jh} which can effectively remove any remaining fragmentation-photon component
would be helpful to further eliminate theoretical uncertainties.
Third, it would be also important that future photon results provide the full covariance matrix, 
just as recent LHC measurements of electroweak boson~\cite{Aad:2011dm} and inclusive 
jet~\cite{Aad:2011fc} production have done, as this would improve the %usefulness 
quantitative treatment of these datasets in PDF analyses.\\

On the theoretical side, it would be useful to carry out a more quantitative study of the
%The main drawback of the present study is that we have not considered the 
possible impact of scale variations on the photon cross sections. It is conceivable that part of the impact of
the photon data is washed out once higher-order uncertainties are accounted for. This situation is common to 
all hadronic production data from colliders (notably, inclusive jets) included in global fits. 
One possibility to reduce this problem would be to include threshold resummation corrections. 
In any case, such a scale variation analysis should be consistently carried out 
together with all the other datasets. %included in the global analyses.
Once photon data are included in PDF analyses, it could be also
important to accurately and consistently determine the strong coupling 
in the framework of a PDF fit, just as inclusive jet production
is now instrumental in that regard~\cite{Lionetti:2011pw,Ball:2011us,Thorne:2011kq} as compared
to DIS--only fits. This is so because the gluon is loosely constrained
in DIS--only fits, where it has a runaway direction.\\

As a conclusion, we have shown that the available isolated-photon data provides constraints on the gluon PDFs
and thus on many relevant LHC processes, most importantly Higgs production in gluon-gluon fusion. 
Given that even more precise data as well as theoretical improvements  will be available in the next
future, we see no objection why isolated-photon data should not become integral part of future global QCD
analyses.\\

\vspace{0.5cm}
{\bf\noindent  Acknowledgments \\}
D.~d'E. is grateful to Fran\c{c}ois Arleo and Jean-Philippe~Guillet for many discussions and for 
implementing various improvements in the \jetphox\ code which facilitated significantly the 
computation of many thousands of theoretical spectra. D.~d'E. expresses his gratitude to 
Rapha\"{e}lle Ichou for her help on the compilation of the experimental isolated-photon data. %, and to Graeme~Watt for very useful exchanges. 
Valuable exchanges with Graeme Watt on PDF and NLO calculations are also acknowledged.
The research of J.~R. has been supported by a Marie Curie Intra--European Fellowship of the European 
Community's 7th Framework Programme under contract number PIEF-GA-2010-272515.

\bigskip

%%%%%%%%%%%%%%%%%%%%%%%%%%%%%%%%%%%%%%%%%%%%%%%%%%%%%%%%%%%%%%%%%%%%%%%%%%%%%%%%%%%%%%%%

%\bibliography{photonsNNPDF}

\end{document}